\newcommand*{\ATLASLATEXPATH}{}
\documentclass[cernpreprint,UKenglish,texlive=2011,txfonts]{\ATLASLATEXPATH atlasdoc}
\pdfoutput=1
\usepackage[backend=biber]{\ATLASLATEXPATH atlaspackage}
\usepackage{\ATLASLATEXPATH atlasbiblatex}

\usepackage{\ATLASLATEXPATH atlasphysics}
\graphicspath{{logos/}{figures/}{plots/}}

\definecolor{todoboxcolor}{rgb}{1.0,0.6,0.6}

\def\pythiaeight{P{\textsc{ythia}}\,8\xspace}
\def\herwig{H{\textsc{erwig}}\xspace}

\def\herwigpp{H{\textsc{erwig}}++\xspace}
\def\herwigseven{H{\textsc{erwig}}\,7\xspace}
\def\sherpa{S{\textsc{herpa}}\xspace}

\newcommand{\ztautau}{\ensuremath{Z\to \tau^+\tau^-}\xspace}

\newcommand{\zpt}{\ensuremath{p_\text{T}(\ell^+\ell^-)}\xspace}
\newcommand{\zptee}{\ensuremath{p_\text{T}(e^+ e^-)}\xspace}
\newcommand{\zptmumu}{\ensuremath{p_\text{T}(\mu^+ \mu^-)}\xspace}

\newcommand{\bt}{\ensuremath{\mathcal{B}}\xspace}

\newcommand{\nch}{\ensuremath{N_{\mathrm{ch}}}}

\newcommand{\sumpt}{\ensuremath{\sum\pt}\xspace}

\usepackage{mathtools}
\usepackage{rotating}
\usepackage{siunitx}

\hypersetup{pdftitle={Measurement of event-shape obervables sensitive to the underlying event in 
$Z \to \ell^{+} \ell^{-}$ events produced in proton-proton collisions at $\sqrt{s}=7$~TeV with 
the ATLAS detector},pdfauthor={The ATLAS Collaboration}}
\AtlasTitle{Measurement of event-shape observables in $Z \to \ell^{+} \ell^{-}$ events
in $pp$ collisions at $\sqrt{s}=7$~{\TeV} with the ATLAS detector at the LHC}

\author{The ATLAS Collaboration}

\AtlasJournal{Eur.\ Phys.\ Jour.\ C}

\AtlasAbstract{
Event-shape observables measured using charged particles in inclusive $Z$-boson
events are presented, using the electron and muon decay modes of the $Z$ bosons.
The measurements are based on an integrated luminosity of $1.1~{\rm fb}^{-1}$
of proton--proton collisions recorded by the ATLAS detector at the LHC at a
centre-of-mass energy $\sqrt{s}=7$~{\TeV}.
Charged-particle distributions, excluding the lepton--antilepton pair from
the $Z$-boson decay, are measured in different ranges of transverse momentum
of the $Z$ boson. Distributions include multiplicity, scalar sum of transverse
momenta, beam thrust, transverse thrust, spherocity, and $\mathcal{F}$-parameter,
which are in particular sensitive to properties of the underlying event at
small values of the $Z$-boson transverse momentum. The measured observables 
are compared with predictions from \pythiaeight, \sherpa, and \herwigseven.
Typically, all three Monte Carlo generators provide
predictions that are in better agreement with the data at high $Z$-boson
transverse momenta than at low $Z$-boson transverse momenta, and for the
observables that are less sensitive to the number of charged particles in
the event.
}

\begin{document}

\maketitle

%\tableofcontents

% List of contributors - print here or after the Bibliography
%\AtlasPrintContribute{0.3}
%\clearpage

%-------------------------------------------------------------------------------
%-------------------------------------------------------------------------------
\section{Introduction}
\label{sec:intro}
%-------------------------------------------------------------------------------

\noindent
The Large Hadron Collider (LHC) was primarily built to explore the mechanism 
of electroweak symmetry breaking and to search for new physics beyond the Standard 
Model (SM) in proton--proton collisions characterised by parton--parton scatterings
with a high momentum transfer.
These parton--parton scatterings are unavoidably accompanied by interactions between 
the proton remnants which are often called the ``underlying event'' (UE) and have to be 
modelled well in order to be able to measure high-momentum-transfer processes to high 
accuracy.

\noindent
Since the UE is dominated by low-scale strong-force interactions, in which 
the strong coupling strength diverges and perturbative methods of quantum 
chromodynamics (QCD) lose predictivity, it is extremely difficult to predict 
UE-sensitive observables from an ab-initio calculation in QCD. 
As a result, one has to rely on models implemented in general-purpose Monte 
Carlo (MC) event generators. Generators such as \herwigseven~\cite{Bellm:2015jjp}, 
\pythiaeight~\cite{Sjostrand:2007gs}, and \sherpa~\cite{Gleisberg:2008ta} 
contain multiple partonic interactions (MPI) as well as QCD radiation in the 
initial and final state to describe the UE. Certain aspects of the UE, e.g. 
the average transverse momenta of charged particles as a function of the 
charged-particle multiplicity, are better modelled by introducing in addition 
a mechanism of colour reconnection as in the event generators 
\pythiaeight and \herwigpp~\cite{Bahr:2008pv,Gieseke:2011na}/\herwigseven.
Such a mechanism is also implemented in \sherpa, but not activated 
by default and not used in ATLAS simulations using \sherpa.
It is impossible to unambiguously separate the UE from the hard scattering 
process on an event-by-event basis. However, distributions can be measured 
that are particularly sensitive to the properties of the UE.
Such measurements have been performed 
in proton--antiproton collisions in jet and in Drell--Yan production by the 
CDF experiment~\cite{Acosta:2004wqa,Aaltonen:2010rm} at centre-of-mass energies 
$\sqrt{s} =$ 1.8 and 1.96~TeV, 
and in proton--proton collisions at $\sqrt{s}=900$~GeV and 7~TeV by the
ATLAS experiment~\cite{Aad:2010fh,Aad:2011qe,Aad:2012jba,Aad:2013bjm,Aad:2014hia,Aad:2014jgf}, 
the ALICE experiment~\cite{ALICE:2011ac} and the 
CMS experiment~\cite{Khachatryan:2010pv,Chatrchyan:2011id,Chatrchyan:2012tb}. 

\noindent
This paper presents an analysis of event-shape observables sensitive to UE properties 
in 7\,\TeV{} proton--proton collisions at the LHC. The dataset of $1.1~{\text{fb}}^{-1}$ 
integrated luminosity was collected by the ATLAS detector~\cite{Aad:2008zzm} during 
data-taking in 2011, and events were selected by requiring a $Z$-boson candidate decaying 
to an $e^{+}e^{-}$ or $\mu^{+}\mu^{-}$ pair. Since the $Z$ boson is an object without 
colour charge, it does not affect hadronic activity in the collision and the observables 
were calculated using charged particles excluding the $Z$-boson decay products.
The charged-particle event-shape observables beam thrust, transverse thrust, spherocity, 
and $\mathcal{F}$-parameter as defined in Section~\ref{sec:eventshapes} were measured 
in inclusive $Z$ production. This paper contains information about aspects of the 
UE which were not explored by previous studies. The transverse thrust event-shape 
variable was measured by the CMS experiment~\cite{Chatrchyan:2013tna} in $Z$ events with 
at least one hard jet, with the goal of testing predictions from perturbative QCD. 
Since different hard process scales have different sensitivities to different aspects of 
the UE modelling, the observables were measured in the present paper in different ranges of 
the transverse momentum\footnote{The ATLAS reference system is a Cartesian right-handed 
coordinate system, with the nominal collision point at the origin. The anticlockwise beam 
direction defines the positive $z$-axis, while the positive $x$-axis is defined as pointing 
from the collision point to the centre of the LHC ring and the positive $y$-axis points upwards. 
The azimuthal angle $\phi$ is measured around the beam axis, and the polar angle $\theta$ is 
measured with respect to the $z$-axis. The pseudorapidity is given by $\eta = -\ln{\tan(\theta/2)}$ 
and the $\eta$--$\phi$ distance between two objects $i$ and $j$ by 
${\Delta R_{i,j}} = \sqrt{(\eta_{i}-\eta_{j})^2 + (\phi_{i}-\phi_{j})^2}$. The transverse 
momentum $p_\text{T}$ is defined relative to the beam axis.} of the $Z$-boson candidate, 
$\zpt$.\footnote{It is implicitly understood that the lepton--antilepton pair is produced
from a $Z$ boson or virtual photon $\gamma^{*}$.} At small $\zpt$ values, events are
expected to have low jet activity from the hard process and hence high sensitivity to UE 
characteristics. At high $\zpt$ values, the event is expected to contain at least one jet 
of high transverse momentum recoiling against the $\ell^{+}\ell^{-}$ system, which is 
expected to be reasonably described by perturbative calculations of the hard process.

\noindent 
The measured distributions have been corrected for the effects of pile-up (PU),
which are additional proton--proton interactions in the same LHC bunch crossing, 
for detector effects, and for the dominant background contribution from multijet events. 
The results are compared with the predictions of the MC event generators \pythiaeight, 
\herwigseven, and \sherpa.

\noindent
The paper is organised as follows:
Section~\ref{sec:eventshapes} introduces the event-shape observables and defines
the particle-level phase space used in this measurement.
Sections~\ref{sec:detector} and~\ref{sec:montecarlo} describe the ATLAS detector 
and the Monte Carlo event generators relevant to this analysis, which is described
in detail in~Section~\ref{sec:analysis}. The results are presented and discussed in
Section~\ref{sec:result} and summarised in~Section~\ref{sec:conclusion}.

\section{Event-shape observables}
\label{sec:eventshapes}

\noindent
The observables were calculated for primary charged particles with 
transverse momenta $p_\text{T}>0.5$~{\GeV} and pseudorapidities $|\eta| <2.5$.
Primary particles are defined as those with a decay distance $c\tau$ of at least 10 mm, 
either stemming from the primary proton--proton interactions or from the decays 
of shorter-lived particles from the primary proton--proton interactions.

\noindent
Distributions $f_{\mathcal{O}}={1}/{N_\text{ev}} \cdot{\text{d} N}/{\text{d} \mathcal{O}}$ were measured 
for all selected events, $N_\text{ev}$, for the following observables $\mathcal{O}$:
\begin{itemize}
    \item The charged-particle multiplicity, \nch.
    \item The scalar sum of transverse momenta of selected charged particles, $\sum\limits_{i} p_{\text{T},i} = \sumpt$.
    \item The beam thrust, \bt, as proposed in Ref.~\cite{Stewart:2009yx,Stewart:2010pd,Berger:2010xi}. 
          This is similar 
          to \sumpt except that in the sum over all charged particles the transverse momentum 
          of each particle is weighted by a factor depending on its pseudorapidity, $\eta$:
          \begin{equation}\label{eq:beamthrust}
             \bt = \sum_i p_{\text{T},i} \cdot {\text e}^{\: -|\eta_{i}|}.
          \end{equation}
          As a result, contributions from particles in the forward and backward direction
          (large values of $|\eta|$) are suppressed with respect to particles emitted at
          central pseudorapidities ($\eta\approx0$). The $\sumpt$ and \bt observables have 
          different sensitivities to hadronic activity from initial-state radiation. 
    \item The transverse thrust, $\mathcal{T}$, as proposed in Ref.~\cite{Banfi:2010xy}:
          \begin{equation}\label{eq:thrust}
          \mathcal{T}= \max_{\vec{n}_{\text T}}\frac{\sum\limits_{i} \left|\vec{p}_{\text{T},i} \cdot\vec{n}_{\text T}\right|}{\sum\limits_{i} p_{\text{T},i}}
          \end{equation}
          where the sum runs over all charged particles, and the thrust axis, $\vec{n}_{\text T}$,
          maximises the expression. The solution for $\vec{n}_{\text T}$ is found iteratively
          following the algorithm described in Ref.~\cite{Sjostrand:2006za} where one starts
          with a direction $\vec{n}_{\text T}^{(0)}$ and obtains the $j+1$ iteration as 
          \begin{equation}
          \vec{n}_{\text T}^{(j+1)}= \frac{\sum\limits_{i} \epsilon\left(\vec{n}_{\text T}^{(j)} \cdot \vec{p}_{\text{T},i} \right) \vec{p}_{\text{T},i}}{\left| \sum\limits_{i} \epsilon\left(\vec{n}_{\text T}^{(j)} \cdot \vec{p}_{\text{T},i} \right) \vec{p}_{\text{T},i} \right|}
          \end{equation}
          where $\epsilon(x)=1$ for $x>0$ and $\epsilon(x)=-1$ for $x<0$.
    \item The spherocity, $\mathcal{S}$, as proposed in Ref.~\cite{Banfi:2010xy}: 
          \begin{equation}\label{eq:spherocity}
          \mathcal{S}=\frac{\pi^2}{4}\underset{\vec{n}=(n_x, n_y,0)^\top}{\min} 
          \left(
          \frac{\sum\limits_{i} \left|\vec{p}_{\text{T},i}\times\vec{n}\right|}{\sum\limits_{i} p_{\text{T},i}}
          \right)^2
          \end{equation}
          where the sum runs over all charged particles and the vector $\vec{n}$ minimises the expression.
          In contrast to the closely related sphericity observable~\cite{Banfi:2010xy,Cuautle:2014yda}, 
          which is computed via a tensor diagonalisation, spherocity is simple to calculate since $\vec{n}$ 
          always coincides with one of the transverse momentum vectors $\vec{p}_{\text{T},i}$ ~\cite{Banfi:2010xy}.
    \item The $\mathcal{F}$-parameter defined as the ratio of the smaller and larger eigenvalues, 
          $\lambda_1$ and $\lambda_2$,
          \begin{equation}\label{eq:f-parameter}
          \mathcal{F}=\frac{\lambda_{1}}{\lambda_{2}}
          \end{equation}
          of the transverse momentum tensor
          \begin{equation}\label{eq:transversemomentumtensor}
          M^\text{lin} = \sum_i\frac{1}{p_{\text{T},i}}
          \left(\begin{matrix}
          p_{x,i}^2 & p_{x,i}p_{y,i} \\
          p_{x,i}p_{y,i} & p_{y,i}^2
          \end{matrix}\right)
          \end{equation}
          where the sum runs over the charged particles in an event.
\end{itemize}
\noindent
Pencil-like events, {e.g.} containing two partons emitted in opposite directions
in the transverse plane,
are characterised by values of $\mathcal{S}$, $\mathcal{T}$, and $\mathcal{F}$ 
close to 0, 1, and 0 respectively. The corresponding values of these observables 
for spherical events, e.g. containing several partons emitted isotropically, 
are close to 1, $2/\pi$, and 1 respectively. While the event-shape observables 
$\mathcal{S}$, $\mathcal{T}$, and $\mathcal{F}$ show very high correlations
among themselves, they are weakly correlated with \nch, \sumpt, and beam thrust.

\noindent
The observables were calculated after removing the $Z$-boson decay products.  
The fiducial $Z$-boson phase-space region requires a decay into a pair of 
oppositely charged leptons, either electrons or muons,\footnote{If not stated 
explicitly, ``electrons'' and ``muons'' denote both the corresponding lepton 
and antilepton.} where each lepton must have $p_\text{T} > 20$~GeV and 
$|\eta| < 2.4$, with a lepton--antilepton invariant mass in the interval 
$[66,116]$~GeV. This mass window contains the $Z$-resonance peak and is wide 
enough to allow the multijet background to be determined from the sideband regions.
        
\noindent
Each observable was determined in the following ranges of the transverse momentum 
of the $Z$ boson, $\zpt$, calculated from the four-momenta of the lepton and 
antilepton: 0--6\;\GeV, 6--12\;\GeV, 12--25\;\GeV, and $\ge 25$~\GeV. As mentionned 
in Section~\ref{sec:intro}, events at small $\zpt$ are expected to be particularly 
sensitive to the UE activity, while events with large $\zpt$ values ($\ge 25$~\GeV) 
are expected to contain significant contributions from jet production coming from 
the hard scattering process. The lowest $\zpt$ range (0--6\;\GeV) was chosen 
accordingly as a compromise between small bin size and minimising migration effects.
The ranges at higher $\zpt$ were each defined so as to contain about the same number 
of events as the 0--6\;{\GeV} range.

\noindent
In simulated events, particle-level leptons are defined as so-called dressed leptons, 
obtained by adding to the stable lepton four-momentum the four-momenta of any photons 
within a cone of ${\Delta R}_{\ell,\gamma} = 0.1$~\cite{Aad:2011qv} and which do not
stem from hadron or $\tau$ decays.

%-------------------------------------------------------------------------------
\section{ATLAS detector}
\label{sec:detector}
%-------------------------------------------------------------------------------

\noindent
The ATLAS detector, described in detail in Ref.~\cite{Aad:2008zzm},
covers almost the full solid angle around the collision point. 
The components relevant to this analysis are the tracking 
detectors, the liquid-argon (LAr) electromagnetic sampling calorimeters
(ECAL) and the muon spectrometer (MS).

\noindent
The inner tracking detector (ID), consisting of a silicon pixel detector 
(pixel), a silicon microstrip tracker (SCT) and a straw-tube transition radiation 
tracker (TRT), covers the full azimuthal angle $\phi$ and the pseudorapidity 
range $|\eta| \le 2.5$.
These individual tracking detectors are placed from inside to outside at a 
radial distance $r$ from the beam line of 50.5--150 mm, 299--560 mm and 563--1066 mm 
respectively, within a 2 T axial magnetic field generated by a solenoid surrounding 
the ID. The inner detector barrel (end-caps) consists of 3 ($2 \times 3$) pixel layers, 
4 ($2 \times 9$) layers of double-sided SCT silicon microstrip modules, and 73 ($2 \times 160$) 
layers of TRT straw-tubes. The typical position resolutions of these subdetectors are 
$10~\si{\micro\metre}$, $17~\si{\micro\metre}$ and $130~\si{\micro\metre}$ respectively
for the $r$--$\phi$ coordinates. 
The pixel and SCT detectors provide $r$--$z$ coordinate measurements with typical resolutions 
of $115~\si{\micro\metre}$ and $580~\si{\micro\metre}$ respectively. 
The TRT covers $|\eta| \le 2.0$. A charged particle traversing the barrel part of the ID 
leads typically to 11 silicon hits (3 pixel clusters and 8 microstrip clusters) and 
more than 30 straw-tube hits.

\noindent
A high-granularity lead/liquid-argon electromagnetic sampling calorimeter~\cite{Aad:2010ai}
covers the pseudorapidity range $|\eta| \le 3.2$. Hadronic calorimetry in the range 
$|\eta| \le 1.7$ is provided by an iron/scintillator-tile calorimeter, consisting of 
a central barrel and two smaller extended barrel cylinders, one on either side of the 
central barrel. In the end-caps ($|\eta| \ge 1.5$), the acceptance of the LAr hadronic 
calorimeters matches the outer $|\eta|$ limits of the end-cap electromagnetic 
calorimeters. The LAr forward calorimeters provide electromagnetic and hadronic 
energy measurements, and extend the coverage to $|\eta| \le 4.9$.

\noindent
The muon spectrometer measures the deflection of muons in large superconducting 
air-core toroid magnets in the pseudorapidity range $|\eta| \le 2.7$. It is instrumented 
with separate trigger and high-precision tracking chambers. Over most of the $\eta$ range, 
a precision measurement of the track coordinates 
is provided by monitored drift tubes. Cathode strip 
chambers with higher granularity are used in the innermost plane over the range 
$2.0 \le |\eta| \le 2.7$, where particle fluxes are higher.

\noindent
The trigger system utilises two stages: a hardware-based Level-1 trigger followed by 
a software-based high-level trigger, consisting of the Level-2 and Event Filter~\cite{Aad:2012xs} 
stages. In the Level-1 trigger, electron candidates are selected by requiring that 
the signal in adjacent electromagnetic calorimeter trigger towers exceed a certain 
transverse energy, $E_{\text T}$, threshold, depending on the detector $\eta$. 
The Event Filter uses the offline reconstruction and identification algorithms to apply 
the final electron selection in the trigger. The $Z \rightarrow e^{+}e^{-}$ events were 
selected in this analysis by using a dielectron trigger in the region $|\eta| \le 2.5$ 
with an electron transverse energy threshold of 12 GeV for each electron. 

The muon trigger system, which covers the pseudorapidity range $|\eta| \le 2.4$, 
uses the signals of resistive-plate chambers in the barrel ($|\eta|<1.05$) and 
thin-gap chambers in the end-cap regions ($1.05 < |\eta| < 2.4$). 
The $Z \rightarrow \mu^{+}\mu^{-}$ events in this analysis were selected with a 
trigger that requires the presence of at least one muon candidate reconstructed 
in the muon spectrometer with transverse momentum of at least 11 GeV at Level-1 
and 18 GeV at the Event Filter stage.

%-------------------------------------------------------------------------------
\section{Monte Carlo simulations}
\label{sec:montecarlo}
%-------------------------------------------------------------------------------

\noindent
Monte Carlo simulated samples for the signal and the various background processes 
were generated at particle level before being passed through a 
G{\textsc{eant}}4-based~\cite{Agostinelli:2002hh} simulation of the ATLAS detector 
response ~\cite{SOFT-2010-01} followed by the detector reconstruction. 
These samples were used to correct the measured observables for detector effects 
and to estimate related systematic uncertainties. 

\noindent
The signal process was simulated with two different event generators in order to 
quantify the model uncertainty in the correction of the measured distributions to 
particle level: the leading-order (LO) generator P{\textsc{ythia}} 8.150 using 
the CTEQ6L1~\cite{Pumplin:2002vw} parton distribution functions (PDFs), and the LO generator
\sherpa 1.3.1 using the CT10 next-to-leading-order (NLO) PDF set~\cite{Lai:2010vv}. 

\noindent
For the \pythiaeight samples, inclusively produced $Z \rightarrow \ell^{+} \ell^{-}$ 
events were generated. The \pythiaeight generator uses a leading-logarithm 
$p_\text{T}$-ordered parton shower (PS) model which is matched to LO matrix element 
calculations. Multiple partonic interactions are phenomenologically modelled by 
perturbative QCD parton--parton scattering processes down to an effective $p_\text{T}$ 
threshold (Sj\"ostrand--van Zijl model~\cite{Sjostrand:2004pf}) accompanied by the mechanism 
of colour reconnection of colour strings. The phenomenological description of hadronisation 
is implemented using the Lund string model~\cite{Andersson:2002ap}.
The \pythiaeight samples were generated with model parameters tuned to Tevatron and
earlier LHC data (4C tune~\cite{ATL-PHYS-PUB-2012-003}).

\noindent
For the \sherpa signal samples, tree-level matrix elements for 
$pp \rightarrow Z + X, Z \rightarrow \ell^{+} \ell^{-}$ were used with up to five
additional final-state partons. 
The model used for MPI in \sherpa is also based on the Sj\"ostrand--van Zijl model, 
but the mechanism of colour reconnection is not activated. Hadronisation modelling 
uses a cluster hadronisation scheme.

\noindent
The background processes ($t\bar{t}$, $Z \rightarrow \tau^{+} \tau^{-}$, $ZZ$, 
and $WZ$ production) relevant to the analysis were generated with \sherpa version 
1.4.0 in the case of $Z \rightarrow \tau^{+} \tau^{-}$, $ZZ$, and $WZ$ production, 
and with version 1.3.1 in the case of $t\bar{t}$ production using in both cases the 
CT10 NLO PDF set. The default parameter tuning performed by the \sherpa authors was used.

\noindent
The events of the MC signal samples were generated with and without overlaid simulated 
pile-up events in order to validate the data-driven PU correction method with simulated 
events. The \pythiaeight generator (version 8.150 with the CTEQ6L1~\cite{Pumplin:2002vw} 
PDF and 4C tune) was used to simulate the pile-up events. The number of PU events 
overlaid was chosen to reproduce the average number of proton--proton collisions per 
bunch crossing observed in the data analysed.

\noindent
For comparison with corrected distributions, three different, recent versions
of MC event generators were used to provide predictions for the signal at particle level:
\sherpa 2.2.0 with up to two additional partons at NLO and with three additional 
partons at LO and taking the NLO matrix element calculations for virtual contributions 
from OpenLoops~\cite{Cascioli:2011va} with the NNPDF 3.0 NNLO PDF set~\cite{Ball:2014uwa};
P{\textsc{ythia}} 8.212 with LO matrix element calculations using the 
NNPDF2.3 LO PDF set~\cite{Ball:2013hta}; and \herwig 7.0~\cite{Bellm:2015jjp} taking 
the NLO matrix element calculations for real emissions from MadGraph~\cite{Alwall:2014hca} 
and for virtual contributions from OpenLoops using the 
MMHT2014 PDF set~\cite{Harland-Lang:2014zoa}. 
The \herwigseven event generator implements a cluster hadronisation 
scheme with parton showering ordered by emission angle. All the parameters relevant to 
the UE modelling were set to values chosen by the corresponding MC generator authors: 
while these were the default values in \sherpa and \herwigseven, for \pythiaeight the 
Monash 2013 tune to LHC data was chosen for the settings of the UE parameters~\cite{Skands:2014pea}. 
The A14 \pythiaeight tune of the ATLAS collaboration~\cite{ATL-PHYS-PUB-2014-021} gives 
predictions for the event-shape observables which are very close to, and differ by at 
most $5\%$ from, the ones obtained by the Monash 2013 tune.

\noindent
The treatment of QED radiation is generator-specific and modelled differently in 
\pythiaeight compared to \sherpa and \herwigseven. The latter radiate more 
soft-collinear and wide-angle photons than \pythiaeight, as a result of their usage 
of a YFS-based model~\cite{Yennie:1961ad} for QED emissions.

%-------------------------------------------------------------------------------
\section{Analysis}
\label{sec:analysis}
%-------------------------------------------------------------------------------

\noindent
Since the track-based observables are sensitive to pile-up effects, 
the analysis was restricted to a subsample of $1.1~{\text{fb}}^{-1}$ 
integrated luminosity of the 2011 dataset, in which the mean number of 
$pp$ collisions per bunch crossing was typically only around five and 
not larger than seven. 
With this dataset the results are in most cases already dominated by 
systematic uncertainties. After the event and track selection the 
event-shape observables were corrected first for PU and then for 
background contributions, and finally corrected for detector effects.

\subsection{Event selection}
\label{sec:eventselection}

\noindent
Only events containing a ``primary vertex'' (PV) as defined below were processed,
to reject events from cosmic-ray muons and other non-collision background. 
A reconstructed vertex must have at least one track with a minimum $p_\text{T}^\text{trk}$ 
of 400 MeV from the region inside the detector where the collisions take place.
The PV is defined as the vertex with the highest $\sum ({p_\text{T}^\text{trk}})^2$ 
value of tracks associated with the vertex.

\noindent
Selected electrons and muons were required to have a $p_{\text{T}}$ 
of at least 20~{\GeV} and a pseudorapidity $|\eta| < 2.4$. In the case of electrons, the $\eta$ 
range $1.37 < |\eta| < 1.52 $ was excluded in order to avoid large amounts of passive detector 
material in the region between the barrel and end-cap ECAL. Electron candidates were identified 
using information from the shower shape in the ECAL, from the association between ID tracks 
and ECAL energy clusters, and from the number of transition radiation hits in the TRT~\cite{PERF-2013-03}. 
Muon candidates were built from track segments in the MS matched to tracks in the ID~\cite{PERF-2014-05}. 
Electron candidates were required to have a transverse impact parameter with respect to the PV
of $|d_0| < 5$~mm and muon candidates of $|d_0| < 3 \times \sigma_{d_0}$, with $\sigma_{d_0}$ being 
the transverse impact parameter resolution of the muon candidate. In addition, muon candidates 
had to pass the longitudinal impact parameter requirement $|z_0| < 10$~mm.
While no isolation criterion was required for muon candidates, the selection requirements for 
electron candidates contain implicitly some isolation cuts. Only events containing exactly one 
pair of oppositely charged leptons passing the selection cuts as described above were considered. 
These were treated as $Z \rightarrow \ell^{+} \ell^{-}$ signal events if the $\ell^{+}\ell^{-}$ 
invariant mass was in the region $m_{\ell^{+}\ell^{-}} \in[66,116]$~\GeV. After all selection 
requirements, about $2.6 \times 10^{5}$ electron--positron events (``electron channel analysis'') 
and $4.1 \times 10^{5}$ muon--antimuon events (``muon channel analysis'') remained.

\subsection{Track selection}
\label{sec:trackselection}

\noindent
To calculate the event-shape observables for charged particles, tracks fulfilling the following 
criteria, identical to those used in Ref.~\cite{Aad:2010ac}, were selected:
\begin{enumerate}
    \item at least one hit in the pixel subdetector;
    \item a hit in the innermost pixel layer if the reconstructed trajectory
          traversed an active pixel module;
    \item at least six SCT hits;
    \item the transverse momentum of the track $p_\text{T}^\text{trk} > 0.5$~{\GeV};
    \item the pseudorapidity of the track $|\eta^\text{trk}| < 2.5$;
    \item the transverse impact parameter of the track with respect to the PV $|d_0| < 1.5$~mm;
    \item the longitudinal impact parameter of the track 
          with respect to the PV $|z_0| \sin\theta < 1.5$~mm;
    \item a goodness-of-fit probability greater than 0.01 for tracks with
          with $p_\text{T}^\text{trk} > 10$~\GeV.
\end{enumerate}
The first two requirements greatly reduce the number of tracks from non-primary particles,
which are those originating from particle decays and interactions with material in the inner 
detector.
The third one imposes an indirect constraint on the minimum track length and hence on the 
precision of the track parameters. The kinematic requirements (4. and 5.) imposed on the 
track selection are driven by the $\eta$-acceptance of the inner detector and the need
for an approximately constant reconstruction efficiency as a function of $p_\text{T}^\text{trk}$. 
The impact parameter requirements (6. and 7.) aim to suppress tracks not originating from 
the PV of the event.
The cut on the goodness-of-fit probability reduces the fraction of mismeasured tracks at 
high $p_\text{T}^\text{trk}$ values. 
With these requirements except for 4., the track reconstruction efficiency rises in the 
$|\eta^\text{trk}|<1.0$ range from $80\%$ at $p_\text{T}^\text{trk}=$ 400 MeV to around 
$90\%$ at $p_\text{T}^\text{trk}=$ 5 GeV and then stays constant. 
For higher $|\eta^\text{trk}|$ values the efficiency variation is stronger: 
at $|\eta^\text{trk}|=2.5$ the efficiency rises from around $50\%$ at 
$p_\text{T}^\text{trk}= 400$ MeV to around $80\%$ at 5 GeV.

\subsection{Lepton track removal}
\label{sec:leptontrackremoval}

\noindent
Since this analysis aims to measure charged-particle distributions, the decay products 
of the $Z$-boson were removed from the set of tracks used to calculate the observables. 
Electrons can interact with the material in front of the ECAL leading to multiple 
tracks as a result of bremsstrahlung and photon conversion.
Hence, tracks were not used in the calculation of each event-shape variable
if they fell inside a cone of ${\Delta R}_{e,\mathrm{trk}} = 0.1$ around 
any selected electron or positron. 
In order to treat the electron and muon channel analyses as similarly as possible,
this approach was also applied to the muon channel.
It was checked that the observables changed in data and in simulated signal 
samples in the same way within statistical uncertainties when the cone size 
was varied within a factor of two.

\subsection{Pile-up correction}
\label{sec:pileupcorrection}

\noindent
If another proton--proton interaction is spatially close to the primary interaction
where the $Z$-boson is produced, it is possible that the vertex algorithm 
assigns tracks from the PU interaction to the reconstructed primary vertex. 
The PU correction used in this analysis is based on the ``Hit Backspace Once More'' 
(HBOM) approach~\cite{Monk:2011pg}, which relies on recursively applying a smearing 
effect to a measured distribution, in this case the effect from the contamination 
by tracks selected from pile-up. 
An event-shape distribution without pile-up tracks, $f_{\mathcal{O}}^{0}$, 
is changed to an event-shape distribution, $f_{\mathcal{O}}^{1}$, when 
pile-up tracks that are passing the selection cuts are taken into account 
in the calculation of the event-shape observables. By adding once more 
pile-up tracks one obtains a distribution, $f_{\mathcal{O}}^{2}$. This 
procedure can be repeated $k$ times, resulting in the distribution 
$f_{\mathcal{O}}^{k}$. Knowing $f_{\mathcal{O}}^{k}$ as a function of $k$ 
allows one to extrapolate from the PU-contaminated distribution 
$f_{\mathcal{O}}^{k=1}$ to $f_{\mathcal{O}}^{k=0}$, hence to the 
distribution without PU contamination. In the analysis, the $k$-th 
application of the PU effect on an event-shape observable was parameterised 
by an $n^{\mathrm {th}}$-order polynomial function, $P(k)$, in the
following called HBOM parametrisation. 
The procedure was carried out in each individual bin of the event-shape 
observables using the Professor toolkit~\cite{Buckley:2009bj} to determine 
the parameters of $P(k)$ by means of a singular value 
decomposition~\cite{Brown:1972:COM:1480083.1480176}.

\noindent
The PU effect on the observables was estimated by constructing a library of 
``pseudo-vertices'' containing tracks passing the track selection requirements 
with respect to vertices that are well isolated from the PV and any other vertex 
(see Section~\ref{sec:eventselection}). Typically, these vertices originate from 
PU and are therefore called PU vertices in the following.
In addition to the track parameters, the library also stores the position of 
the corresponding PU vertex along the beam-line, $z_\text{vtx}^{\text{PU}}$. 
All vertices of events passing the nominal event selection were potential candidates 
for the library. However, to safeguard against cases in which a single vertex is falsely 
reconstructed as two or more vertices close in $z$ (``split vertices'') it was required 
that the selected vertices have a minimum distance along the beam line from any other 
vertex, $\Delta z_\text{min}^\text{vtx}$, of 60~mm. In the process of building a 
pseudo-vertex at $z_\text{vtx}^{\text{PU}}$, tracks were required to satisfy
\begin{equation}\label{eq:trackselwindow}
   \left| \left( z_\text{vtx}^\text{PU} - z_{0,\text{trk}} \right) \, \sin\theta_\text{trk} \right| < 3~\text{mm}.
\end{equation}
\noindent
This selection window is larger than the nominal track selection window with respect to the PV
in order to account for the possibility that the PV marginally overlaps with a pseudo-vertex. 
Parameters of each track fulfilling the requirements above were stored to form the pseudo-vertex.

\noindent
The effect of the pile-up contamination was then quantified as follows:
\begin{enumerate}
    \item For each event, draw a random number, $N_\text{rdm}$, from the distribution of the 
          number of vertices per event.
    \item Obtain $N_\text{rdm}$ random vertex positions, $z_{\text{rdm}, i}$ ($i=1, ..., N_\text{rdm}$), 
          from the distribution of reconstructed pile-up vertices fulfilling the
          $\Delta z_\text{min}^\text{vtx}$ requirement, and for each of those, a random pseudo-vertex 
          from the library entry corresponding to $z_{\text{rdm}, i}$, each containing an independent 
          number of tracks.
    \item Any track $j$ belonging to such a selected pseudo-vertex $i$ with a longitudinal impact 
          parameter with respect to the pseudo-vertex $z_{0,ij}^\text{PU} \, \sin\theta_\text{trk}^{ij}$ 
          is then added to the list of an event's signal tracks if it falls in the signal track selection 
          window
\begin{equation}\label{eq:trackselwindow_PU} 
\left|\,\, \left( z_{\text{rdm}, i} + {z_{0,ij}^\text{PU}} - z_\text{PV}  \right) \, \sin\theta_\text{trk}^{ij} \,\, \right| < 1.5~\text{mm}.
\end{equation}
\end{enumerate}
\noindent
With these additional tracks each observable was then re-calculated to 
determine $f_{\mathcal{O}}^{k}$ for $k=2, ..., 11$. The HBOM parameterisation 
for $f_{\mathcal{O}}^{k}$ as a function of $k$ was parameterised by a 
third-order polynomial used to extrapolate to $k=0$.

\noindent
The PU correction varies when changing the random seed of the selection. 
To reflect the statistical nature of the PU correction, ten different
statistically independent versions of the PU correction were determined.
The final PU correction was the mean of these ten PU corrections.

\noindent
Using a library of pseudo-vertices built from detector-simulated PU events 
(see Section~\ref{sec:montecarlo}), four tests were performed to validate 
the PU correction method.
\begin{enumerate}
\item In the first ``forward-closure'' test, the effect of PU contamination
      in the event-shape observables as modelled by the HBOM parameterisation 
      was applied to a simulated sample without PU events overlaid by adding 
      to each event-shape observable $f_{\mathcal{O}}$ binwise the term 
      $P(1)-P(0)$. It was found that event-shape observables obtained in this 
      way were in very good agreement with those obtained where PU events were overlaid. 
      Only in the charged-multiplicity bin $N_{\mathrm{trk}}=0$ was a sizeable 
      non-closure of the order of 10\% (22\%) to 20\% (34\%) in the muon (electron) 
      channel observed. This effect is likely 
      caused by an unavoidable bias in the vertex selection for the PU library 
      and was considered as a systematic uncertainty.
\item In the second ``backward-closure'' test, the HBOM parameterisation was used 
      to correct event-shape observables in simulated samples containing PU events 
      to distributions without PU effect. The results were found to be in very 
      good agreement with the corresponding samples without PU events overlaid. 
      As in the ``forward-closure'' test, the only non-closure was observed in 
      the charged-multiplicity bin $N_{\mathrm{trk}}=0$. 
\item In the third test, the selection cuts defining the PU library were varied 
      and no significant deviations beyond the systematic uncertainties assigned 
      to the HBOM method were observed.
\item The $z_{\text{rdm}}$ distribution of the pseudo-vertices in the library is 
      similar but not identical to the $z^\text{PU}_\text{vtx}$ distribution of
      all PU vertices. 
      In the fourth test, the $z^\text{PU}_\text{vtx}$ distribution was used 
      instead of the $z_{\text{rdm}}$ distribution and again the PU-corrected 
      result was found to be in very good agreement with the corresponding 
      samples without PU events overlaid.
\end{enumerate}
While for $\nch$ the PU correction varied from $20\%$ at low multiplicities to 
$40\%$ at high multiplicities, the PU corrections for all other event-shape 
observables were at most $15$--$20\%$ for both the electron and the muon channel.

\subsection{Background treatment}
\label{sec:backgroundtreatment}

\noindent
In addition to $Z \rightarrow \ell^{+} \ell^{-}$ events the following background 
sources were assumed to contribute to the signal region: events from multijet 
production with misidentified lepton candidates or leptons from decays of hadrons, 
production of $t \bar{t}$ quark pairs, production of $Z$ bosons decaying into a 
$\tau^{+} \tau^{-}$ pair with subsequent decays to electrons or muons, and diboson 
production $ZZ$ and $WZ$ with gauge-boson decays into leptons.

\noindent
All background contributions were found to be small compared to the number of 
$Z \rightarrow \ell^{+} \ell^{-}$ events, with the most prominent contribution coming 
from multijet events. 
While the effect of multijet events was estimated from data and corrected for, 
no explicit correction was made for the other background sources because their 
contribution was found to be very small: using MC simulation the background fraction 
from $t \bar{t}$, $Z \rightarrow \tau^{+} \tau^{-}$, and diboson production $WZ$ and 
$ZZ$ was estimated to be about $0.25\%$ for the complete $Z$-boson transverse momentum 
phase space. 
About $70\%$ of these background contributions ($ZZ$ production as well as \ztautau events) 
had event-shape distributions very similar to the ones of the signal process. 
The fraction of $t \bar{t}$ ($WZ$) background, showing significantly different 
event-shape distributions in the MC simulation compared to the signal process, 
was found to be $0.04$--$0.05\%$ ($0.03\%$) in the full $\zpt$ spectrum. 
Since these background fractions are very small and other systematic uncertainties 
significantly larger, no correction for $t \bar{t}$ and $WZ$ background was applied.

\noindent
In both lepton channels, the relative number of multijet events as well as their 
event-shape observables were estimated from data as described below. The measured, 
PU-corrected event-shape observables $f_{\mathcal{O}}^\text{meas}$ were then 
corrected by applying bin-wise the multiplicative factor 
$1-{f_{\mathcal{O}}^\text{multijet}}/{f_{\mathcal{O}}^\text{meas}}$ where
$f_{\mathcal{O}}^\text{multijet}$ represents the estimate of the event-shape 
observable for multijet events.
 
\noindent
Modified event and/or lepton selections for the electron and muon channels, as 
described below, were performed to obtain the dilepton invariant mass distributions, 
$m_{\ell\ell}^\text{multijet}$, dominated by contributions from multijet events. 
These distributions were fitted using a linear function, 
$g^\text{multijet}(m_{\ell\ell})$, omitting the peak region $m_{\ell\ell}^\text{multijet}\in[77,97]$~{\GeV} 
to avoid a fit bias from remaining peaking signal contributions.
Assuming that only multijet events contribute to these samples, the integral, $I^\text{multijet}$, 
of the fit function over the whole signal window ($m_{\ell\ell}^\text{multijet}\in[66,116]$~\GeV) 
was used to estimate the amount of multijet background entering the signal region. 
The event-shape distributions obtained with the modified selection criteria were used 
as an estimate of the corresponding multijet background shape and were then scaled so 
as to match the total amount of the multijet background, $I^\text{multijet}$.
This procedure was performed for all $\zpt$ ranges separately since the amount 
of the multijet background depends on $\zpt$ and rises with increasing 
$\zpt$. For the fully inclusive distributions, it amounted to $0.7\%$ in the 
electron channel and to $1.9\%$ in the muon channel.

\noindent                                                                                   
In the electron case, two different samples with either different event selection 
criteria or different lepton selection criteria were considered in estimating the 
number of multijet events and the distributions of their event-shape observables.
In the first sample, the lepton-pair selection was changed from opposite-sign to 
same-sign charged electrons (i.e. an electron--electron or positron--positron pair). 
Drell--Yan contributions to this multijet-enriched sample were estimated to be of 
the order $15\%$. This sample was used to estimate the number of multijet events 
and their event-shape observables as described above, assuming the same selection 
efficiency for multijet events in the opposite-sign and same-sign electrons 
selection. In addition, opposite-sign and same-sign electron events were selected 
with significantly looser electron selection requirements to obtain a second
multijet-enriched sample. With the second sample, it was verified that the 
opposite-sign and same-sign requirements select nearly equal numbers of multijet 
events and that the event-shape distributions for multijet background agree 
for the opposite-sign and same-sign electron selections. 
The multijet background correction factors for the electron channel were found 
to be very close to one, where the largest change in the event-shape observables 
was not more than $3\%$.

\noindent
In the muon case, an isolation criterion, which is based on the scalar sum of transverse 
momenta of tracks found in a cone in $\eta$--$\phi$ space around the muon, was introduced 
to obtain a sample with a much smaller multijet background contribution. The fraction of 
multijet background was then determined by subtracting the $m_{\mu\mu}$ distribution for 
the isolated muon selection, assuming negligible contributions from multijet events, from 
the one for the standard muon selection, since the two have very similar 
$Z \rightarrow \mu^{+}\mu^{-}$ selection efficiencies. Contributions from signal events 
to this multijet-dominated distribution were estimated to be of the order of $5\%$. 
The event-shape distributions of multijet background were estimated accordingly by subtracting 
the event-shape distributions for the isolated muon selection from the one of the standard 
selection. Compared to the electron channel, the multijet background correction factors in 
the muon channel were found to deviate significantly more from one and to show more functional 
dependence in the event-shape distributions.

\noindent
As a cross-check of the background subtraction procedure the reconstructed event-shape 
distributions were measured for smaller $m_{\ell\ell}$ signal window widths of $30$, 
$20$, and $10$ GeV while using the background estimate from the standard $m_{\ell\ell}$ 
selection applied to the narrower $m_{\ell\ell}$ signal window.
By narrowing the $m_{\ell\ell}$ window, the signal-to-background ratio is increased and 
as a result the effect from background becomes smaller. Differences seen in some 
individual bins were found to be much smaller than the systematic uncertainties, and no 
systematic dependence of the event-shape distributions as a function of the $m_{\ell\ell}$ 
window size was observed.

\subsection{Unfolding}
\label{sec:unfolding}

\noindent
The observables were measured in different $\zpt$ ranges and corrected for contributions 
from non-primary particles, detector efficiency and resolution effects using an unfolding 
technique. 

\noindent
The bin sizes for the distributions of the event-shape observables were chosen taking into 
account two aspects: to have a fine enough binning to best see the shape of each distribution, 
and to have enough events in each bin, particularly in the tails of the distributions.
It was explicitly checked with unfolding closure tests as described below that the bin sizes 
were not too small compared to the experimental resolution.

\noindent
For the unfolding of the measured observables a Bayesian approach was 
applied~\cite{2010arXiv1010.0632D}. The unfolding procedure requires an input distribution
(called the prior distribution), which was taken from MC signal samples, and the detector 
response matrix $M_{ij}$. The matrix, $M_{ij}$, determined using simulated signal samples, 
quantifies the probability that an event with the event-generator value (at particle level) 
in bin $i$ of a distribution is reconstructed in bin $j$. 
Since the unfolding result depends on the prior distribution, the Bayesian unfolding is performed 
in an iterative way until convergence, minimising the dependence on the prior distribution. 
For the iterative Bayesian unfolding the I{\textsc{magiro}} framework~\cite{Wynne:2012jb} 
was used, with improvements, as proposed in Ref.~\cite{Adye:2011gm}, to the error calculation 
in the orginal work described in Ref.~\cite{2010arXiv1010.0632D}. The number of iteration 
steps in the I{\textsc{magiro}} framework is obtained in an automatised way. Distributions 
of \pythiaeight events at reconstruction level were unfolded with a detector response matrix 
obtained with simulated \sherpa events and vice versa. The level of agreement of the unfolded 
distributions with the particle distributions of the corresponding event generator was 
quantified by a $\chi^2$ test and a Kolmogorov--Smirnov (KS) test. 
The optimal number of iteration steps was set to the number of iteration steps 
for which the minimum (maximum) of the $\chi^2$ (KS) test statistic was observed in 
the simulation. In general, the optimal number of iteration steps was found to be two,  
except for $\sum p_\text{T}$ in the $\zpt$ bin 12--25\;\GeV, in which case it was three.

\noindent
Since corrections were made for the effect of pile-up on the observables before unfolding,
the simulated signal samples used for the prior distribution and the detector response matrix 
did not contain pile-up events. Signal samples generated with either \pythiaeight or with
\sherpa were used to determine the prior distribution and the detector response matrix.
The results of the unfolding obtained with these two simulations were then averaged.

\noindent
The complete analysis chain was tested on reconstructed MC signal samples simulated with 
either \pythiaeight or \sherpa with overlaid pile-up events generated by \pythiaeight. 
The event-shape observables were corrected for pile-up using the same strategy as in data. 
The resulting distributions were then unfolded using detector response matrices and priors 
obtained from the MC signal samples without pile-up. In general, the unfolding results
showed good closure: the corrected MC distributions were found to be in very good agreement 
with the particle-level distributions. This was also the case when events generated by 
\pythiaeight were unfolded with \sherpa prior distributions and \sherpa detector response 
matrices and vice versa.

\subsection{Systematic uncertainties}
\label{sec:systematicuncertainties}

\noindent
Several categories of systematic uncertainties that influence the distributions after
corrections and unfolding were quantified. 

\begin{itemize}
\item {\bf Lepton selection:}\\
\noindent
Uncertainties in the lepton selection affect not only the selected events but also 
the reconstructed $\zpt$ in data and simulation, and hence are important for the 
unfolding where the subdivision of the data into different $\zpt$ ranges is performed. 
Variations were performed for each source of systematic uncertainty and were 
propagated through the unfolding to estimate their effect on the results.

\noindent
For the electron channel, systematic uncertainties in the energy resolution, 
the energy scale, and the trigger, reconstruction and identification efficiencies 
were quantified~\cite{PERF-2013-03,PERF-2010-04}. 
The largest effect on the event-shape observables was observed from the electron energy 
scale systematic uncertainties. The total effect was typically in the subpercent range 
and therefore much smaller than the statistical and other systematic uncertainties. 

\noindent
For the muon channel, systematic uncertainties in the observables from the efficiencies 
(reconstruction and trigger) as well as from the calibration of the reconstructed muon 
transverse momentum~\cite{PERF-2014-05} were also typically below the percent level.

\item {\bf Track reconstruction:}\\
\noindent
In order to estimate the effect of uncertainty in the track reconstruction efficiency 
on the observables, the data distributions were unfolded with a modified detector 
response matrix taking into account variations of the track reconstruction efficiencies.

\noindent
The relative track reconstruction efficiency systematic uncertainties were estimated 
as a function of $\pt^{\text{trk}}$ and $|\eta^{\text{trk}}|$:
\begin{itemize}
    \item For tracks with $|\eta^{\text{trk}}|<2.1$ the relative uncertainty was estimated to be
          $1.5\%$ for tracks with $p_\text{T}^{\text{trk}}$ in the range 500--800\;{\MeV} 
          and $0.7\%$ for all tracks with $p_\text{T}^{\text{trk}} >800$~\MeV~\cite{Aad:2010ac}.
    \item For tracks with $|\eta^{\text{trk}}| \ge 2.1$ several effects were assessed to
          quantify the systematic uncertainty~\cite{Aad:2014xca}: uncertainties 
          in the modelling of the detector material in particular in the vicinity 
          of service structures and cooling pipes (4--7\%), systematic uncertainties 
          in the track selection related to the requirements on the transverse impact 
          parameter and on the innermost pixel layer to suppress charged particles 
          stemming from interactions with the detector material (1\%), the fraction 
          of mismeasured tracks for transverse momenta above 10 GeV (1.2\% between 
          10 and 15 GeV, up to 80\% above 30 GeV at high $|\eta^{\text{trk}}|$ values), and the 
          systematic uncertainty due to the goodness-of-fit probability cut
          to reduce mismeasured tracks above 10 GeV (10\%).
\end{itemize}

\noindent
The systematic uncertainty in the track reconstruction efficiency was generally found 
to be the dominant systematic uncertainty for observables where the number of charged 
particles does not cancel in the definition ($N_\text{ch}$, $\sum p_\text{T}$, 
beam thrust) and reached as high as 10\%. For all other observables, it was typically 
between 1\% and 3\%. The contribution was of the same order when comparing unfolded 
distributions from the electron channel and the muon channel.

\item {\bf Non-primary particles:}\\
\noindent
The effect from non-primary particles, which are those originating from decays and 
interactions with material in the inner detector, was taken into account by the 
unfolding procedure. The fraction and composition of non-primary particles in data 
is not perfectly modelled by the MC simulation, which is able to reproduce the 
fraction in data to an accuracy of about $10$--$20\%$ as a result of a fit to the 
$d_{0}$ distribution~\cite{Aad:2014jgf}.
To estimate the corresponding systematic uncertainty, the requirement on the track 
impact parameter $|d_{0}|$ was varied from the nominal value of 1.5\;mm downward 
to 1.0\;mm and upward to 2.5\;mm, resulting in a 0.5--4\% change in the fraction 
of the non-primary particles~\cite{Aad:2014jgf}. 
The resulting event-shape distributions were unfolded using MC signal samples 
selected with the same impact parameter requirements to test the stability of 
the unfolding result. The maximum residual difference was taken as the systematic 
uncertainty from the impact parameter requirement. The typical relative 
uncertainty was $2\%$ or smaller, except for a few individual bins.

\item {\bf Pile-up correction:}\\
\noindent
The standard deviation of the mean PU correction obtained from the ten independent 
PU corrections was considered as a systematic uncertainty of statistical nature. 

\noindent
The default HBOM parameterisations used third-order polynomials giving a very good 
description of the pile-up effect. Similarly good descriptions were obtained by 
fourth-order polynomials. The differences between using third-order and fourth-order 
polynomials were used to quantify the systematic uncertainty coming from the choice 
of HBOM parameterisation, resulting in systematic uncertainties in the event-shape 
observables typically below 2\%.

\noindent
In contrast to a $\chi^2$ fit, the singular value decomposition used to obtain the 
polynomial parameterisation does not take into account uncertainties. Hence, there 
is no \textit{a priori} goodness-of-fit measure for the parameterisation. If the 
polynomial $P(k)$ provides a good prediction of each HBOM point, $f_{\mathcal O}^{k}$, 
and if each HBOM point fluctuates around $P(k)$ with the same uncertainty $\sigma$, 
then one expects $\sum_{k=1}^{11}(P(k)-f_{\mathcal O}^{k})^2/\sigma^{2}=11$.
This equation was used to estimate the size of such a typical uncertainty $\sigma$ 
for each bin of the observables. The so-determined average uncertainty was then taken 
as a systematic uncertainty for the HBOM extrapolation. This systematic uncertainty 
is similar in size to the variation from third-order to fourth-order polynomials.

\noindent
A further check was made by omitting the $k$-th point when calculating the parameterisations. 
In each bin, the largest deviation of these extrapolations from the nominal extrapolation 
was taken as a systematic uncertainty. This deviation was found to rarely exceed $1\%$ and 
hence is negligible in most bins.

\noindent
To obtain the total uncertainty of the method, the four systematic uncertainties were 
added in quadrature.

\noindent
The $N_\text{trk}=0$ bin showed a bias in the MC tests due to the track and vertex 
selections, leading to a sizeable non-closure for this particular bin. An additional 
correction for this expected non-closure as determined from simulation was performed 
and the full size of the correction was applied as an additional uncertainty.

\noindent
The systematic uncertainty in the pile-up correction propagated through the unfolding 
led to a systematic uncertainty in the event-shape observables of 1\% to 3\% with the 
exception of some bins with few events.
In general, fewer events in a given bin corresponded to a larger systematic uncertainty 
in the PU correction. The PU correction systematic uncertainty was found to have 
negligible dependence on $\zpt$. The results for the electron and muon channel were 
of comparable magnitude.

\item {\bf Multijet background correction:}\\
\noindent
For the electron channel, a systematic uncertainty was assigned to the shape 
of the multijet background event-shapes by taking into account the differences 
between the distributions obtained with the same-sign and opposite-sign events
with the loosened electron selection criteria.

\noindent
In order to estimate the systematic uncertainty in the multijet background in 
the muon channel, the calculation of the multijet background correction factors 
was repeated for several variations of the isolation critera.
The largest difference per bin from the central isolation was taken as the systematic 
uncertainty.

\noindent
The systematic uncertainty in the background correction was found to be negligible 
in almost all bins of all observables. Similar to the pile-up correction systematic 
uncertainty, significant contributions were observed in bins with few events.

\item {\bf Unfolding:}\\
\noindent
The model uncertainty in the unfolding was estimated by using \pythiaeight and 
\sherpa separately for the prior distribution and the detector response matrix. 
The systematic uncertainty corresponding to the unfolding with different priors and 
detector response matrices was taken from the differences between the central value 
and the individual results obtained with \pythiaeight and \sherpa.

\noindent
For most observables, the unfolding model error was of the order of 1\% or below,
except for poorly populated bins in which it can reach up to 15\%. The sizes observed 
in the electron and the muon channels were found to be in good agreement.
\end{itemize}

\noindent
The total systematic uncertainties were constructed by adding the above systematic
uncertainties in quadrature. The systematic uncertainties in the electron channel 
were typically slightly larger than the ones obtained in the muon channel.
They are of the order of 5\% to 10\% for those observables where the track reconstruction 
systematic uncertainties are large ($N_\text{ch}$, $\sum p_\text{T}$, beam thrust).
For all other observables the systematic uncertainties rarely exceed 5\% and are 
typically of the order of 2\%. Tables~\ref{tab:SystematicUncertainties_zpt_0-6}--\ref{tab:SystematicUncertainties_zpt_ge25} 
provide an overview of the range of the relative statistical and systematic uncertainties 
for all six observables separately for the electron channel and the muon channel in the 
four $\zpt$ ranges. All systematic uncertainties except the lepton-specific uncertainties 
are highly correlated between the electron channel and the muon channel.
\begin{table}
\centering
\begin{tabular}{rr|*{9}{c}}
\toprule
Observable    & Channel  & ${\delta^\text{stat}_{\mathcal{O}}}$ & ${\delta^\text{Lepton}_{\mathcal{O}}}$ & ${\delta^\text{Tracking}_{\mathcal{O}}}$ & ${\delta^\text{Non-Prim.}_{\mathcal{O}}}$ & ${\delta^\text{PU}_{\mathcal{O}}}$ & ${\delta^\text{Multijet}_{\mathcal{O}}}$ & ${\delta^\text{Unfold}_{\mathcal{O}}}$ \\
              &          &  $[\%]$                                                & $[\%]$                                                   & $[\%]$                                                     & $[\%]$                                                      & $[\%]$                                               & $[\%]$                                                   & $[\%]$  \\
\midrule
$N_\text{ch}$ & ($e^{+}e^{-}$)     & $1$--$5$     & $0.2$--$0.6$  & $<0.1$--$9$    & $0.1$--$2.5$  & $0.5$--$28$   & $<0.1$--$0.6$  & $0.2$--$8.4$ \\
              & ($\mu^{+}\mu^{-}$) & $0.8$--$4.3$ & $0.1$--$0.5$  & $0.3$--$9.9$   & $0.1$--$2.1$  & $0.2$--$19$   & $<0.1$--$0.4$ & $0.1$--$4.4$ \\
$\sumpt$      & ($e^{+}e^{-}$)     & $1$--$3$     & $0.1$--$0.5$  & $0.3$--$5.5$   & $<0.1$--$1.3$ & $0.13$--$6.8$ & $0.01$--$0.4$  & $<0.1$--$0.8$ \\
              & ($\mu^{+}\mu^{-}$) & $0.8$--$2.4$ & $0.1$--$0.5$  & $0.3$--$5.3$   & $<0.1$--$1.3$ & $0.2$--$3.5$  & $<0.1$--$0.3$  & $<0.1$--$1$ \\
\bt           & ($e^{+}e^{-}$)     & $0.8$--$14$  & $0.1$--$2.4$  & $<0.1$--$6.2$  & $0.1$--$2.1$  & $0.1$--$36$   & $<0.1$--$2.1$  & $0.2$--$2.9$ \\
              & ($\mu^{+}\mu^{-}$) & $0.6$--$9.5$ & $0.1$--$2.0$  & $<0.1$--$5.8$  & $<0.1$--$4.5$ & $0.2$--$14$   & $<0.1$--$1.6$  & $0.1$--$5.9$ \\
$\mathcal{T}$ & ($e^{+}e^{-}$)     & $0.6$--$4.4$ & $0.1$--$0.5$  & $0.2$--$2.2$   & $0.1$--$1.6$  & $0.1$--$4.7$  & $0.1$--$0.3$   & $0.1$--$2.6$ \\
              & ($\mu^{+}\mu^{-}$) & $0.5$--$3.5$ & $0.1$--$0.6$  & $0.1$--$2.0$   & $0.1$--$1.2$  & $0.1$--$4.0$  & $<0.1$         & $0.2$--$2.9$ \\
$\mathcal{S}$ & ($e^{+}e^{-}$)     & $0.6$--$3.8$ & $0.1$--$0.4$  & $0.3$--$2.6$   & $0.1$--$1.4$  & $0.1$--$4.3$  & $0.1$--$0.4$   & $0.1$--$2.2$ \\
              & ($\mu^{+}\mu^{-}$) & $0.5$--$3.0$ & $0.1$--$0.4$  & $0.1$--$1.9$   & $0.1$--$1.8$  & $0.1$--$4.1$  & $<0.1$         & $0.1$--$5.4$ \\
$\mathcal{F}$ & ($e^{+}e^{-}$)     & $0.6$--$3.6$ & $0.1$--$0.5$  & $0.3$--$1.6$   & $0.1$--$1.5$  & $0.1$--$1.7$  & $0.1$--$0.3$   & $0.1$--$2.0$ \\
              & ($\mu^{+}\mu^{-}$) & $0.5$--$2.9$ & $0.1$--$0.3$  & $0.1$--$1.9$   & $0.1$--$1.2$  & $0.1$--$1.6$  & $<0.1$         & $0.1$--$1.9$ \\
\bottomrule
\end{tabular}
\caption{Ranges of the relative uncertainties $\frac{\delta_{\mathcal{O}}}{\mathcal{O}}$ of the event-shape observables 
$\mathcal{O}$ for the electron and muon channels indicated by ($e^{+}e^{-}$) and ($\mu^{+}\mu^{-}$) for the $\zpt$ 
range 0--6\;{\GeV} in percent. The superscripts denote the statistical (`stat') and the individual systematic
uncertainties in the lepton reconstruction and identification (`Lepton'), track reconstruction efficiency (`Tracking'),  
non-primary particles (`Non-prim.'), pile-up correction (`PU'), multijet background (`Multijet'), and the unfolding (`Unfold').}
\label{tab:SystematicUncertainties_zpt_0-6}
\end{table}

\begin{table}
\centering
\begin{tabular}{rr|*{9}{c}}\toprule
Observable    & Channel  & ${\delta^\text{stat}_{\mathcal{O}}}$ & ${\delta^\text{Lepton}_{\mathcal{O}}}$ & ${\delta^\text{Tracking}_{\mathcal{O}}}$ & ${\delta^\text{Non-Prim.}_{\mathcal{O}}}$ & ${\delta^\text{PU}_{\mathcal{O}}}$ & ${\delta^\text{Multijet}_{\mathcal{O}}}$ & ${\delta^\text{Unfold}_{\mathcal{O}}}$ \\
              &          &  $[\%]$                                                & $[\%]$                                                   & $[\%]$                           \
                          & $[\%]$                                                      & $[\%]$                                               & $[\%]$                         \
                            & $[\%]$  \\
\midrule
$N_\text{ch}$ & ($e^{+}e^{-}$)     & $1$--$10$    & $0.1$--$2.2$  & $0.2$--$10$    & $0.2$--$6.6$  & $0.1$--$24$   & $<0.1$--$0.2$   & $<0.1$--$10$ \\
              & ($\mu^{+}\mu^{-}$) & $0.8$--$8.4$ & $0.1$--$1.8$  & $<0.1$--$11.4$ & $0.1$--$4.5$  & $0.6$--$21$   & $<0.1$--$0.4$   & $0.7$--$7.7$ \\
$\sumpt$      & ($e^{+}e^{-}$)     & $1$--$2.3$   & $0.1$--$0.5$  & $0.1$--$5.3$   & $<0.1$--$1.9$ & $0.4$--$2.9$  & $<0.1$--$0.3$   & $<0.1$--$1.8$ \\
              & ($\mu^{+}\mu^{-}$) & $0.8$--$1.8$ & $0.1$--$0.6$  & $<0.1$--$4.9$  & $<0.1$--$1.4$ & $0.1$--$3.2$  & $<0.1$--$0.3$   & $0.1$--$1.7$ \\
\bt           & ($e^{+}e^{-}$)     & $0.7$--$8.8$ & $0.1$--$1.5$  & $0.2$--$4.3$   & $0.1$--$1.5$  & $<0.1$--$19$  & $<0.1$--$1$     & $<0.1$--$2.4$ \\
              & ($\mu^{+}\mu^{-}$) & $0.6$--$6.7$ & $0.1$--$1$    & $0.3$--$3.9$   & $<0.1$--$1.9$ & $0.1$--$10$   & $<0.1$--$0.6$   & $0.1$--$2.4$ \\
$\mathcal{T}$ & ($e^{+}e^{-}$)     & $0.6$--$4.7$ & $0.1$--$0.5$  & $0.2$--$2.2$   & $0.1$--$1.5$  & $0.1$--$2.9$  & $0.1$--$0.5$    & $0.1$--$2.5$ \\
              & ($\mu^{+}\mu^{-}$) & $0.5$--$3.7$ & $0.1$--$1$    & $0.2$--$2.8$   & $0.1$--$1$    & $0.1$--$4.4$  & $<0.1$          & $0.2$--$2.7$ \\
$\mathcal{S}$ & ($e^{+}e^{-}$)     & $0.6$--$3.6$ & $0.1$--$0.3$  & $0.2$--$2.4$   & $0.1$--$1.6$  & $0.1$--$5.0$  & $0.1$--$0.4$    & $0.2$--$3.4$ \\
              & ($\mu^{+}\mu^{-}$) & $0.5$--$2.9$ & $0.2$--$0.7$  & $0.2$--$2.2$   & $0.1$--$1.1$  & $0.1$--$4.4$  & $<0.1$          & $0.1$--$3.1$ \\
$\mathcal{F}$ & ($e^{+}e^{-}$)     & $0.6$--$3.8$ & $0.1$--$0.4$  & $0.1$--$2.0$   & $0.1$--$0.9$  & $0.1$--$7.4$  & $0.1$--$0.4$    & $0.2$--$2.7$ \\
              & ($\mu^{+}\mu^{-}$) & $0.5$--$3.0$ & $0.1$--$0.6$  & $0.1$--$2.4$   & $0.1$--$1.3$  & $0.1$--$1.6$  & $<0.1$          & $0.1$--$3.2$ \\
\bottomrule
\end{tabular}
\caption{Ranges of the relative uncertainties $\frac{\delta_{\mathcal{O}}}{\mathcal{O}}$ of the event-shape observables
$\mathcal{O}$ for the electron and muon channels indicated by ($e^{+}e^{-}$) and ($\mu^{+}\mu^{-}$) for the $\zpt$
range 6--12\;{\GeV} in percent. The superscripts denote the statistical (`stat') and the individual systematic
uncertainties in the lepton reconstruction and identification (`Lepton'), track reconstruction efficiency (`Tracking'),
non-primary particles (`Non-prim.'), pile-up correction (`PU'), multijet background (`Multijet'), and the unfolding (`Unfold').}
\label{tab:SystematicUncertainties_zpt_6-12}
\end{table}

\begin{table}
\centering
\begin{tabular}{rr|*{9}{c}}\toprule
Observable    & Channel  & ${\delta^\text{stat}_{\mathcal{O}}}$ & ${\delta^\text{Lepton}_{\mathcal{O}}}$ & ${\delta^\text{Tracking}_{\mathcal{O}}}$ & ${\delta^\text{Non-Prim.}_{\mathcal{O}}}$ & ${\delta^\text{PU}_{\mathcal{O}}}$ & ${\delta^\text{Multijet}_{\mathcal{O}}}$ & ${\delta^\text{Unfold}_{\mathcal{O}}}$ \\
              &          &  $[\%]$                                                & $[\%]$                                                   & $[\%]$                           
                          & $[\%]$                                                      & $[\%]$                                               & $[\%]$                         
                            & $[\%]$  \\
\midrule
$N_\text{ch}$ & ($e^{+}e^{-}$)     & $1$--$18.8$   & $0.1$--$2.8$   & $0.24$--$9.9$  & $0.14$--$4.5$ & $0.2$--$22$   & $<0.1$--$0.5$  & $0.1$--$4.7$ \\
              & ($\mu^{+}\mu^{-}$) & $0.8$--$14.3$ & $0.1$--$1.9$   & $0.15$--$9.2$  & $0.2$--$1.6$  & $0.1$--$18$   & $<0.1$--$0.6$  & $<0.1$--$3.7$ \\
$\sumpt$      & ($e^{+}e^{-}$)     & $1.2$--$4.8$  & $0.1$--$0.7$   & $0.1$--$4.3$   & $0.1$--$1.9$  & $0.5$--$6$    & $<0.1$--$0.4$  & $<0.1$--$1$ \\
              & ($\mu^{+}\mu^{-}$) & $0.9$--$3.6$  & $0.1$--$1.4$   & $0.1$--$4.6$   & $<0.1$--$1.8$ & $<0.1$--$1.4$ & $<0.1$--$0.3$  & $0.1$--$2$ \\
\bt           & ($e^{+}e^{-}$)     & $0.8$--$5.7$  & $0.1$--$0.8$   & $0.1$--$3.7$   & $0.1$--$1.4$  & $0.1$--$9.1$  & $0.1$--$1.4$   & $0.1$--$2.7$ \\
              & ($\mu^{+}\mu^{-}$) & $0.6$--$4.3$  & $0.14$--$1$    & $<0.1$--$3.9$  & $0.1$--$0.9$ & $0.18$--$4.9$ & $<0.1$--$0.5$  & $<0.1$--$1.7$ \\
$\mathcal{T}$ & ($e^{+}e^{-}$)     & $0.7$--$5.0$  & $0.1$--$0.5$   & $0.1$--$2.8$   & $0.1$--$1.8$  & $0.1$--$5.4$  & $0.1$--$0.8$   & $0.2$--$3.7$ \\
              & ($\mu^{+}\mu^{-}$) & $0.5$--$3.9$  & $0.1$--$0.5$   & $0.1$--$2.3$   & $0.1$--$1.2$  & $0.1$--$4.9$  & $<0.1$         & $0.2$--$3.7$ \\
$\mathcal{S}$ & ($e^{+}e^{-}$)     & $0.7$--$3.2$  & $0.1$--$0.3$   & $0.3$--$2.4$   & $0.2$--$1.3$  & $0.1$--$2.8$  & $0.1$--$0.9$   & $0.1$--$4.7$ \\
              & ($\mu^{+}\mu^{-}$) & $0.5$--$2.4$  & $0.1$--$0.4$   & $0.2$--$2.1$   & $0.1$--$1.1$  & $0.1$--$2.5$  & $<0.1$         & $0.1$--$4.6$ \\
$\mathcal{F}$ & ($e^{+}e^{-}$)     & $0.7$--$3.7$  & $0.1$--$0.3$   & $0.1$--$2.2$   & $0.1$--$1.2$  & $0.2$--$4.4$  & $0.1$--$1$     & $0.1$--$2.1$ \\
              & ($\mu^{+}\mu^{-}$) & $0.5$--$2.8$  & $0.1$--$0.5$   & $0.1$--$1.7$   & $0.1$--$1$    & $0.1$--$2$    & $<0.1$         & $0.1$--$1.8$ \\
\bottomrule
\end{tabular}
\caption{Ranges of the relative uncertainties $\frac{\delta_{\mathcal{O}}}{\mathcal{O}}$ of the event-shape observables
$\mathcal{O}$ for the electron and muon channels indicated by ($e^{+}e^{-}$) and ($\mu^{+}\mu^{-}$) for the $\zpt$
range 12--25\;{\GeV} in percent. The superscripts denote the statistical (`stat') and the individual systematic
uncertainties in the lepton reconstruction and identification (`Lepton'), track reconstruction efficiency (`Tracking'),
non-primary particles (`Non-prim.'), pile-up correction (`PU'), multijet background (`Multijet'), and the unfolding (`Unfold').}
\label{tab:SystematicUncertainties_zpt_12-25}
\end{table}

\begin{table}
\centering
\begin{tabular}{rr|*{9}{c}}\toprule
Observable    & Channel  & ${\delta^\text{stat}_{\mathcal{O}}}$ & ${\delta^\text{Lepton}_{\mathcal{O}}}$ & ${\delta^\text{Tracking}_{\mathcal{O}}}$ & ${\delta^\text{Non-Prim.}_{\mathcal{O}}}$ & ${\delta^\text{PU}_{\mathcal{O}}}$ & ${\delta^\text{Multijet}_{\mathcal{O}}}$ & ${\delta^\text{Unfold}_{\mathcal{O}}}$ \\
              &          &  $[\%]$                                                & $[\%]$                                                   & $[\%]$                           
                          & $[\%]$                                                      & $[\%]$                                               & $[\%]$                         
                            & $[\%]$  \\
\midrule
$N_\text{ch}$ & ($e^{+}e^{-}$)     & $1.1$--$47$   & $0.1$--$2.5$  & $0.3$--$8.9$ & $<0.1$--$15$  & $<0.1$--$34$  & $0.1$--$3.5$    & $<0.1$--$2.1$ \\
              & ($\mu^{+}\mu^{-}$) & $0.9$--$28$   & $0.1$--$3.5$  & $0.2$--$6.9$ & $<0.1$--$5.3$ & $0.14$--$34$  & $<0.1$--$0.2$   & $0.1$--$8.9$ \\
$\sumpt$      & ($e^{+}e^{-}$)     & $1$--$8.9$    & $0.1$--$1.2$  & $0.3$--$4.1$ & $0.1$--$1.2$  & $0.1$--$2.5$  & $<0.1$--$1.2$   & $0.1$--$1.4$ \\
              & ($\mu^{+}\mu^{-}$) & $0.7$--$6.3$  & $<0.1$--$1$   & $0.4$--$4.1$ & $0.1$--$1.7$  & $<0.1$--$3.2$ & $<0.1$--$0.2$   & $0.1$--$2.1$ \\
\bt           & ($e^{+}e^{-}$)     & $1$--$3$      & $0.1$--$0.3$  & $0.2$--$2.7$ & $0.2$--$0.7$  & $0.1$--$2.3$  & $0.1$--$0.9$    & $0.1$--$1.7$ \\
              & ($\mu^{+}\mu^{-}$) & $0.8$--$2.2$  & $0.1$--$0.6$  & $0.3$--$2.9$ & $0.1$--$0.8$  & $0.1$--$1.5$  & $<0.1$--$0.1$   & $<0.1$--$1.6$ \\
$\mathcal{T}$ & ($e^{+}e^{-}$)     & $0.9$--$4.4$  & $0.1$--$0.3$  & $0.1$--$1.5$ & $0.1$--$1.8$  & $0.1$--$5.3$  & $0.1$--$0.8$    & $0.4$--$3.7$ \\
              & ($\mu^{+}\mu^{-}$) & $0.7$--$3.5$  & $0.1$--$0.5$  & $0.1$--$1.6$ & $0.1$--$1.9$  & $0.1$--$3.7$  & $<0.1$          & $0.1$--$4.3$ \\
$\mathcal{S}$ & ($e^{+}e^{-}$)     & $0.9$--$3.9$  & $0.1$--$0.6$  & $0.1$--$1.8$ & $0.1$--$1.1$  & $0.2$--$12.3$ & $0.1$--$0.8$    & $0.1$--$2.7$ \\
              & ($\mu^{+}\mu^{-}$) & $0.7$--$3.1$  & $0.1$--$0.7$  & $0.1$--$1.8$ & $0.1$--$0.5$  & $0.1$--$8$    & $<0.1$          & $0.1$--$6.2$ \\
$\mathcal{F}$ & ($e^{+}e^{-}$)     & $0.9$--$2.8$  & $0.1$--$0.3$  & $0.1$--$0.8$ & $0.1$--$1.1$  & $0.1$--$5.4$  & $0.1$--$0.8$    & $0.1$--$2.4$ \\
              & ($\mu^{+}\mu^{-}$) & $0.7$--$2.1$  & $0.1$--$0.6$  & $0.1$--$0.9$ & $0.1$--$0.9$  & $0.1$--$2.7$  & $<0.1$          & $0.1$--$0.9$ \\
\bottomrule
\end{tabular}
\caption{Ranges of the relative uncertainties $\frac{\delta_{\mathcal{O}}}{\mathcal{O}}$ of the event-shape observables
$\mathcal{O}$ for the electron and muon channels indicated by ($e^{+}e^{-}$) and ($\mu^{+}\mu^{-}$) for $\zpt>25$\;{\GeV}  
in percent. The superscripts denote the statistical (`stat') and the individual systematic
uncertainties in the lepton reconstruction and identification (`Lepton'), track reconstruction efficiency (`Tracking'),
non-primary particles (`Non-prim.'), pile-up correction (`PU'), multijet background (`Multijet'), and the unfolding (`Unfold').}
\label{tab:SystematicUncertainties_zpt_ge25}
\end{table}

%-------------------------------------------------------------------------------
\section{Results}
\label{sec:result}
%-------------------------------------------------------------------------------

\noindent
The results from the electron and muon channels are in good agreement and numerical 
values for each channel are provided in HEPDATA~\cite{Buckley:2010jn}. 
The statistical uncertainties in the muon results are slightly smaller than those 
in the electron results and in general the results are dominated by the 
systematic uncertainties. 
Since the electron- and muon-specific systematic uncertainties are smaller than the 
common dominant systematic uncertainties in the track reconstruction efficiency, 
the PU correction factors, and the unfolding model, the electron and muon results 
were not combined.
\begin{figure}[p]
    \captionsetup[subfigure]{margin=0pt, width=.38\textwidth}
    \centering
    \subfloat[$N_\text{ch}$]{
        \includegraphics[width=.38\textwidth]{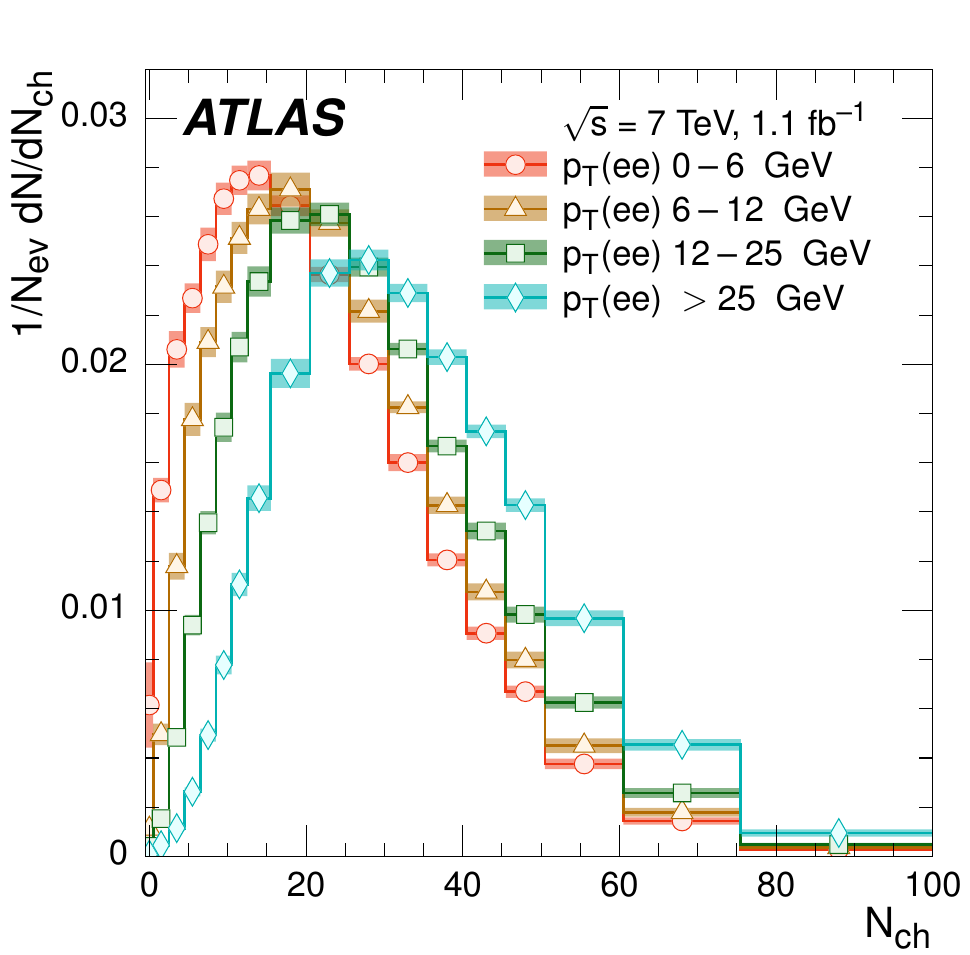}
        \label{fig:datavszptelecnch}
    }\subfloat[$\sum p_\text{T}$]{
        \includegraphics[width=.38\textwidth]{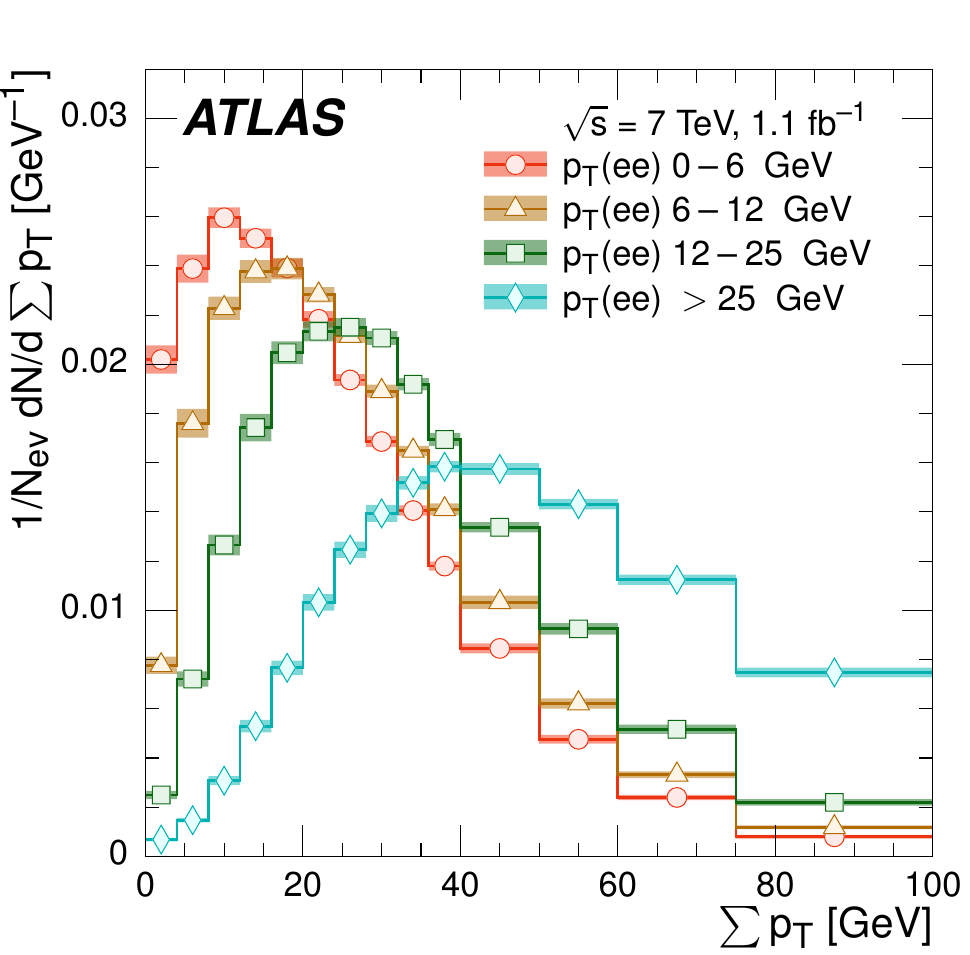}
        \label{fig:datavszptelecsumpt}
    }
    \\
    \subfloat[Beam thrust]{
        \includegraphics[width=.38\textwidth]{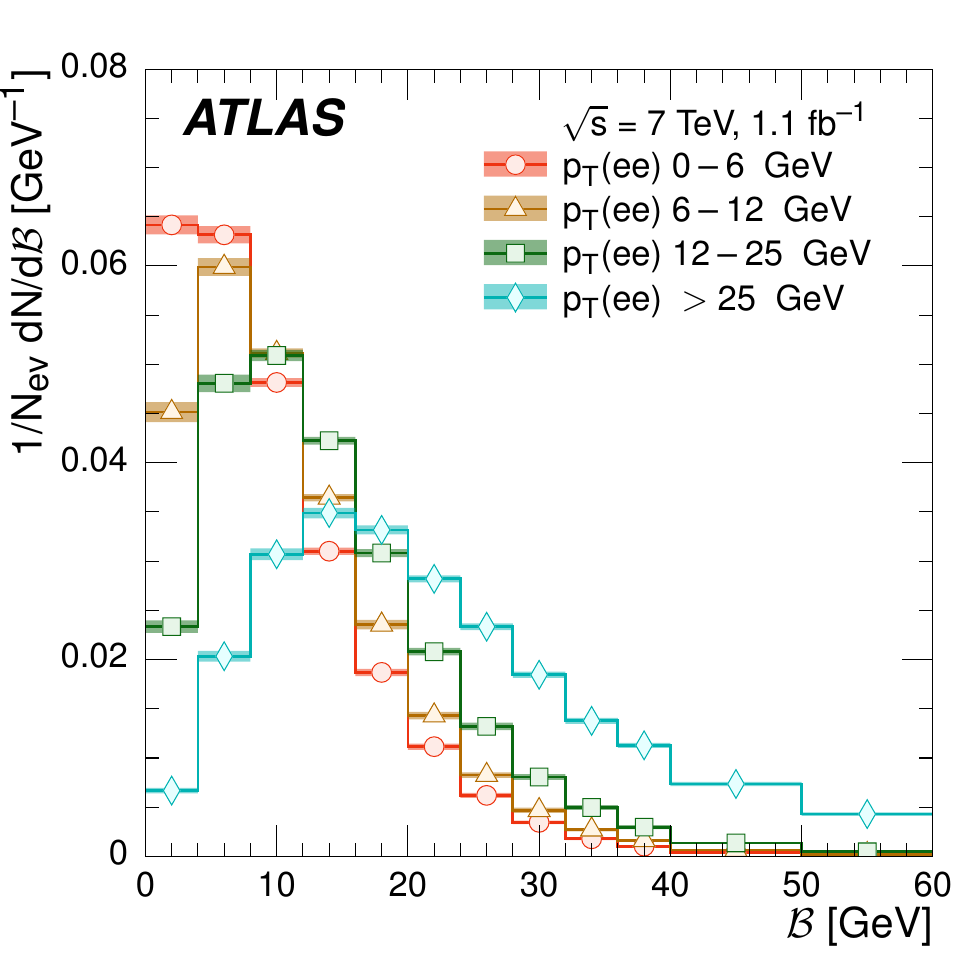}
        \label{fig:datavszptelebt}
    }\subfloat[Transverse thrust]{
        \includegraphics[width=.38\textwidth]{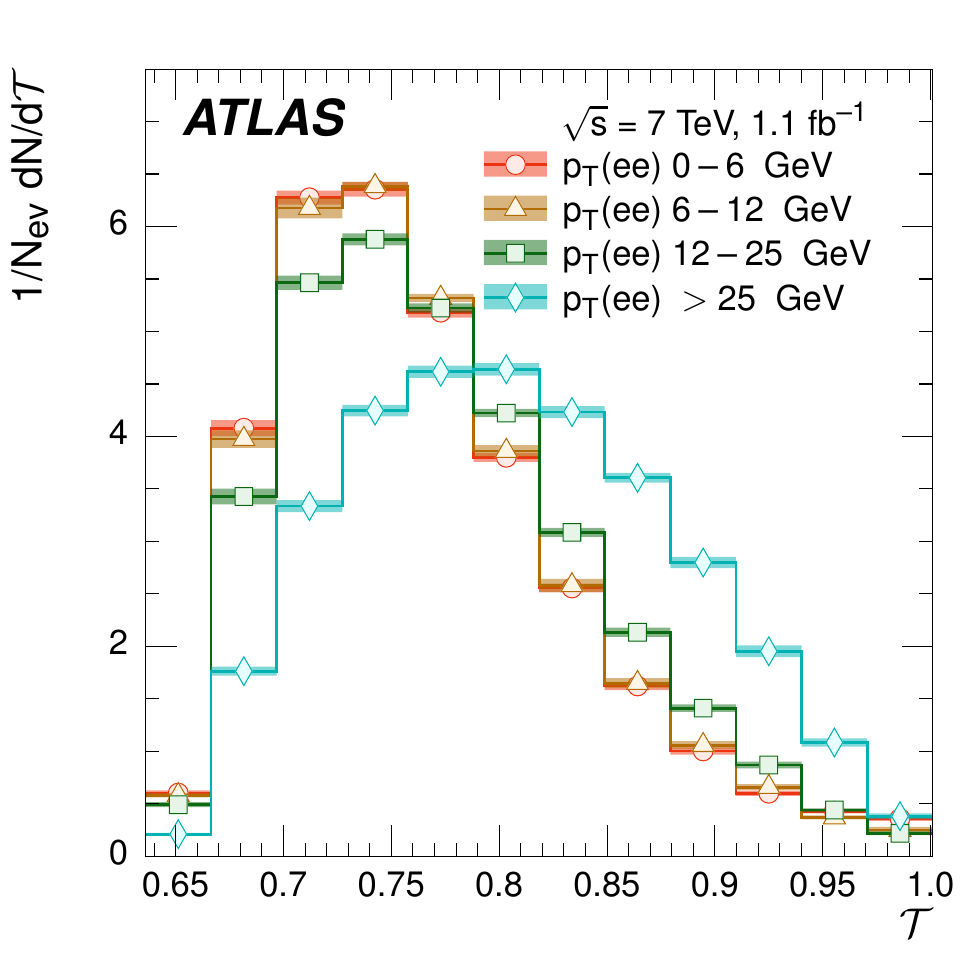}
        \label{fig:datavszptelecthrust}
    }
    \\
    \subfloat[Spherocity]{
        \includegraphics[width=.38\textwidth]{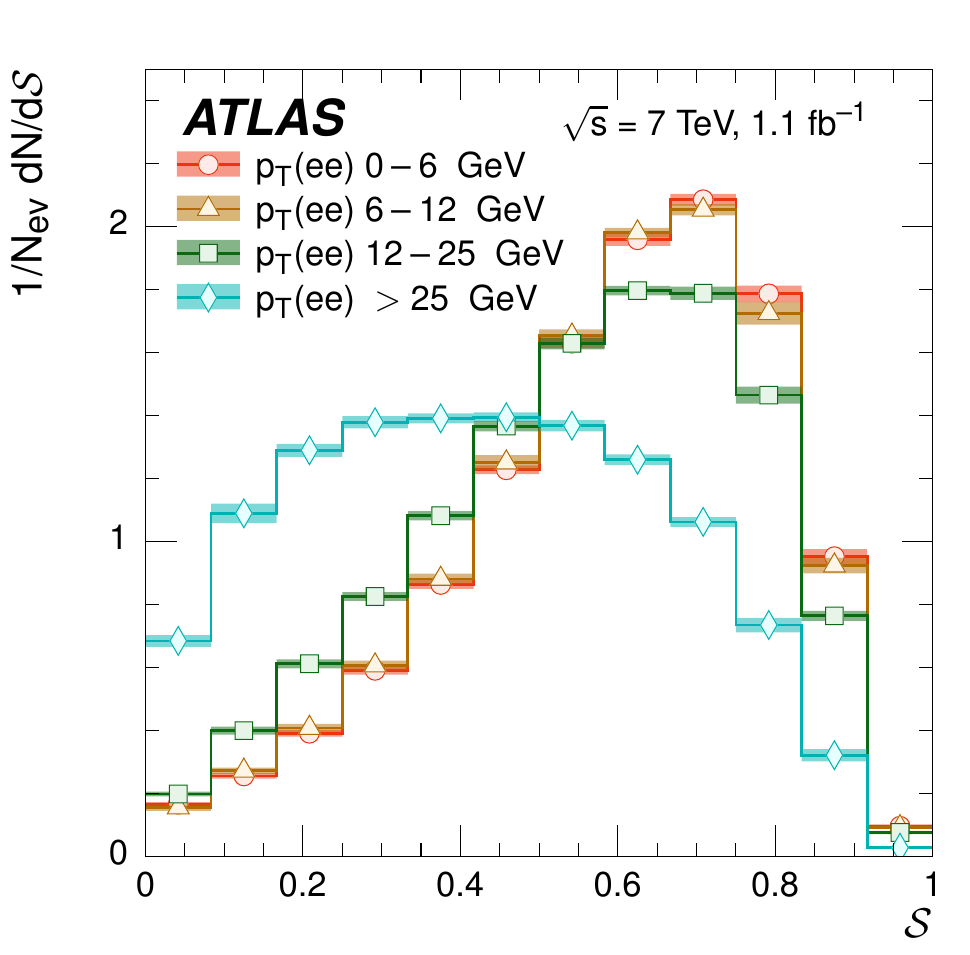}
        \label{fig:datavszptelecspherocity}
    }
    \subfloat[$\mathcal{F}$-parameter]{
        \includegraphics[width=.38\textwidth]{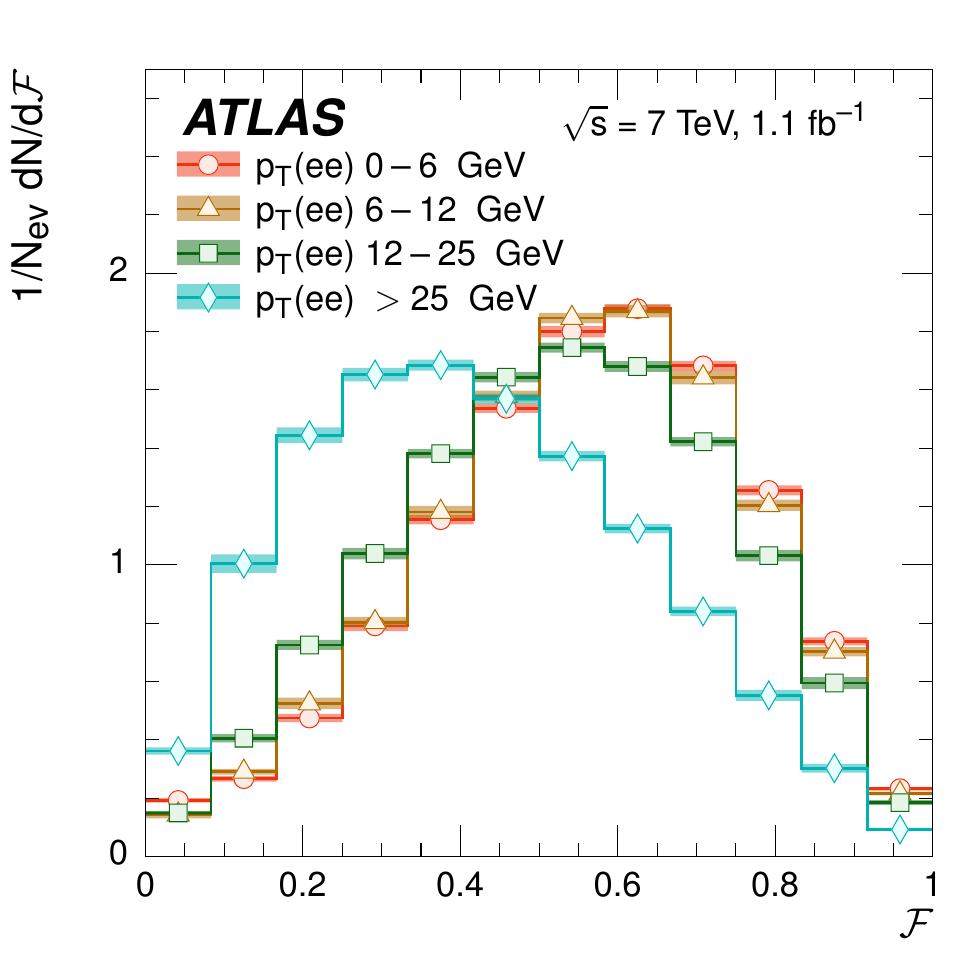}
        \label{fig:datavszptelecfparameter}
    }
    \caption[Unfolded data summary plots]{Distributions of the event-shape variables (a) charged-particle multiplicity $\nch$, 
(b) summed transverse momenta $\sumpt$, (c) beam thrust $\mathcal{B}$, 
(d) transverse thrust $\mathcal{T}$, (e) spherocity $\mathcal{S}$, and (f) $\mathcal{F}$-parameter as defined in Section~\ref{sec:eventshapes}
measured in $Z \rightarrow e^{+} e^{-}$ events for the different ranges of the transverse momentum of the $e^{+} e^{-}$ system, 
$\zptee$ (open circles: 0--6\;\GeV, open triangles: 6--12\;\GeV, open boxes: 12--25\;\GeV, open diamonds: $\ge 25$\;\GeV). 
$N_{\text{ev}}$ denotes the number of events in the $\zptee$ range passing the analysis cuts. 
 The bands show the sum in quadrature of the statistical and all systematic uncertainties.}
    \label{fig:datavszptelec}
\end{figure}
\begin{figure}[p]
    \captionsetup[subfigure]{margin=0pt, width=.38\textwidth}
    \centering
    \subfloat[$N_\text{ch}$]{
        \includegraphics[width=.38\textwidth]{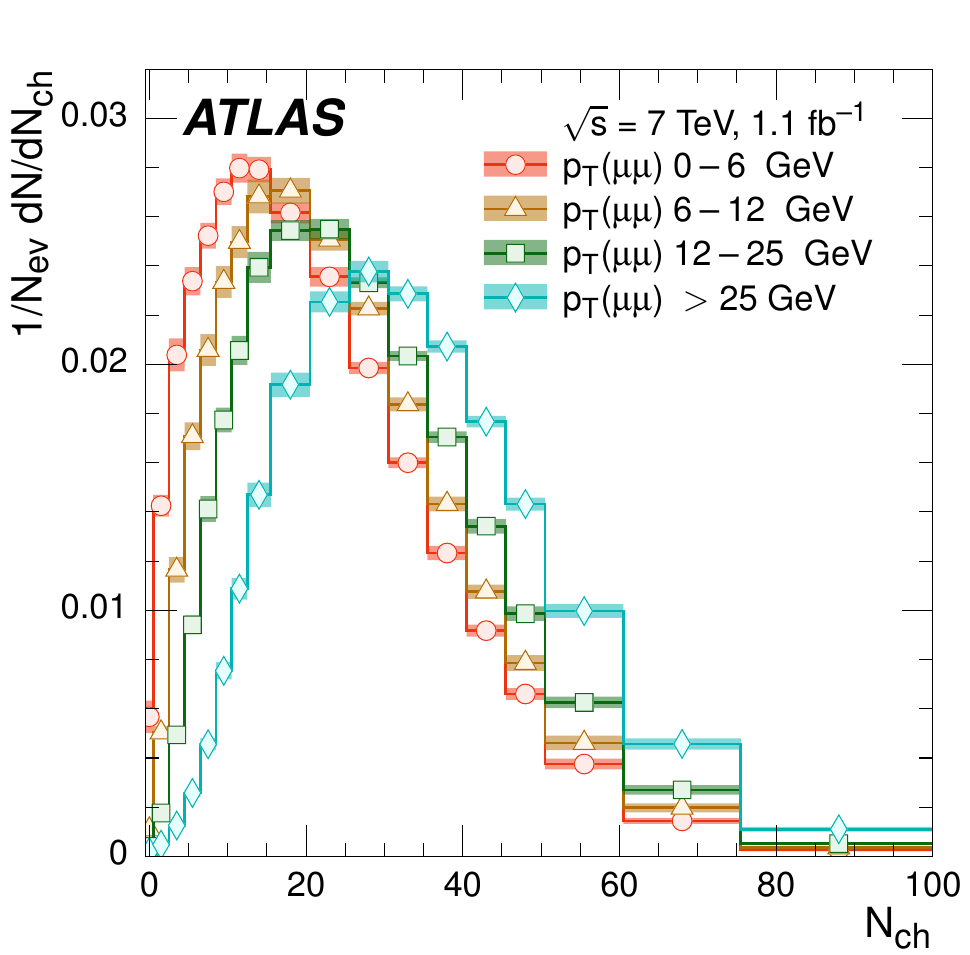}
        \label{fig:datavszptmuonnch}
    }\subfloat[$\sum p_\text{T}$]{
        \includegraphics[width=.38\textwidth]{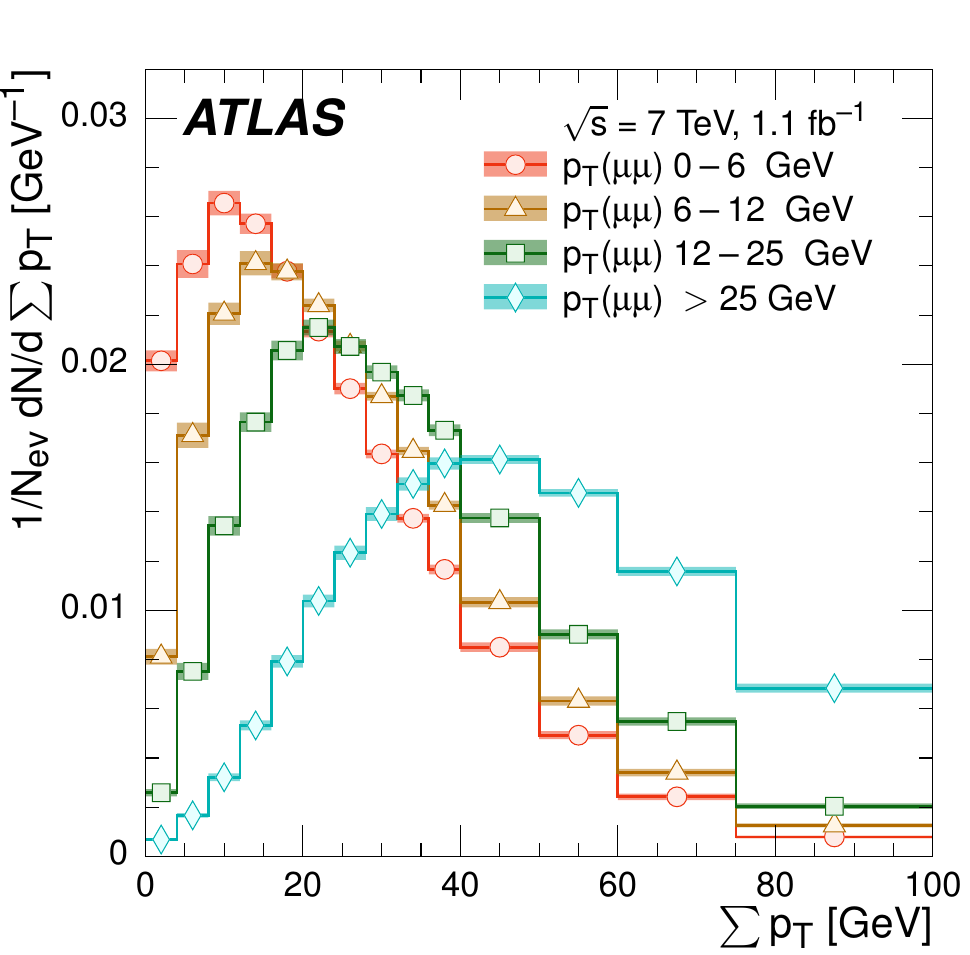}
        \label{fig:datavszptmuonsumpt}
    }
    \\
    \subfloat[Beam thrust]{
        \includegraphics[width=.38\textwidth]{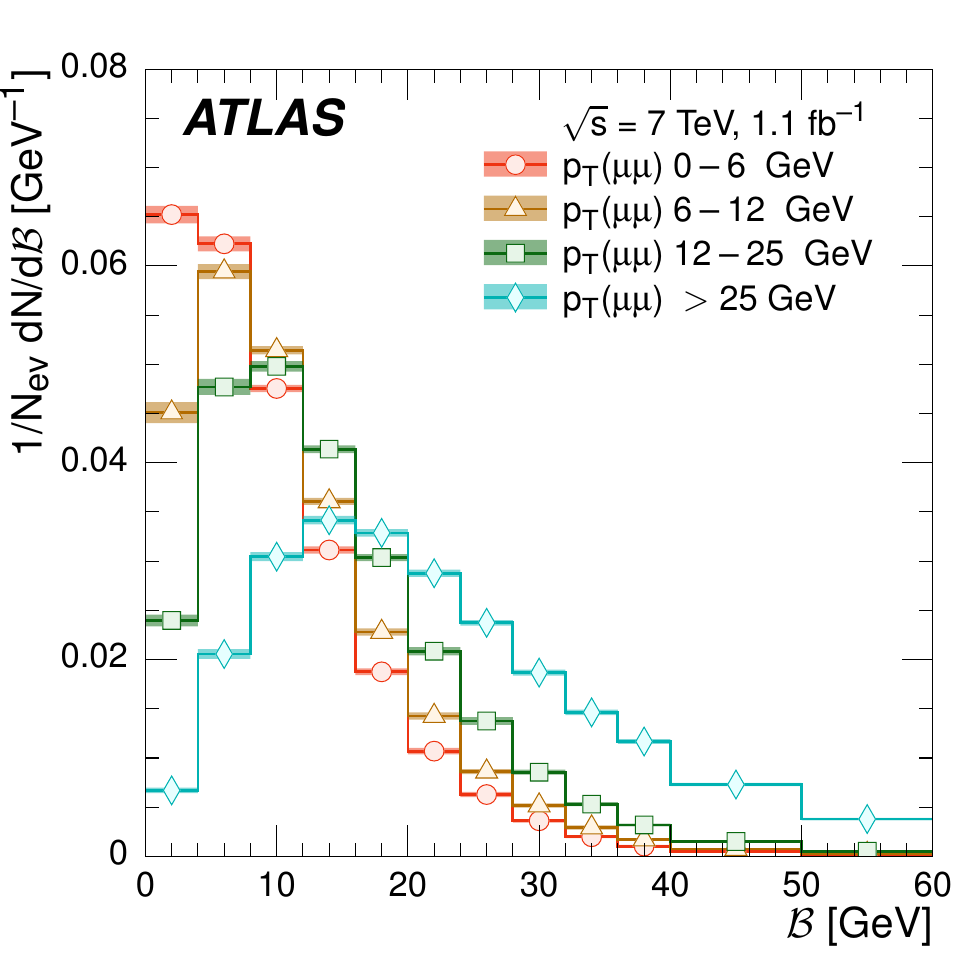}
        \label{fig:datavszptmuonbt}
    }\subfloat[Transverse thrust]{
        \includegraphics[width=.38\textwidth]{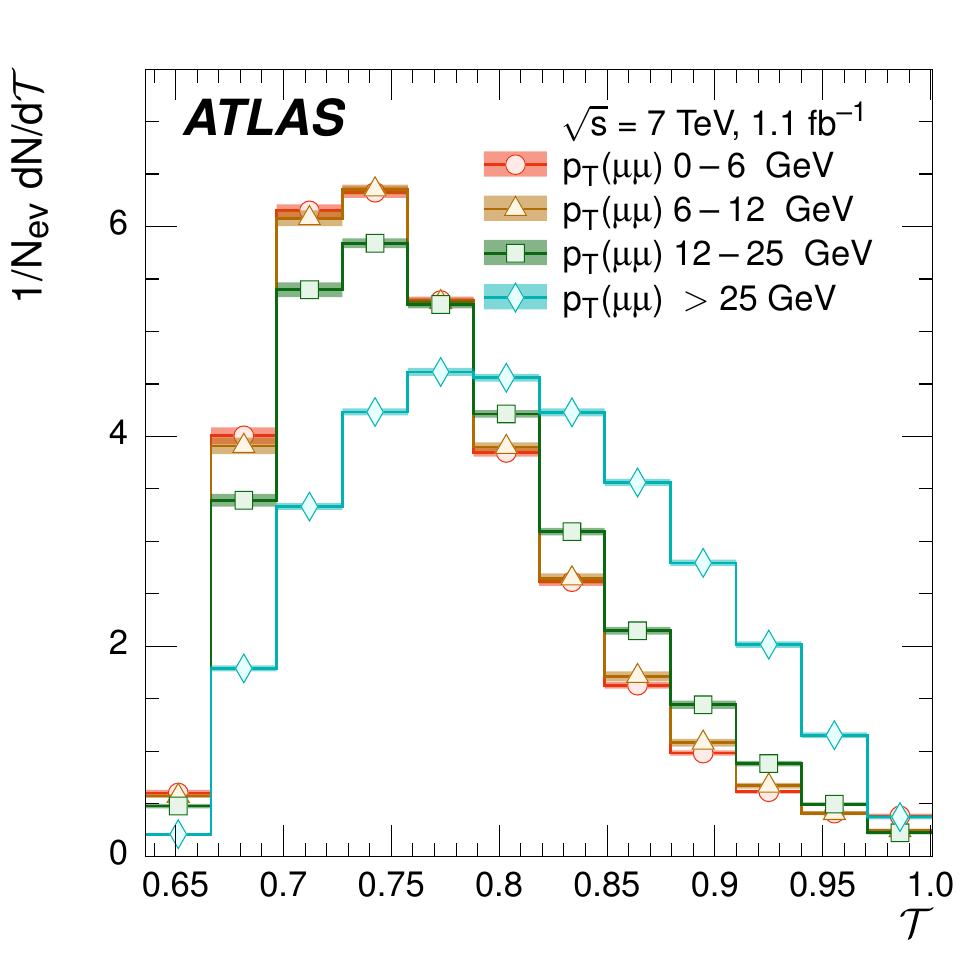}
        \label{fig:datavszptmuonthrust}
    }
    \\
    \subfloat[Spherocity]{
        \includegraphics[width=.38\textwidth]{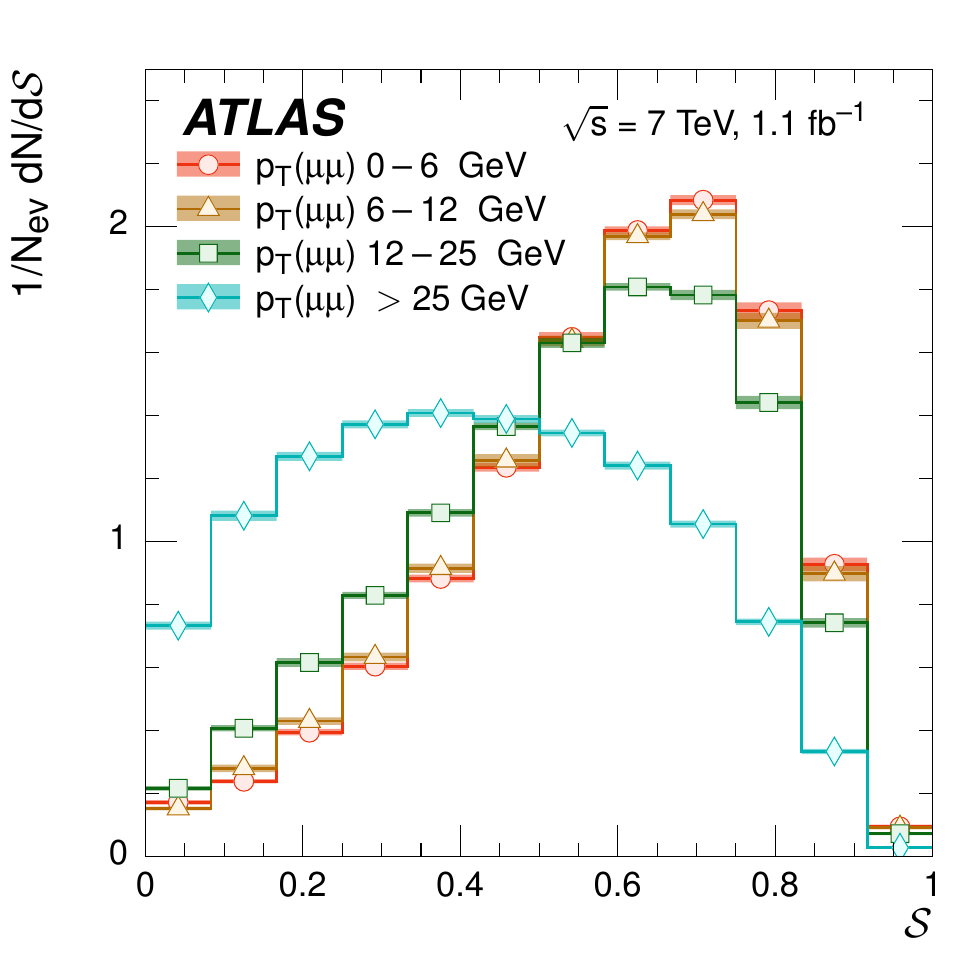}
        \label{fig:datavszptmuonspherocity}
    }
    \subfloat[$\mathcal{F}$-Parameter]{
        \includegraphics[width=.38\textwidth]{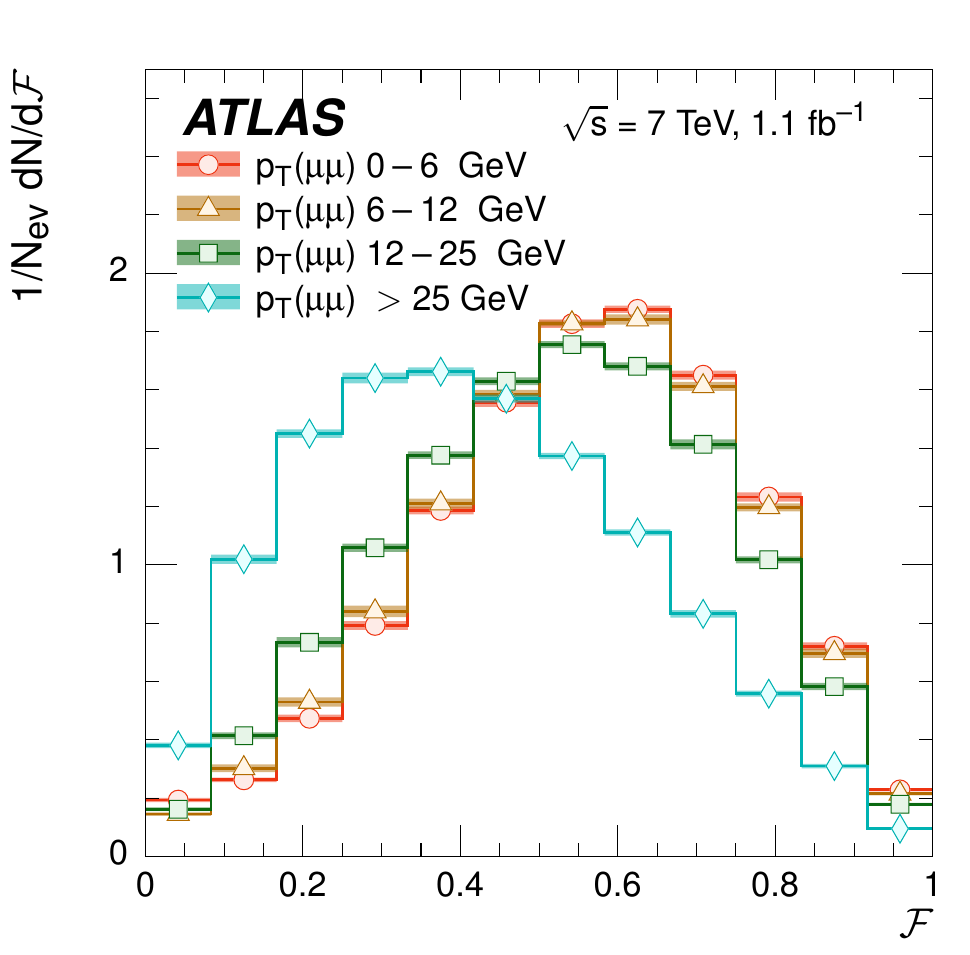}
        \label{fig:datavszptmuonfparameter}
    }
    \caption[Unfolded data summary plots]{Distributions of the event-shape variables (a) charged-particle multiplicity $\nch$, 
(b) summed transverse momenta $\sumpt$, (c) beam thrust $\mathcal{B}$,
(d) transverse thrust $\mathcal{T}$, (e) spherocity $\mathcal{S}$, and (f) $\mathcal{F}$-parameter as defined in Section~\ref{sec:eventshapes}
measured in $Z \rightarrow \mu^{+} \mu^{-}$ events for the different ranges of the transverse momentum of the 
$\mu^{+} \mu^{-}$ system, $\zptmumu$ (open circles: 0--6\;\GeV, open triangles: 6--12\;\GeV, open boxes: 12--25\;\GeV, 
open diamonds: $\ge 25$\;\GeV). $N_{\text{ev}}$ denotes the number of events in the $\zptmumu$ range passing the analysis cuts.
The bands show the sum in quadrature of the statistical and all systematic uncertainties.}
    \label{fig:datavszptmuon}
\end{figure}

\noindent
Figure~\ref{fig:datavszptelec} (\ref{fig:datavszptmuon}) shows the unfolded electron (muon) channel results 
for the six observables in the various $\zpt$ ranges, with the total uncertainty presented as the quadratic 
sum of the statistical and total systematic uncertainties. As $\zpt$ rises, i.e. as recoiling jets emerge, 
the number of produced charged particles $\nch$ increases, as do $\sumpt$ and beam thrust. Correspondingly, 
transverse thrust moves towards higher values and spherocity towards smaller values as a result of the 
increasing jettiness of the events.

\begin{figure}[p]
    \captionsetup[subfigure]{margin=0pt, width=.42\textwidth}
    \centering
    \subfloat[$N_\text{ch}$, $\zptee$: 0--6\;\GeV]{
        \includegraphics[width=.42\textwidth]{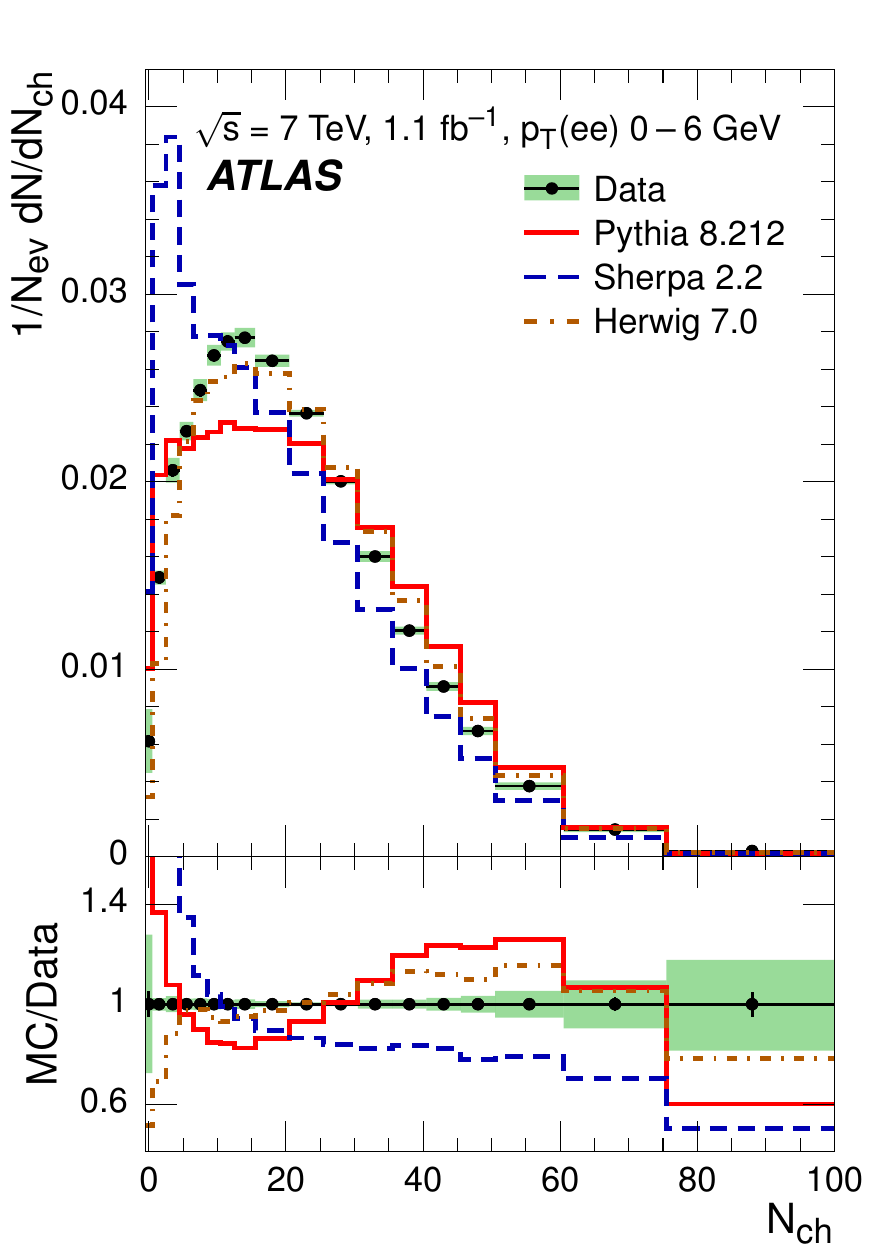}
        \label{fig:results-nch-0_6-Elec-final}
    }
    \subfloat[$N_\text{ch}$, $\zptee$: 6--12\;\GeV]{
        \includegraphics[width=.42\textwidth]{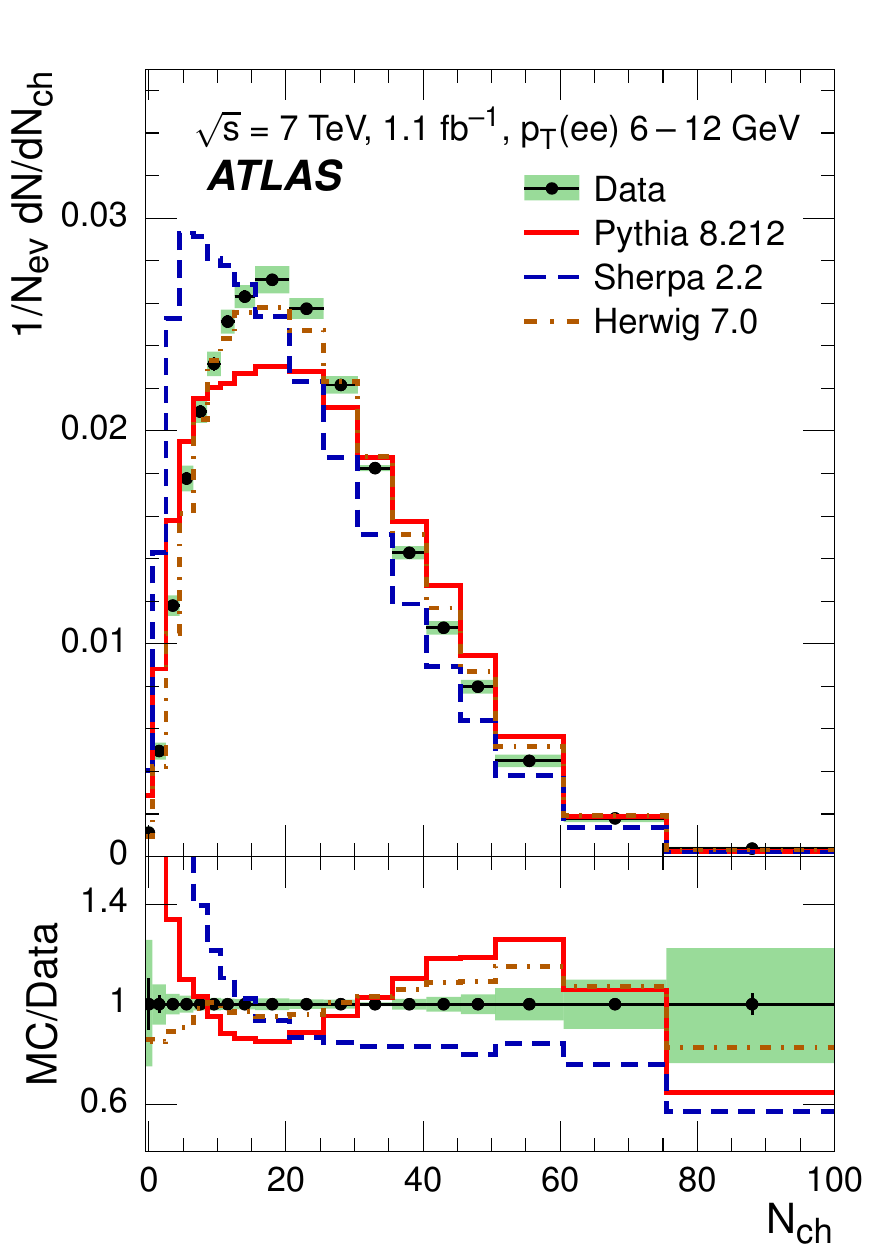}
        \label{fig:results-nch-6_12-Elec-final}
    }
    \\
    \subfloat[$N_\text{ch}$, $\zptee$: 12--25\;\GeV]{
        \includegraphics[width=.42\textwidth]{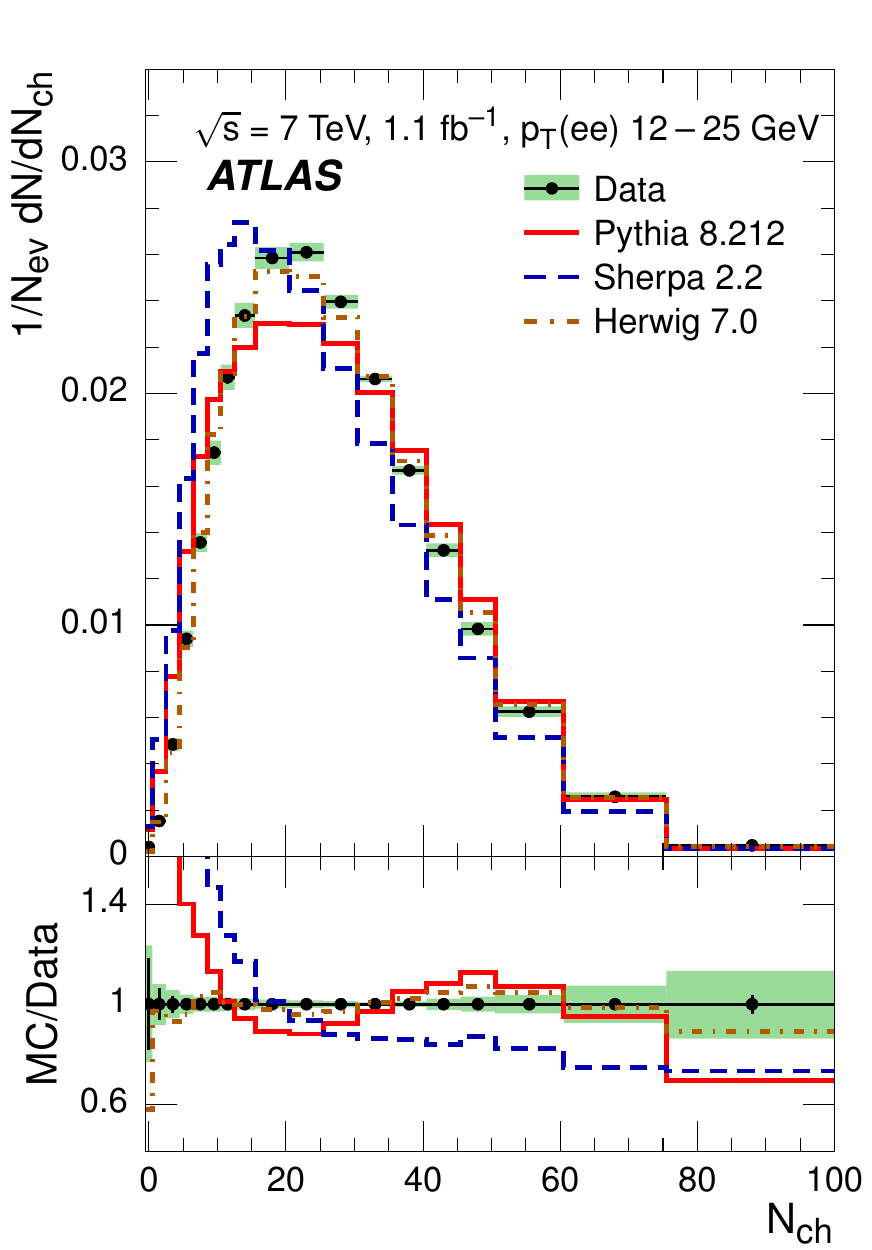}
        \label{fig:results-nch-12_25-Elec-final}
    }
    \subfloat[$N_\text{ch}$, $\zptee \ge 25$~\GeV]{
        \includegraphics[width=.42\textwidth]{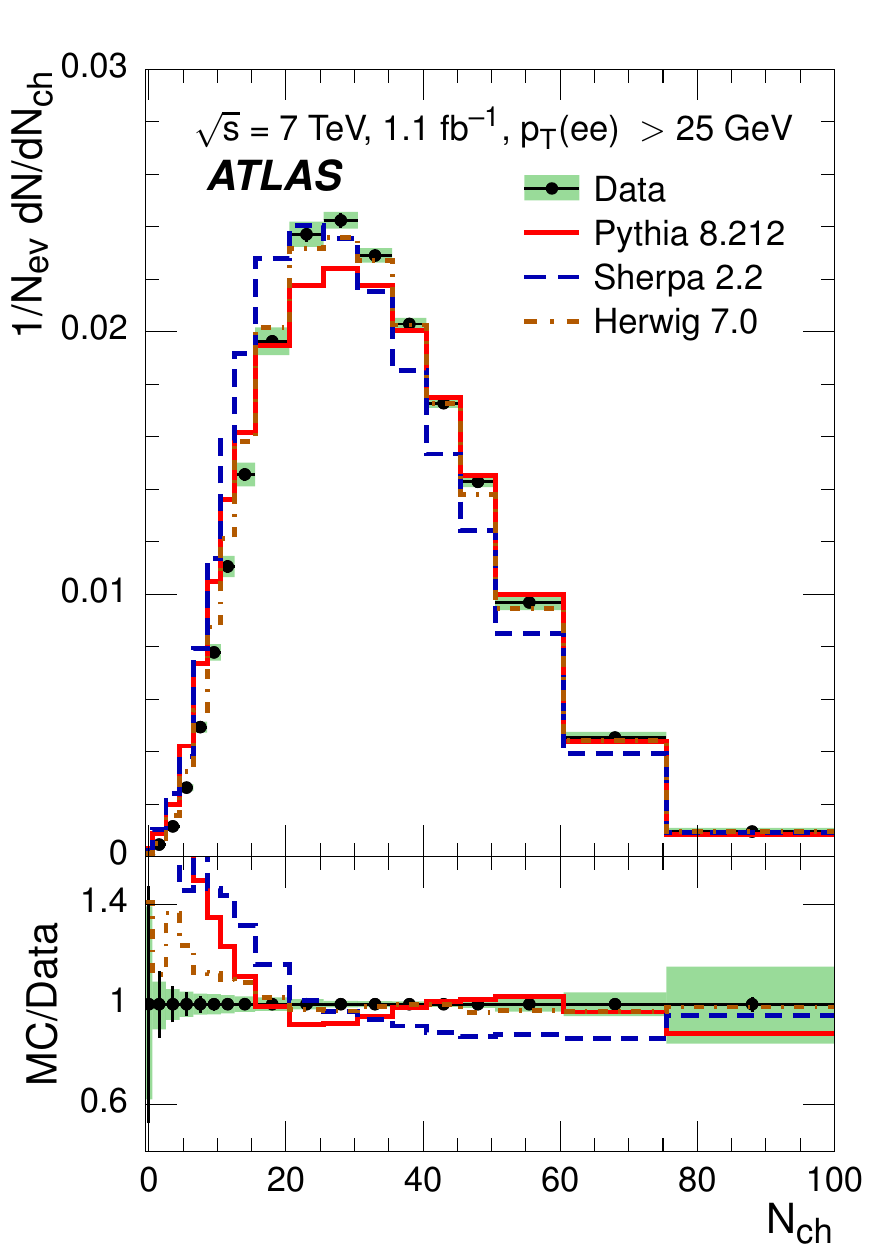}
        \label{fig:results-nch-25_999-Elec-final}
    }
    \caption[Results, $N_\text{ch}$, all phase-spage regions]{Distribution of charged-particle multiplicity, $N_\text{ch}$, 
for $Z \rightarrow e^{+}e^{-}$ with statistical (error bars) and total systematic (band) uncertainties for the
four $\zptee$ ranges ((a): 0--6\;\GeV, (b): 6--12\;\GeV, (c): 12--25\;\GeV, (d): $\ge 25$~\GeV) compared to the predictions 
from the MC generators \pythiaeight (full line), \sherpa (dashed line), and \herwigseven (dashed-dotted line). 
In each subfigure, the top plot shows the observable and the bottom plot shows the 
ratio of the MC simulation to the data.}
    \label{fig:results-nch-Elec-final}
\end{figure}

\begin{figure}[p]
    \captionsetup[subfigure]{margin=0pt, width=.42\textwidth}
    \centering
    \subfloat[$\sumpt$, $\zptee$: 0--6\;\GeV]{
        \includegraphics[width=.42\textwidth]{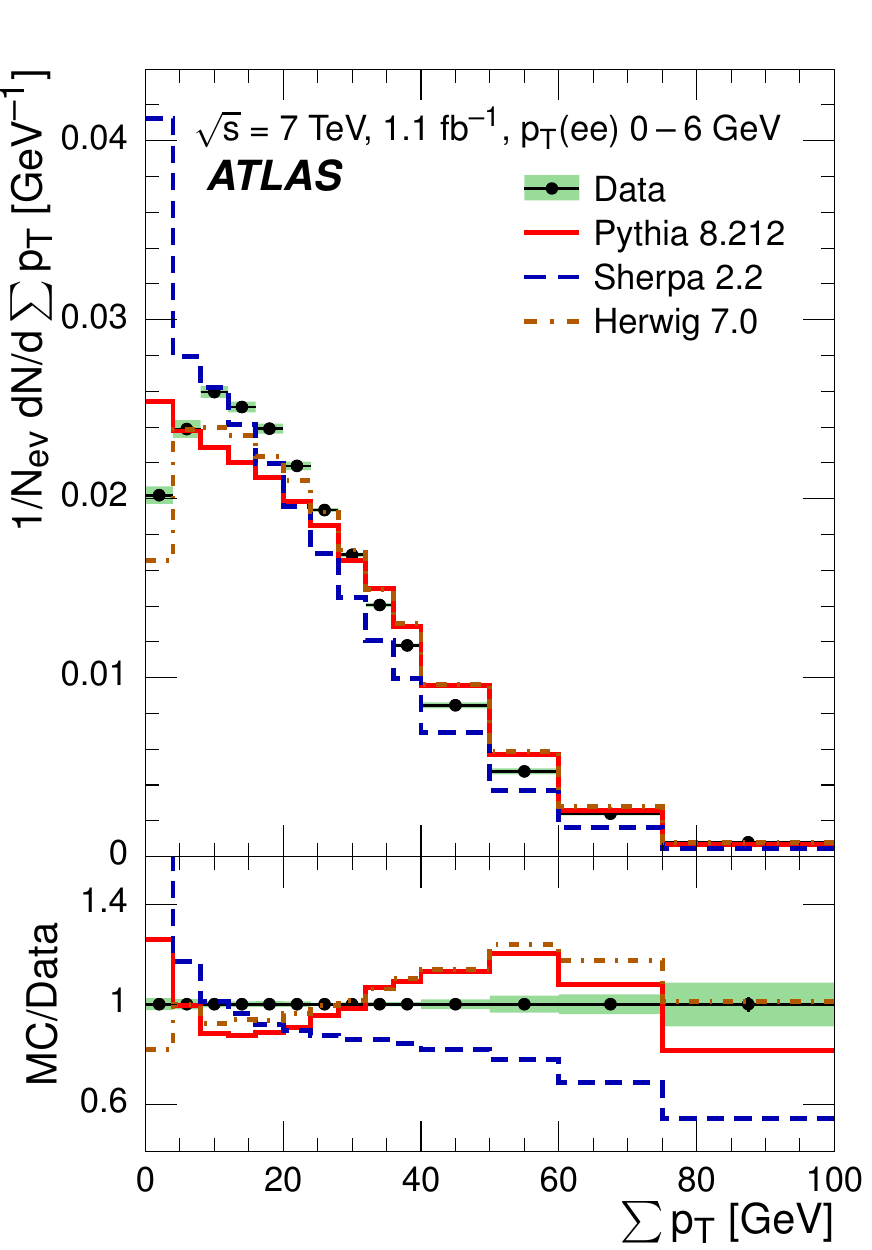}
        \label{fig:results-sumpt-0_6-Elec-final}
    }
    \subfloat[$\sumpt$, $\zptee$: 6--12\;\GeV]{
        \includegraphics[width=.42\textwidth]{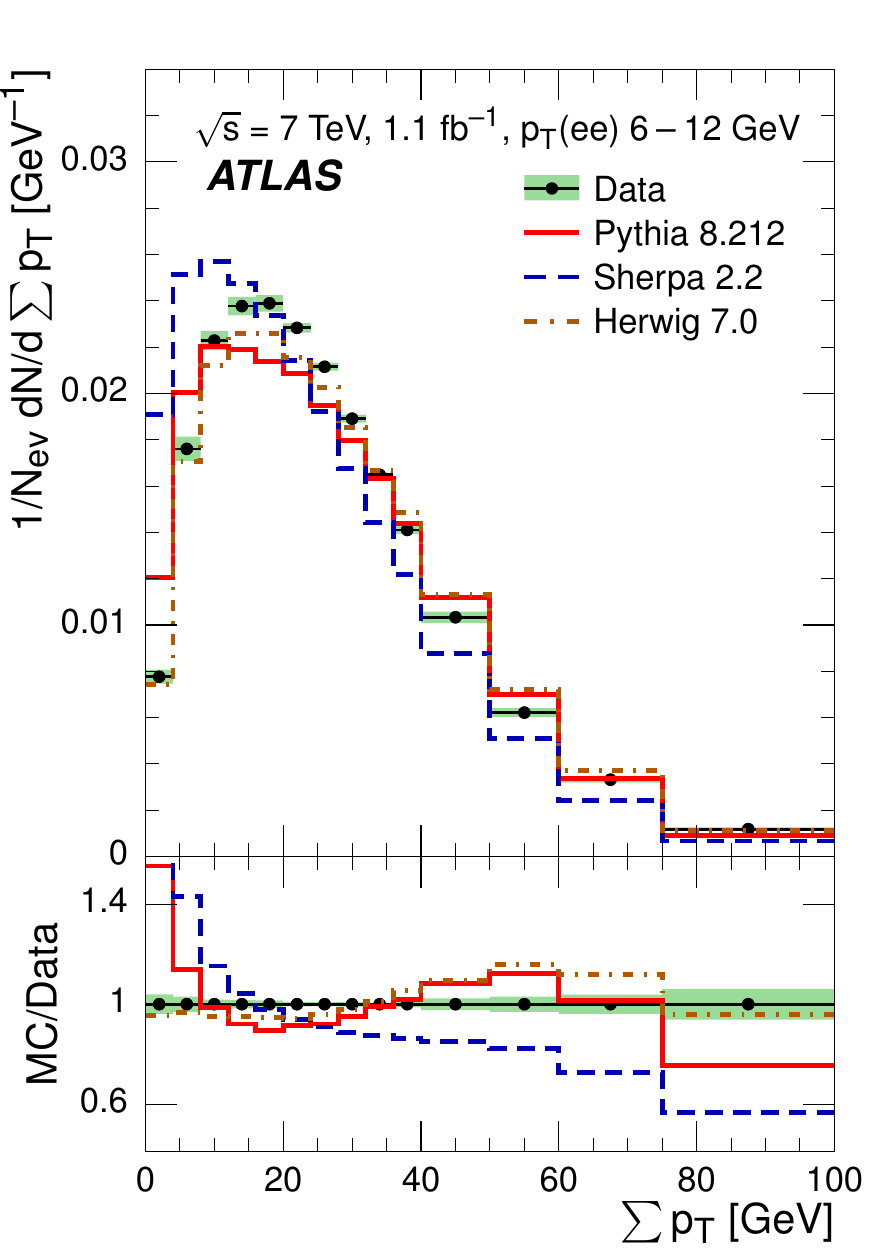}
        \label{fig:results-sumpt-6_12-Elec-final}
    }
    \\
    \subfloat[$\sumpt$, $\zptee$: 12--25\;\GeV]{
        \includegraphics[width=.42\textwidth]{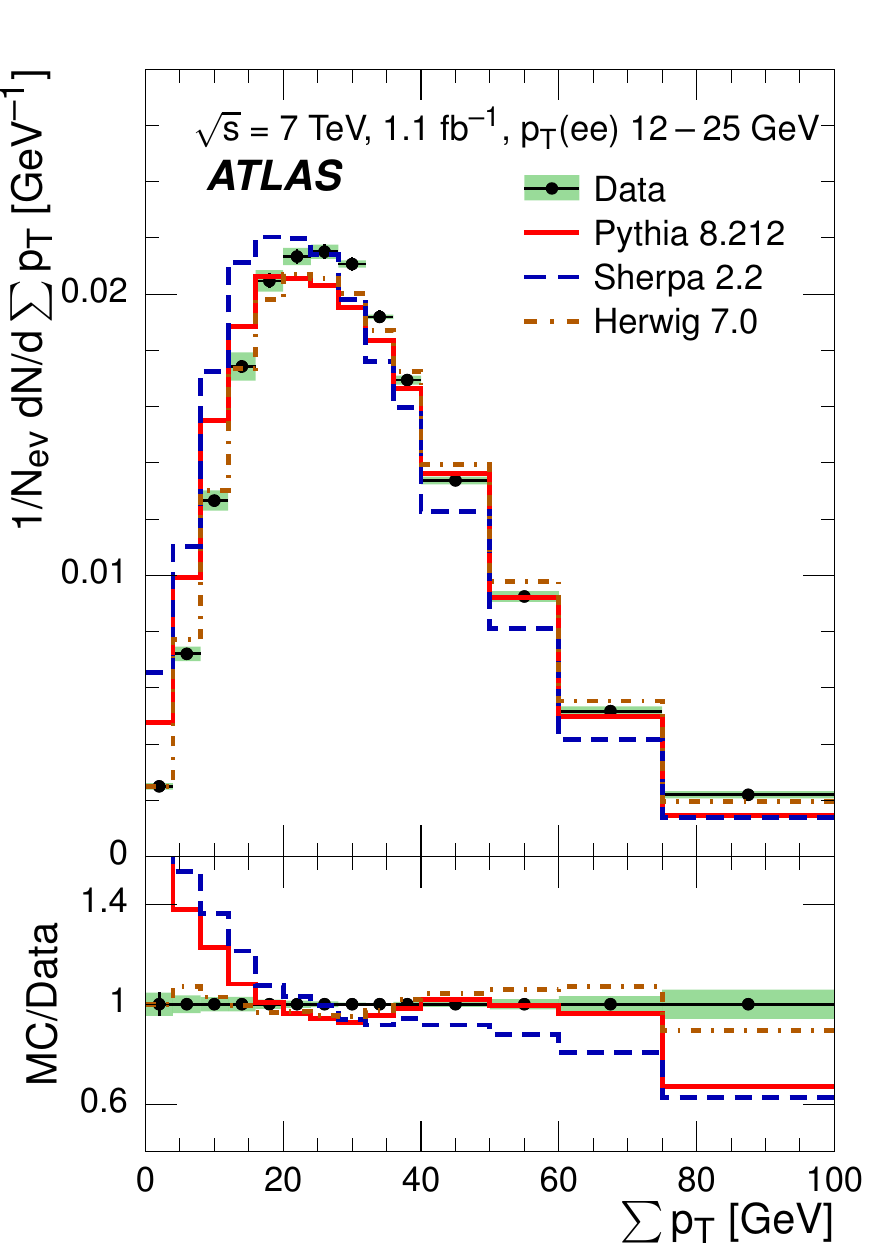}
        \label{fig:results-sumpt-12_25-Elec-final}
    }
    \subfloat[$\sumpt$, $\zptee \ge 25$~\GeV]{
        \includegraphics[width=.42\textwidth]{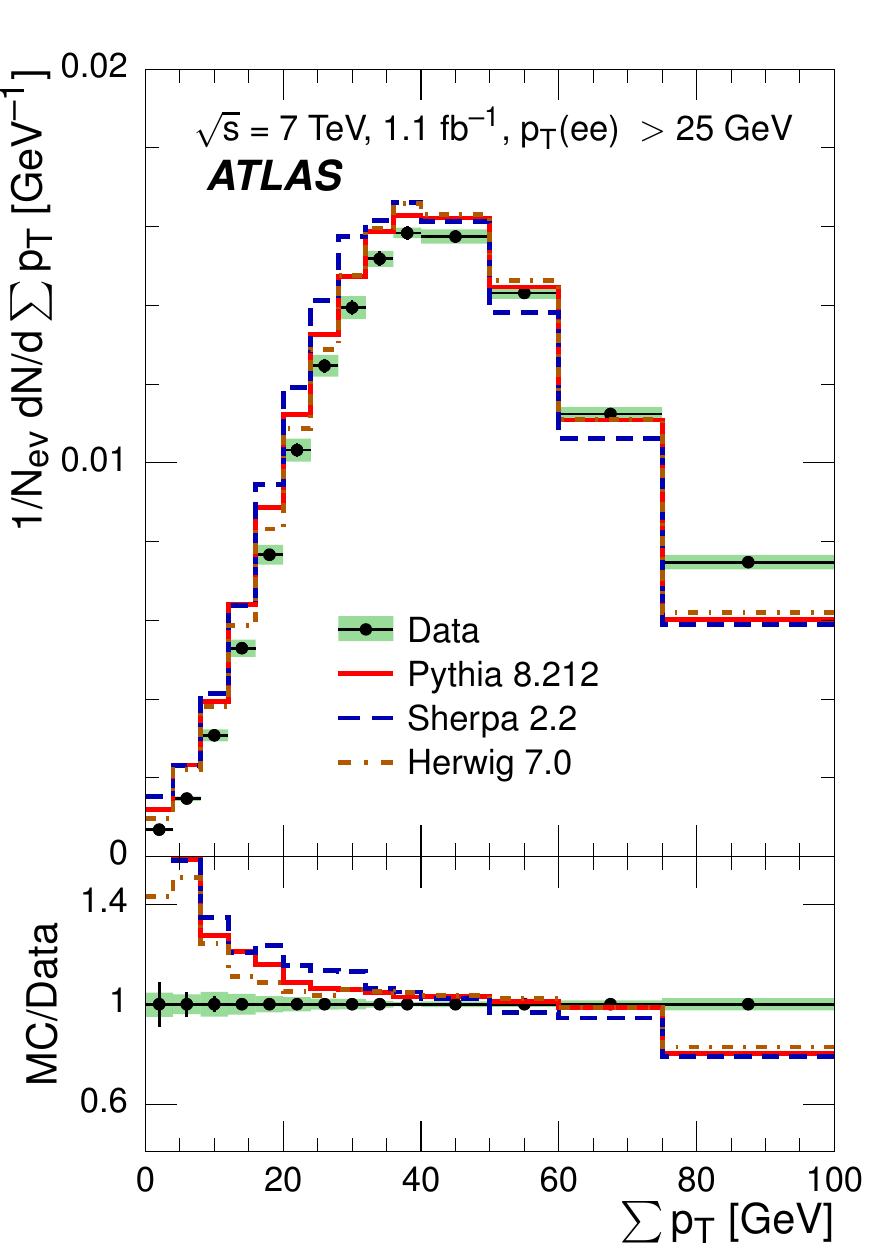}
        \label{fig:results-sumpt-25_999-Elec-final}
    }
    \caption[Results, $\sumpt$, all phase-spage regions]{Summed transverse momenta $\sumpt$ distribution of charged particles
for $Z \rightarrow e^{+}e^{-}$ with statistical (error bars) and total systematic (band) uncertainties for the
four $\zptee$ ranges ((a): 0--6\;\GeV, (b): 6--12\;\GeV, (c): 12--25\;\GeV, (d): $\ge 25$~\GeV) compared to the predictions
from the MC generators \pythiaeight (full line), \sherpa (dashed line), and \herwigseven (dashed-dotted line).
In each subfigure, the top plot shows the observable and the bottom plot shows the ratio of the MC simulation to the data.}
    \label{fig:results-sumpt-Elec-final}
\end{figure}

\begin{figure}[p]
    \captionsetup[subfigure]{margin=0pt, width=.42\textwidth}
    \centering
    \subfloat[Beam thrust, $\zptee$: 0--6\;\GeV]{
        \includegraphics[width=.42\textwidth]{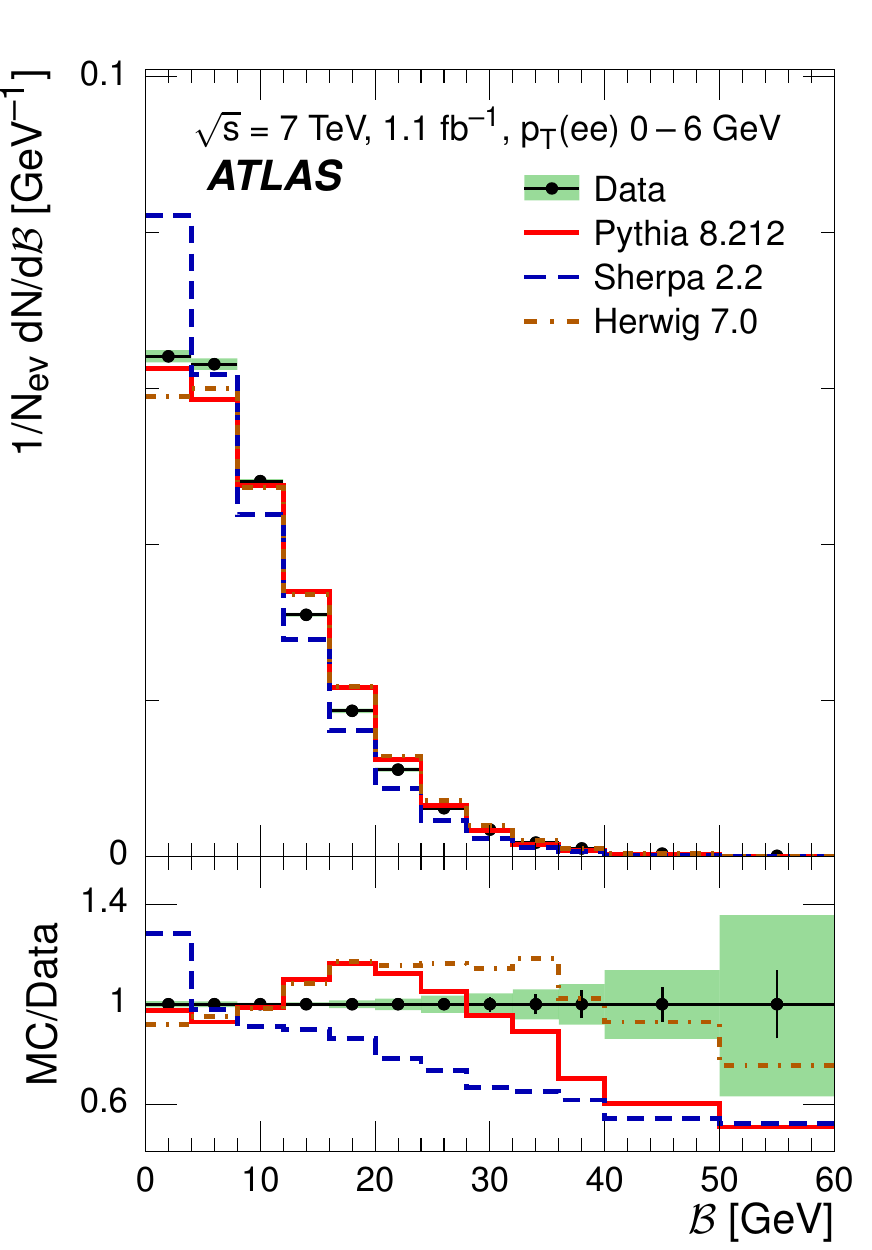}
        \label{fig:results-beamthrust-0_6-Elec-final}
    }
    \subfloat[Beam thrust, $\zptee$: 6--12\;\GeV]{
        \includegraphics[width=.42\textwidth]{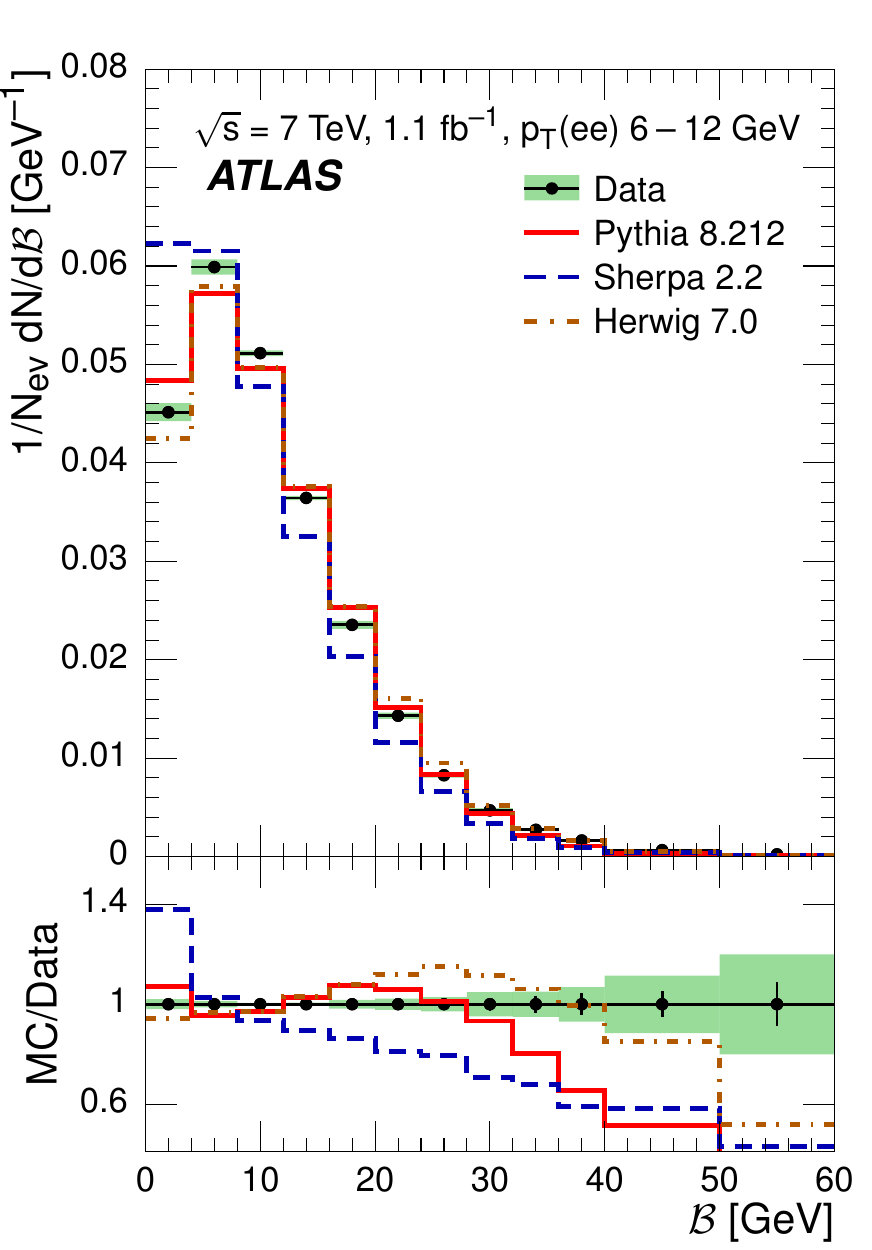}
        \label{fig:results-beamthrust-6_12-Elec-final}
    }
    \\
    \subfloat[Beam thrust, $\zptee$: 12--25\;\GeV]{
        \includegraphics[width=.42\textwidth]{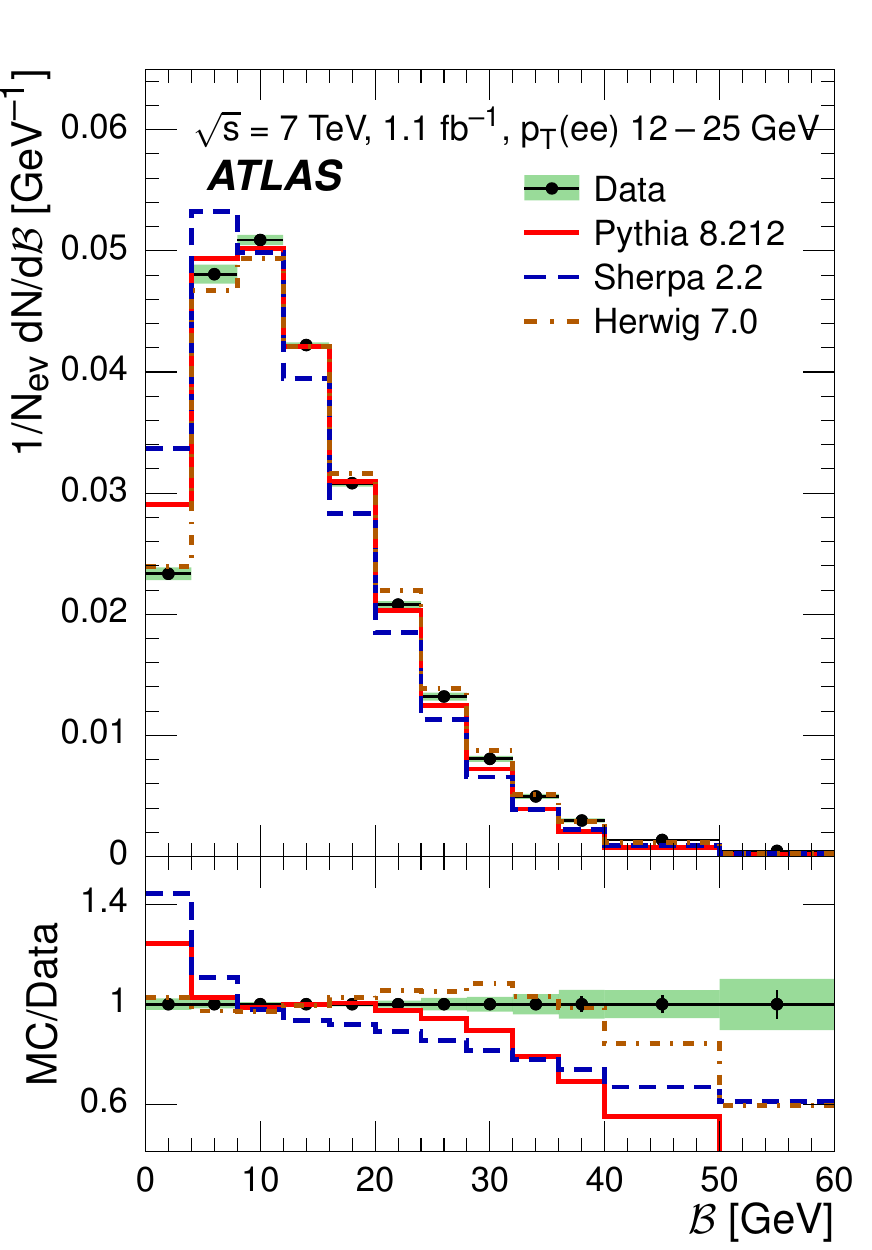}
        \label{fig:results-beamthrust-12_25-Elec-final}
    }
    \subfloat[Beam thrust, $\zptee \ge 25$~\GeV]{
        \includegraphics[width=.42\textwidth]{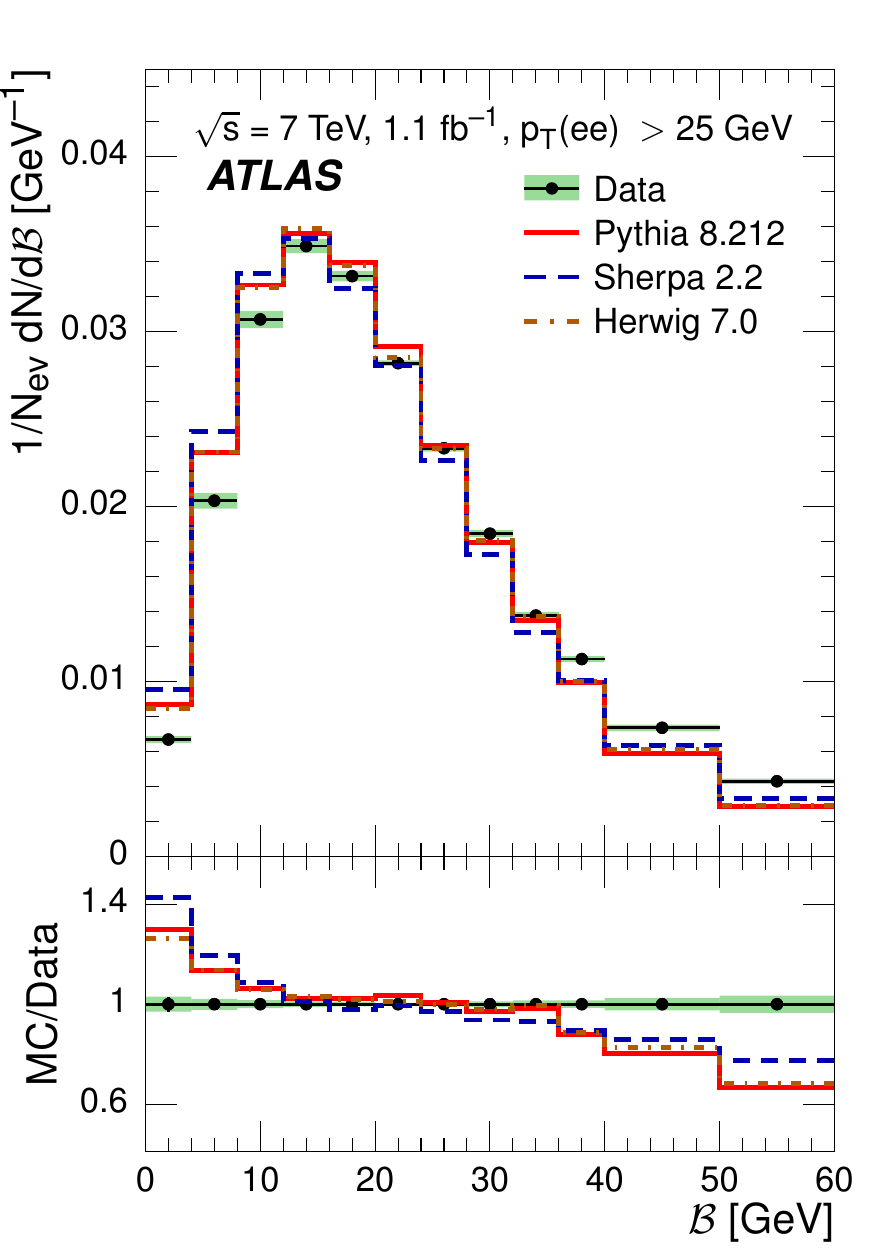}
        \label{fig:results-beamthrust-25_999-Elec-final}
    }
    \caption[Results, beam thrust, all phase-spage regions]{Beam thrust $\mathcal{B}$ distribution of charged particles
for $Z \rightarrow e^{+}e^{-}$ with statistical (error bars) and total systematic (band) uncertainties for the
four $\zptee$ ranges ((a): 0--6\;\GeV, (b): 6--12\;\GeV, (c): 12--25\;\GeV, (d): $\ge 25$~\GeV) compared to the predictions
from the MC generators \pythiaeight (full line), \sherpa (dashed line), and \herwigseven (dashed-dotted line).
In each subfigure, the top plot shows the observable and the bottom plot shows the ratio of the MC simulation to the data.}
    \label{fig:results-beamthrust-Elec-final}
\end{figure}

\begin{figure}[p]
    \captionsetup[subfigure]{margin=0pt, width=.42\textwidth}
    \centering
    \subfloat[Transverse thrust, $\zptee$: 0--6\;\GeV]{
        \includegraphics[width=.42\textwidth]{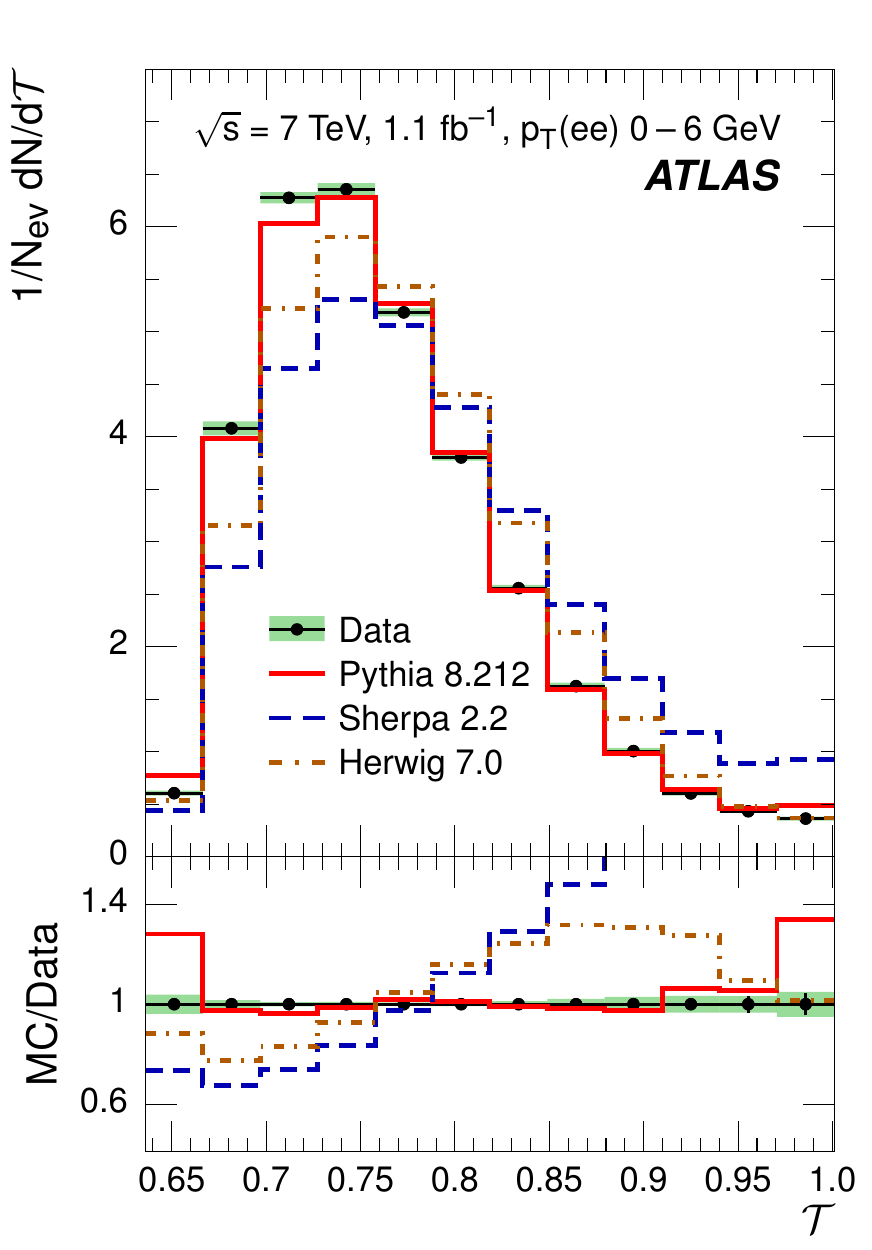}
        \label{fig:results-thrust-0_6-Elec-final}
    }
    \subfloat[Transverse thrust, $\zptee$: 6--12\;\GeV]{
        \includegraphics[width=.42\textwidth]{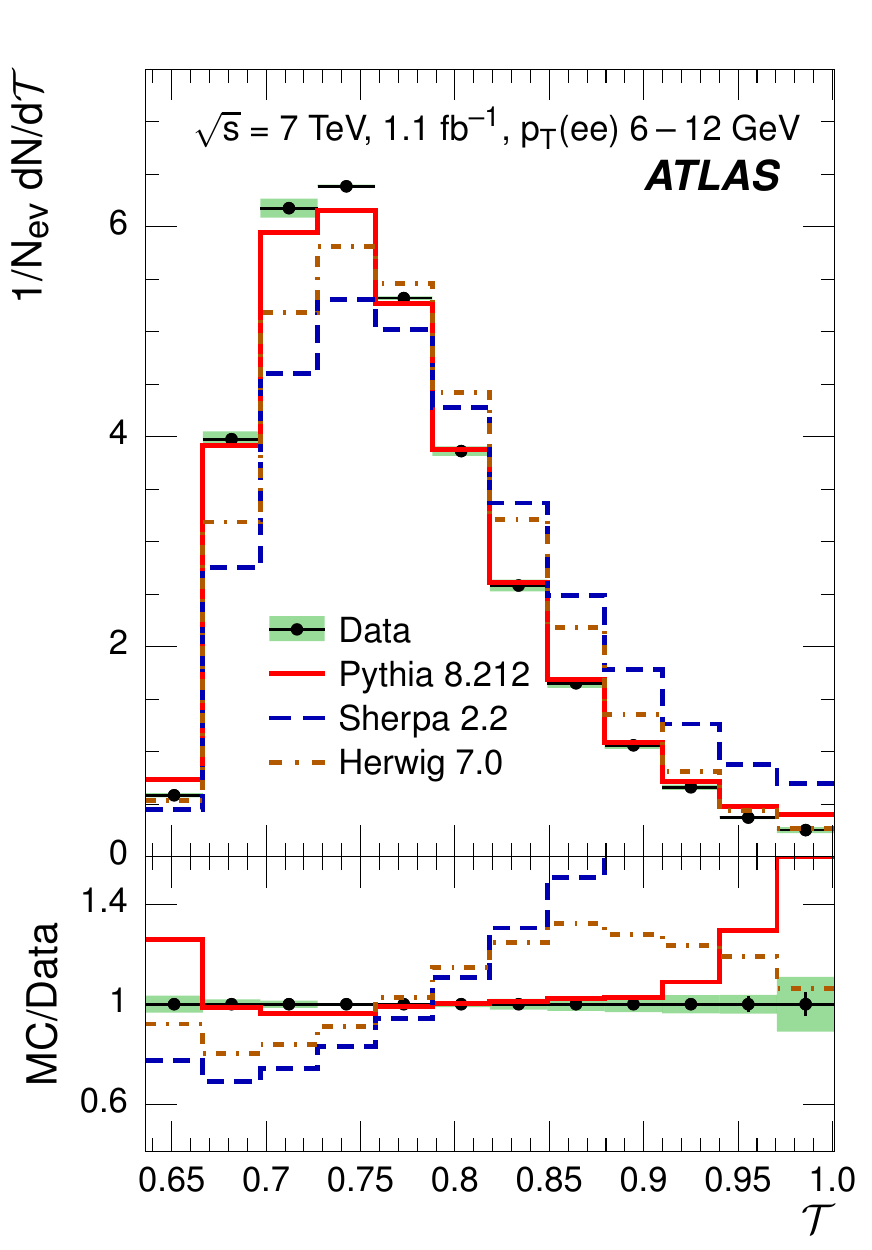}
        \label{fig:results-thrust-6_12-Elec-final}
    }
    \\
    \subfloat[Transverse thrust, $\zptee$: 12--25\;\GeV]{
        \includegraphics[width=.42\textwidth]{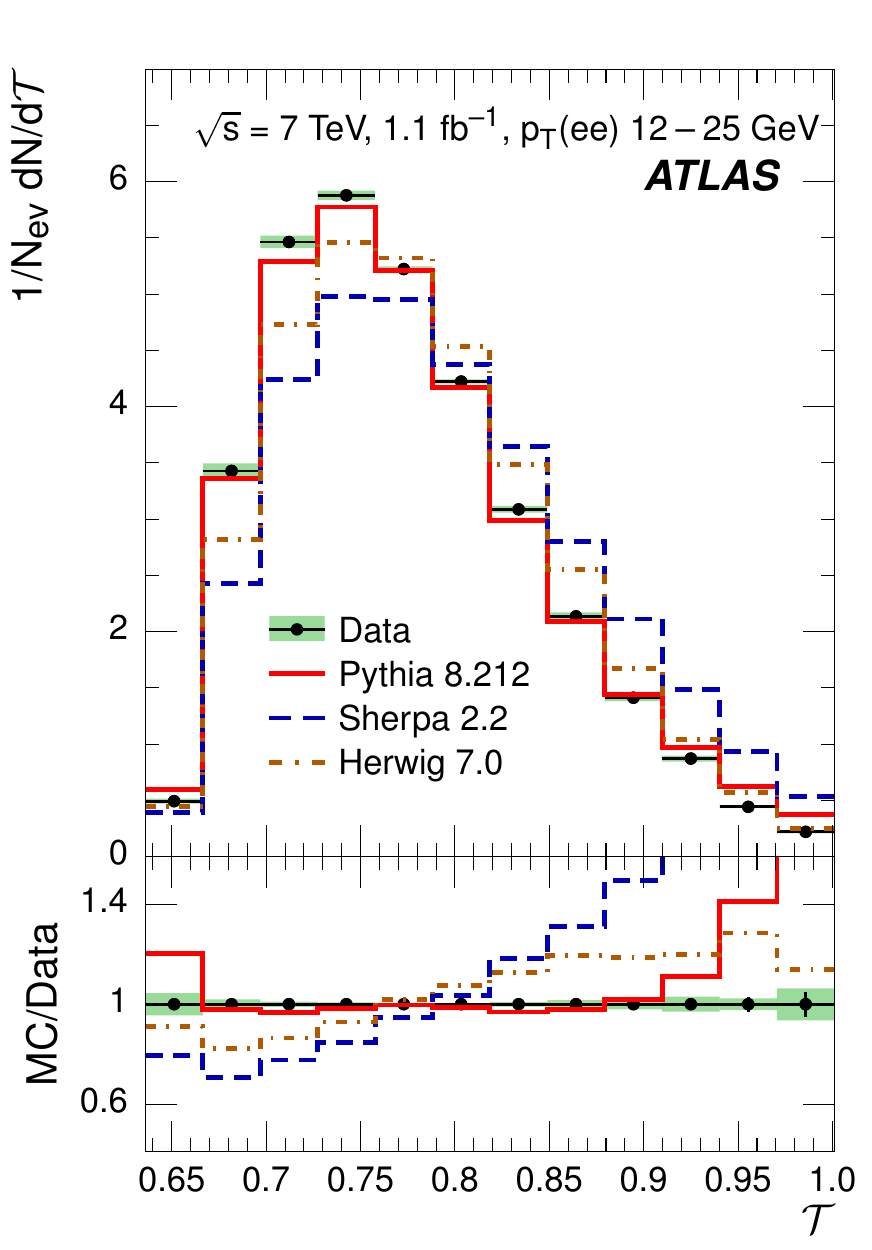}
        \label{fig:results-thrust-12_25-Elec-final}
    }
    \subfloat[Transverse thrust, $\zptee \ge 25$~\GeV]{
        \includegraphics[width=.42\textwidth]{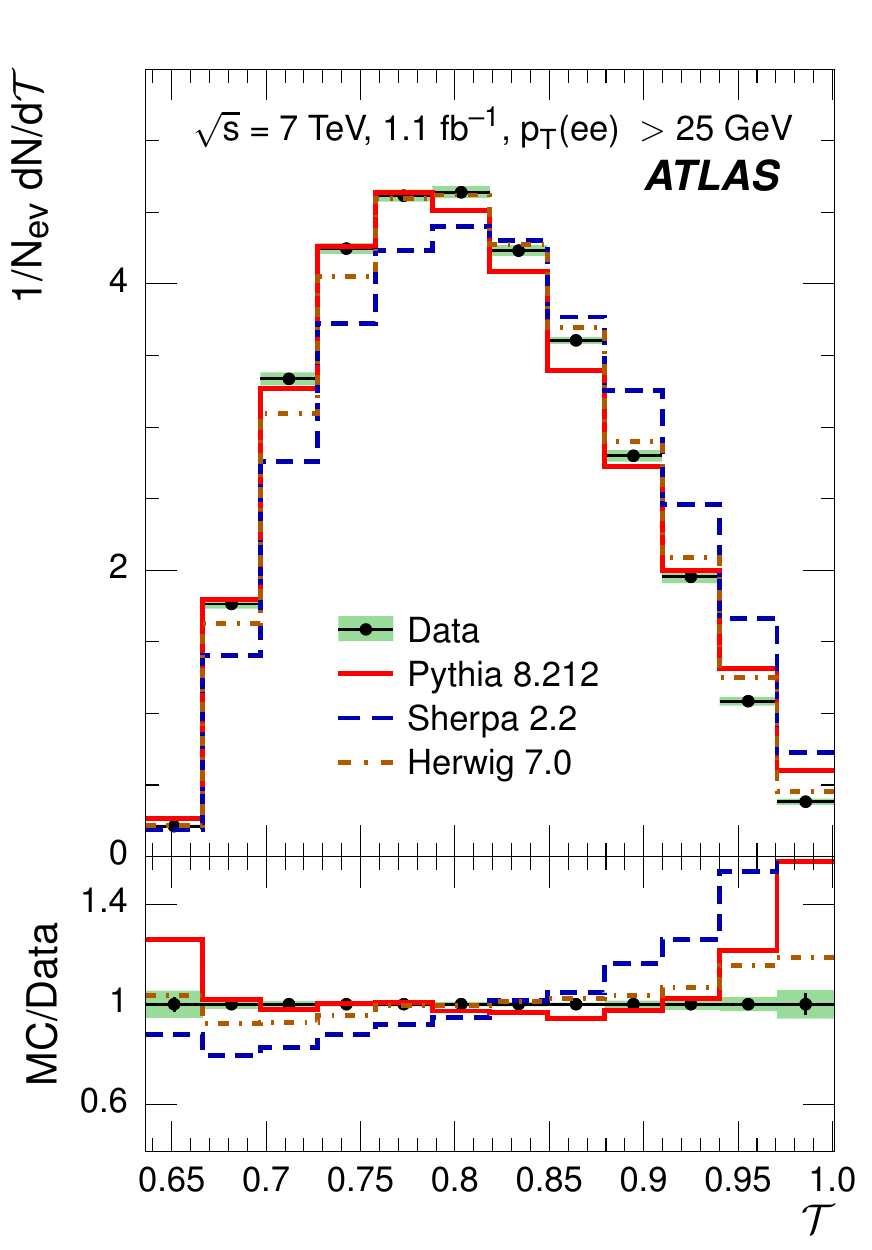}
        \label{fig:results-thrust-25_999-Elec-final}
    }
    \caption[Results, transverse thrust, all phase-spage regions]{Transverse thrust $\mathcal{T}$ distribution of charged particles
for $Z \rightarrow e^{+}e^{-}$ with statistical (error bars) and total systematic (band) uncertainties for the
four $\zptee$ ranges ((a): 0--6\;\GeV, (b): 6--12\;\GeV, (c): 12--25\;\GeV, (d): $\ge 25$~\GeV) compared to the predictions
from the MC generators \pythiaeight (full line), \sherpa (dashed line), and \herwigseven (dashed-dotted line).
In each subfigure, the top plot shows the observable and the bottom plot shows the ratio of the MC simulation to the data.}
    \label{fig:results-thrust-Elec-final}
\end{figure}

\begin{figure}[p]
    \captionsetup[subfigure]{margin=0pt, width=.42\textwidth}
    \centering
    \subfloat[Spherocity, $\zptee$: 0--6\;\GeV]{
        \includegraphics[width=.42\textwidth]{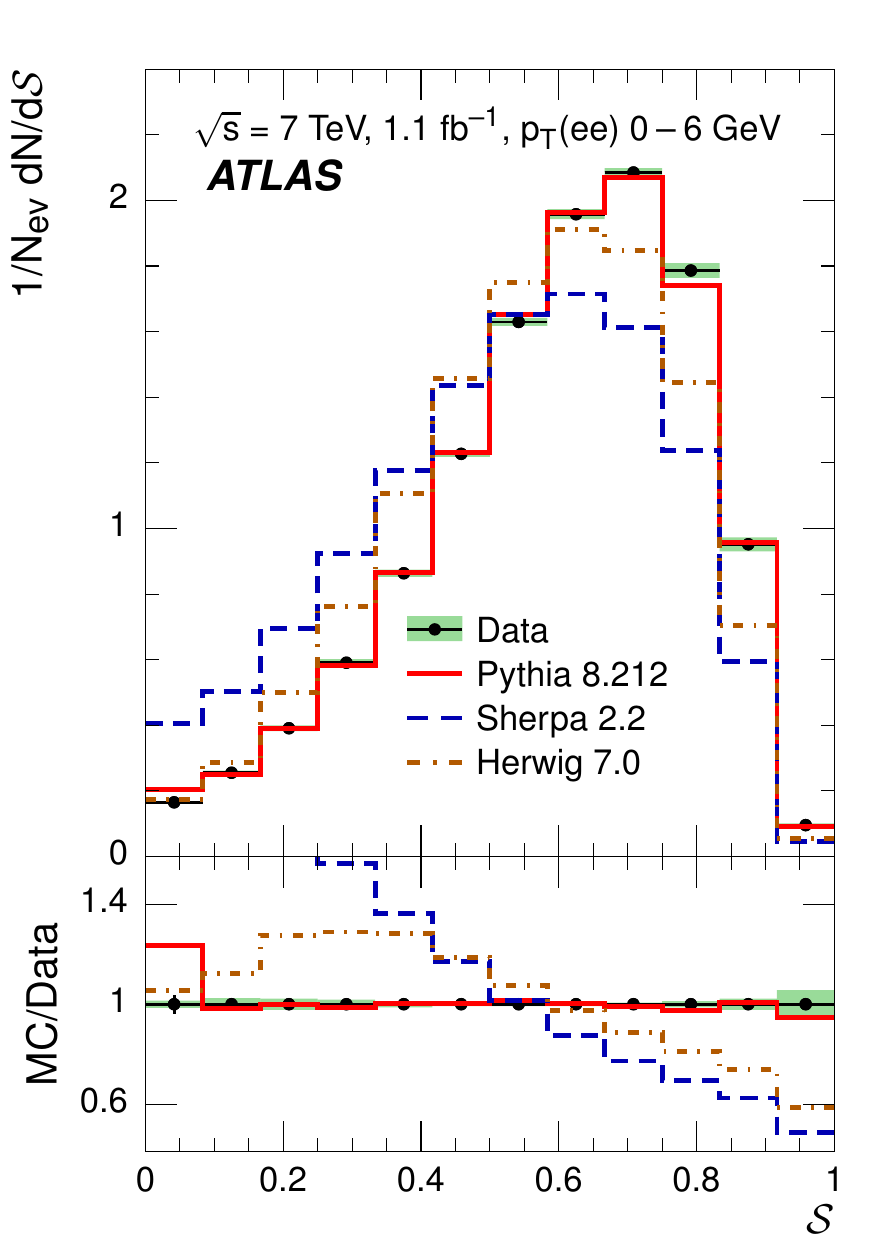}
        \label{fig:results-spherocity-0_6-Elec-final}
    }
    \subfloat[Spherocity, $\zptee$: 6--12\;\GeV]{
        \includegraphics[width=.42\textwidth]{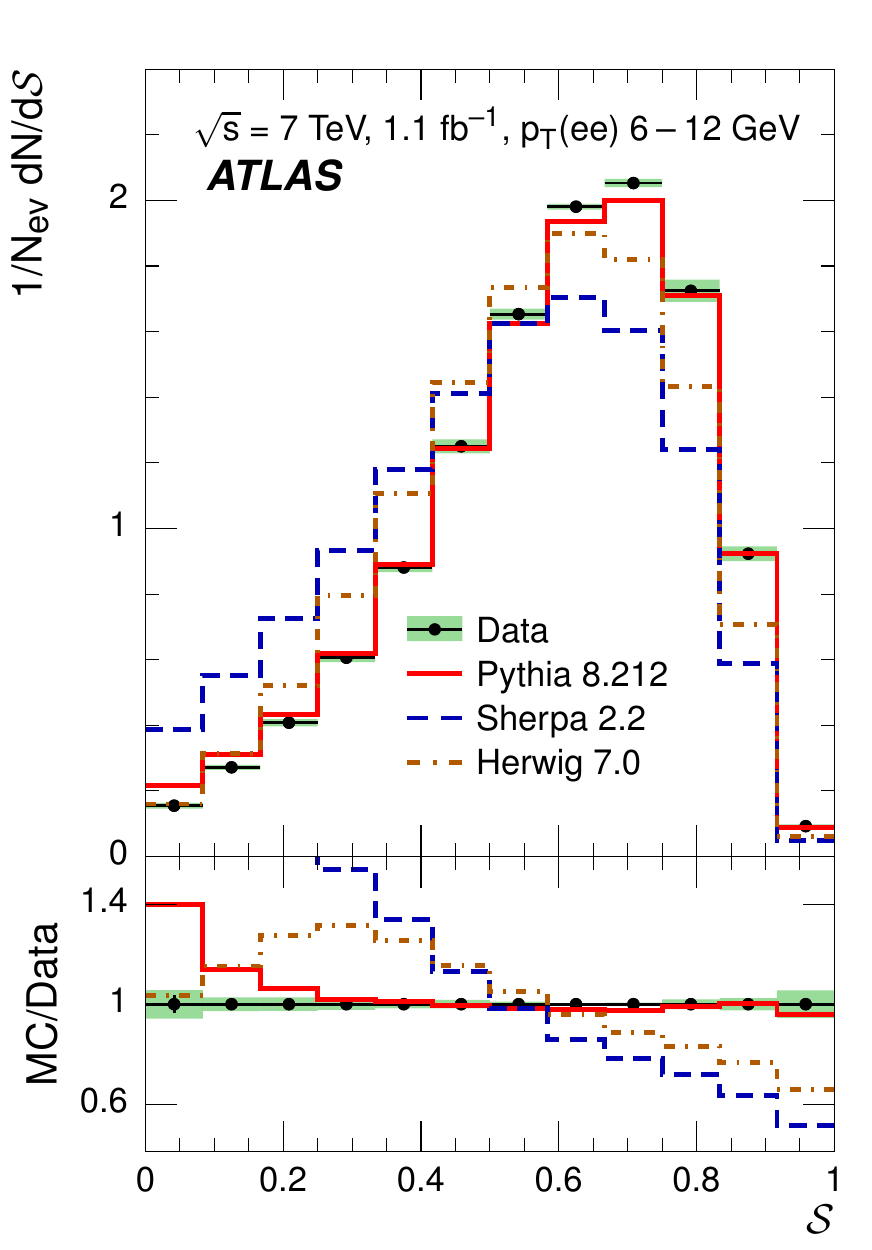}
        \label{fig:results-spherocity-6_12-Elec-final}
    }
    \\
    \subfloat[Spherocity, $\zptee$: 12--25\;\GeV]{
        \includegraphics[width=.42\textwidth]{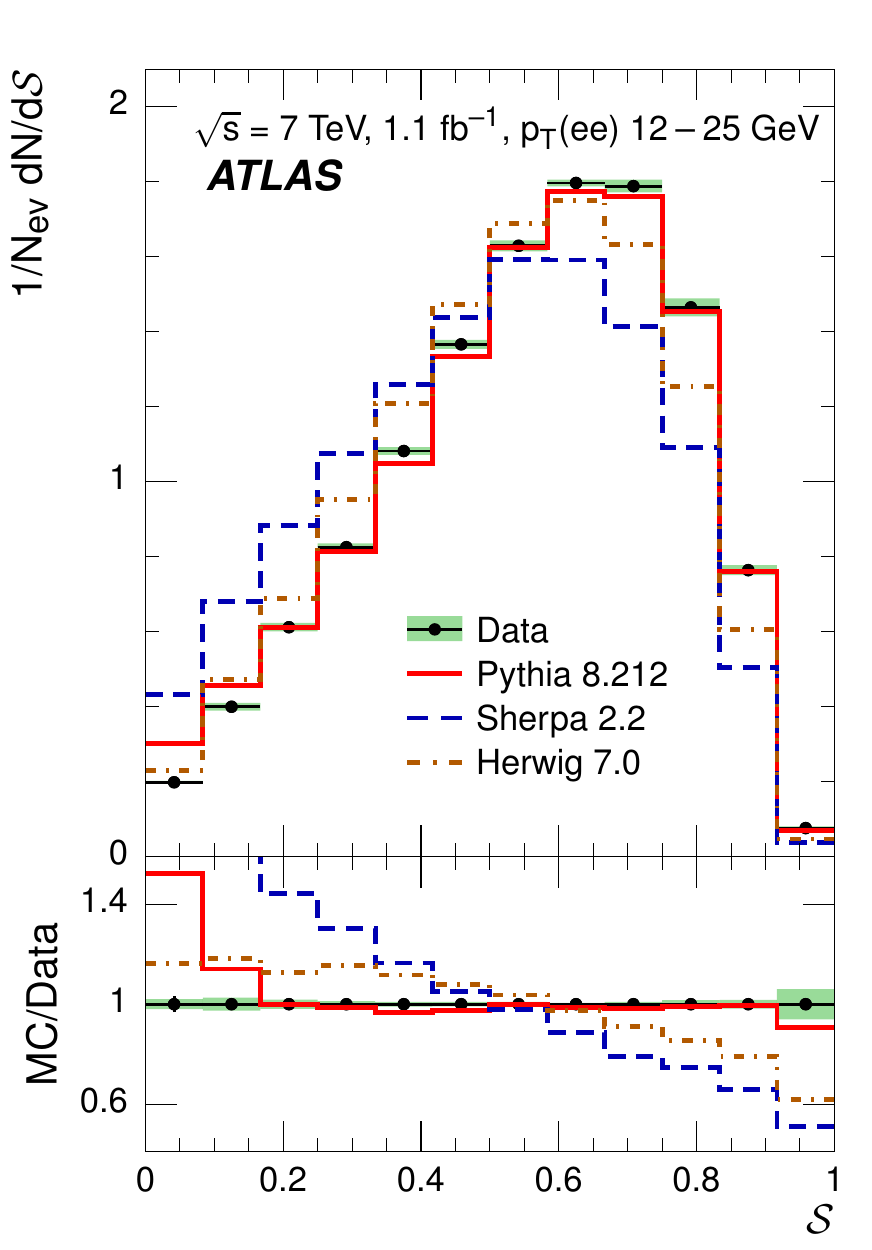}
        \label{fig:results-spherocity-12_25-Elec-final}
    }
    \subfloat[Spherocity, $\zptee \ge 25$~\GeV]{
        \includegraphics[width=.42\textwidth]{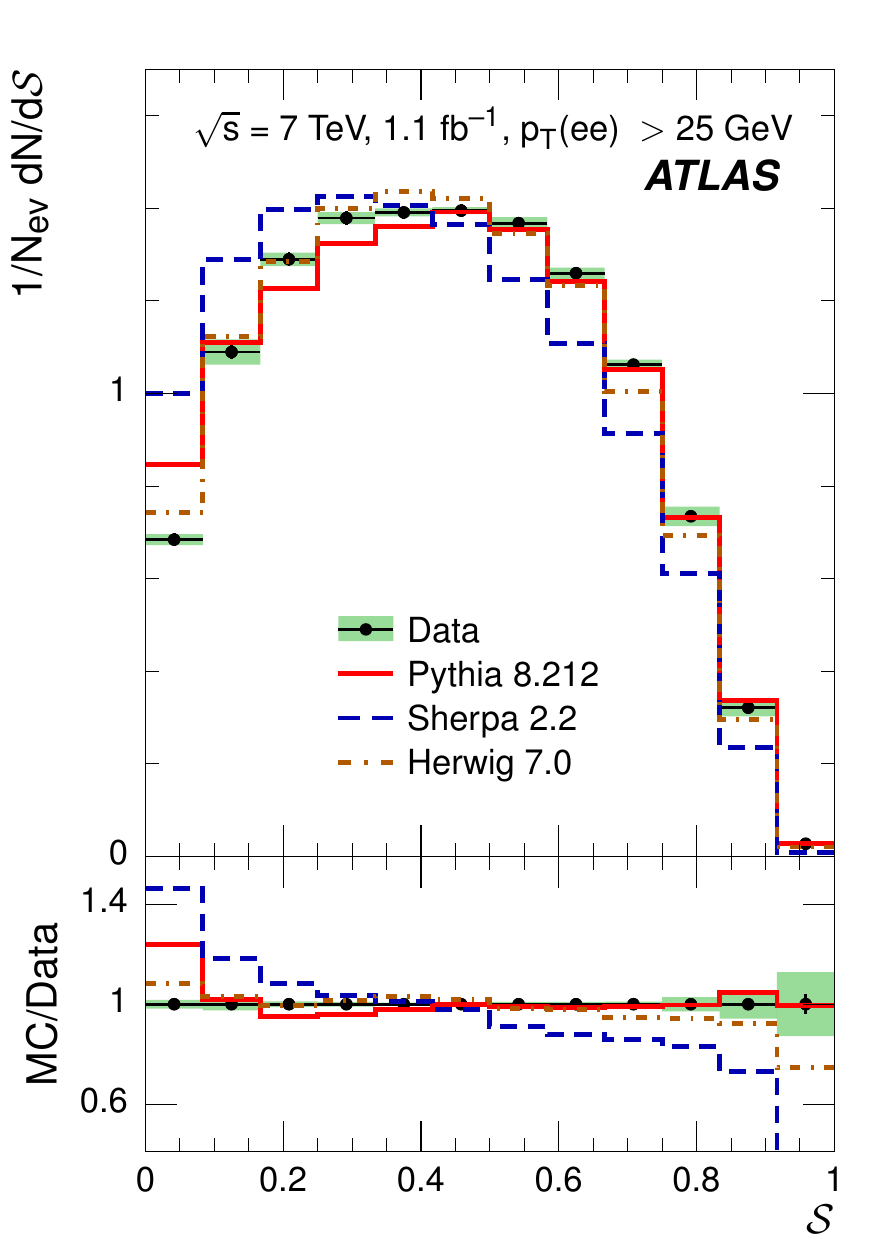}
        \label{fig:results-spherocity-25_999-Elec-final}
    }
    \caption[Results, spherocity, all phase-spage regions]{Spherocity $\mathcal{S}$ distribution of charged particles
for $Z \rightarrow e^{+}e^{-}$ with statistical (error bars) and total systematic (band) uncertainties for the
four $\zptee$ ranges ((a): 0--6\;\GeV, (b): 6--12\;\GeV, (c): 12--25\;\GeV, (d): $\ge 25$~\GeV) compared to the predictions
from the MC generators \pythiaeight (full line), \sherpa (dashed line), and \herwigseven (dashed-dotted line).
In each subfigure, the top plot shows the observable and the bottom plot shows the ratio of the MC simulation to the data.}
    \label{fig:results-spherocity-Elec-final}
\end{figure}

\begin{figure}[p]
    \captionsetup[subfigure]{margin=0pt, width=.42\textwidth}
    \centering
    \subfloat[$\mathcal{F}$-parameter, $\zptee$: 0--6\;\GeV]{
        \includegraphics[width=.42\textwidth]{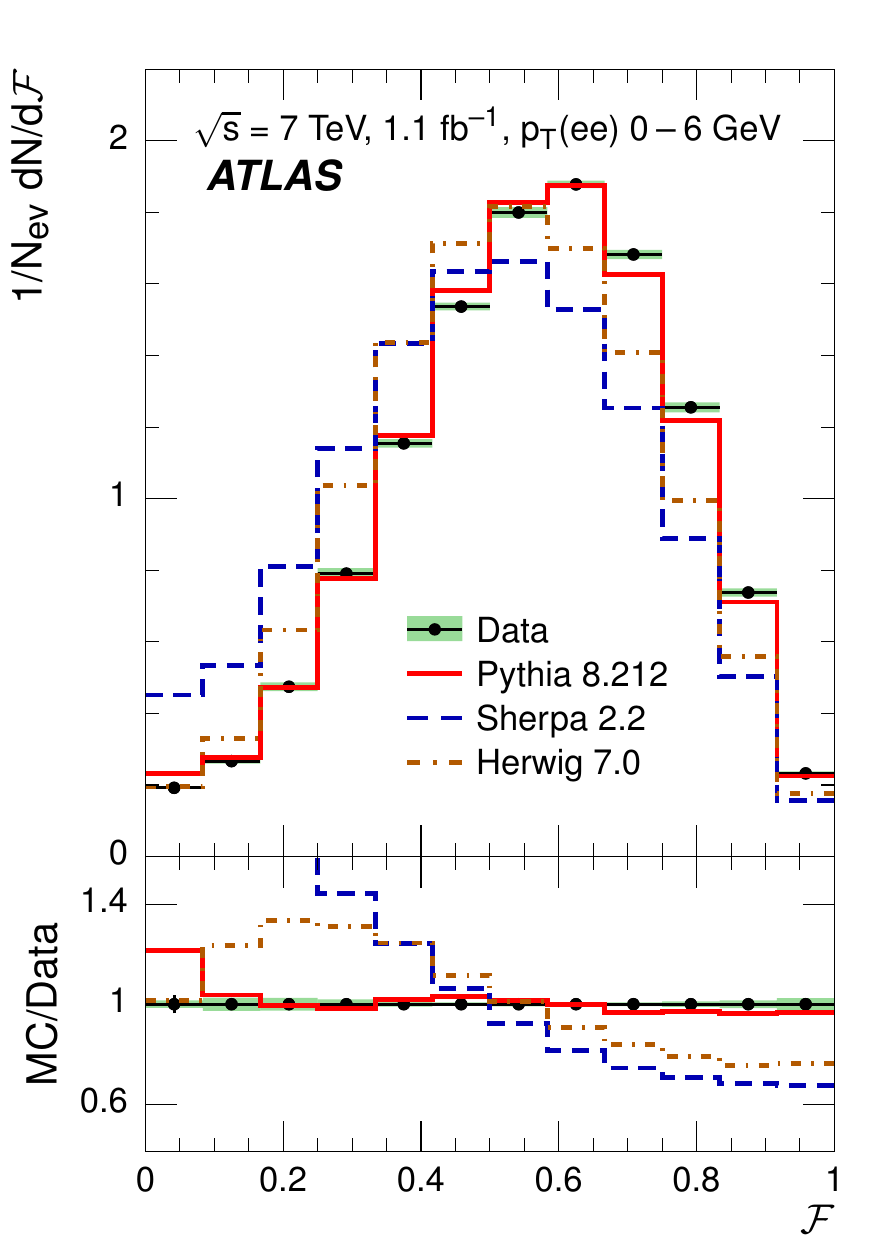}
        \label{fig:results-fparameter-0_6-Elec-final}
    }
    \subfloat[$\mathcal{F}$-parameter, $\zptee$: 6--12\;\GeV]{
        \includegraphics[width=.42\textwidth]{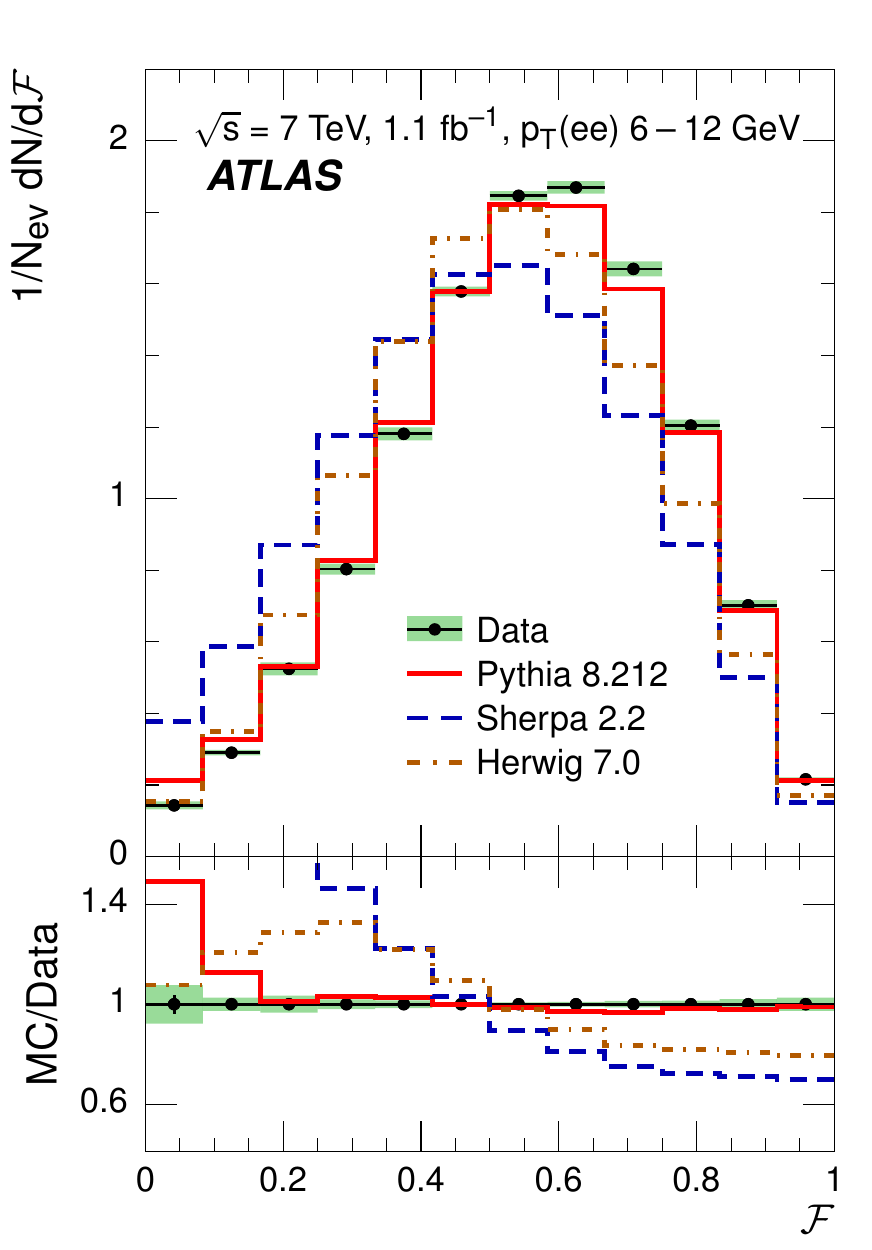}
        \label{fig:results-fparameter-6_12-Elec-final}
    }
    \\
    \subfloat[$\mathcal{F}$-parameter, $\zptee$: 12--25\;\GeV]{
        \includegraphics[width=.42\textwidth]{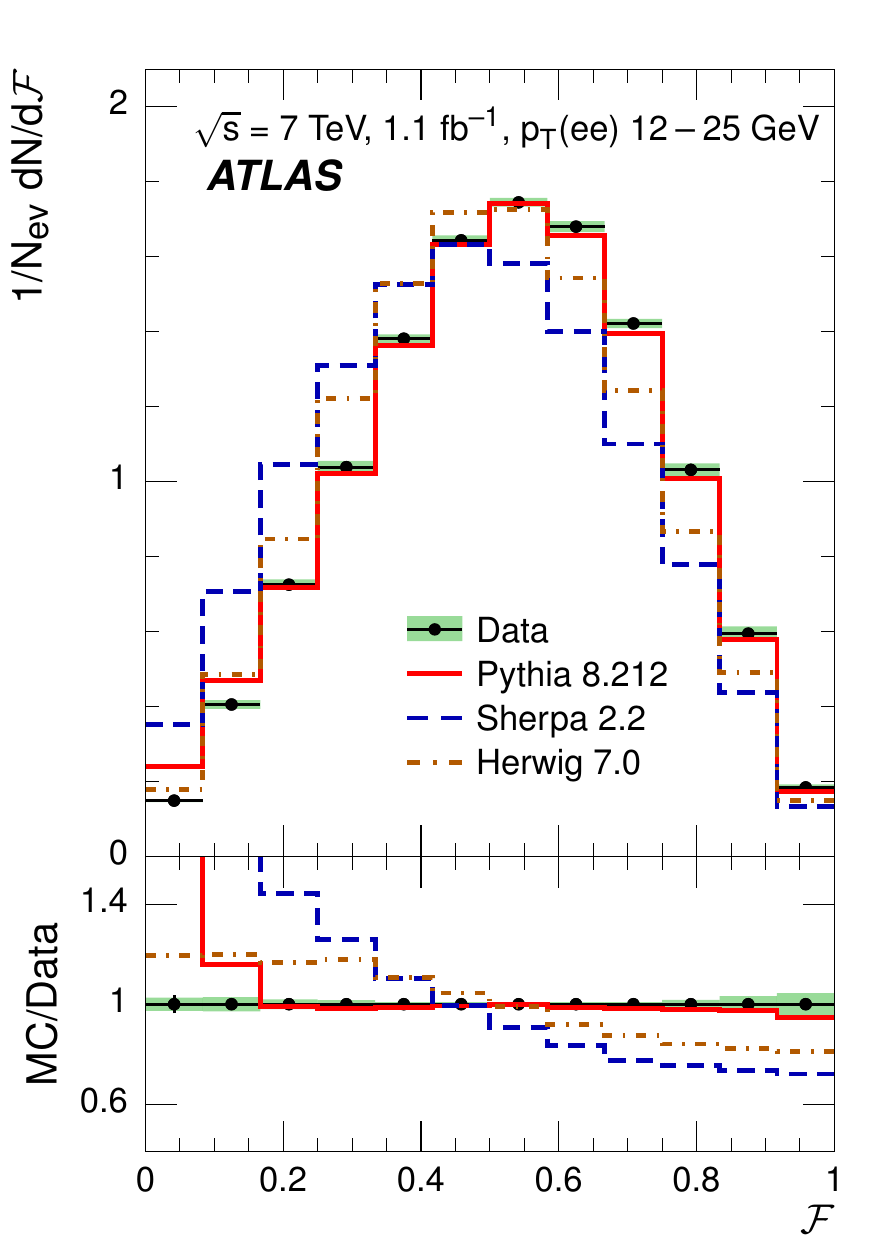}
        \label{fig:results-fparameter-12_25-Elec-final}
    }
    \subfloat[$\mathcal{F}$-parameter, $\zptee \ge 25$~\GeV]{
        \includegraphics[width=.42\textwidth]{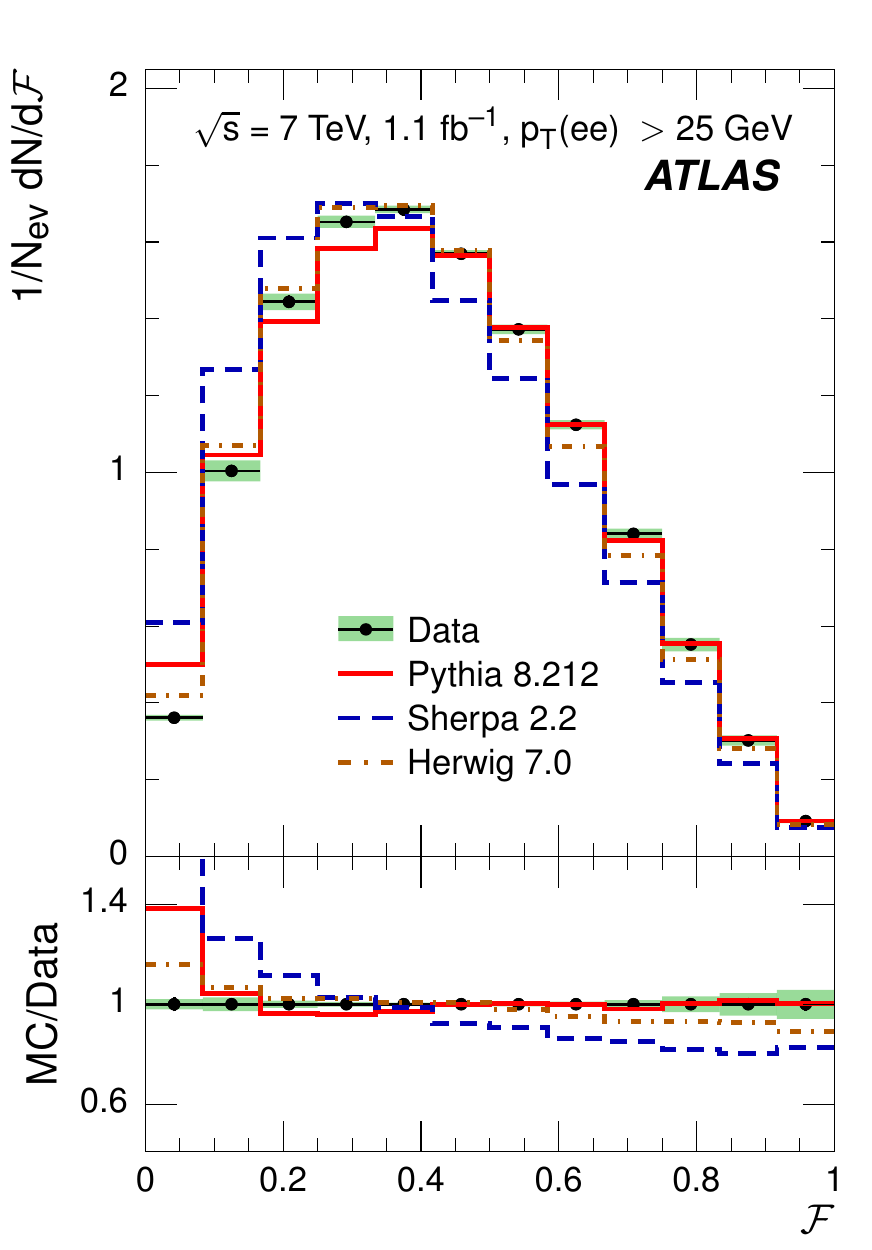}
        \label{fig:results-fparameter-25_999-Elec-final}
    }
    \caption[Results, $\mathcal{F}$-parameter, all phase-spage regions]{$\mathcal{F}$-parameter distribution of charged particles
for $Z \rightarrow e^{+}e^{-}$ with statistical (error bars) and total systematic (band) uncertainties for the
four $\zptee$ ranges ((a): 0--6\;\GeV, (b): 6--12\;\GeV, (c): 12--25\;\GeV, (d): $\ge 25$~\GeV) compared to the predictions
from the MC generators \pythiaeight (full line), \sherpa (dashed line), and \herwigseven (dashed-dotted line).
In each subfigure, the top plot shows the observable and the bottom plot shows the ratio of the MC simulation to the data.}
    \label{fig:results-fparameter-Elec-final}
\end{figure}

\begin{figure}[p]
    \captionsetup[subfigure]{margin=0pt, width=.42\textwidth}
    \centering
    \subfloat[$N_\text{ch}$, $\zptmumu$: 0--6\;\GeV]{
        \includegraphics[width=.42\textwidth]{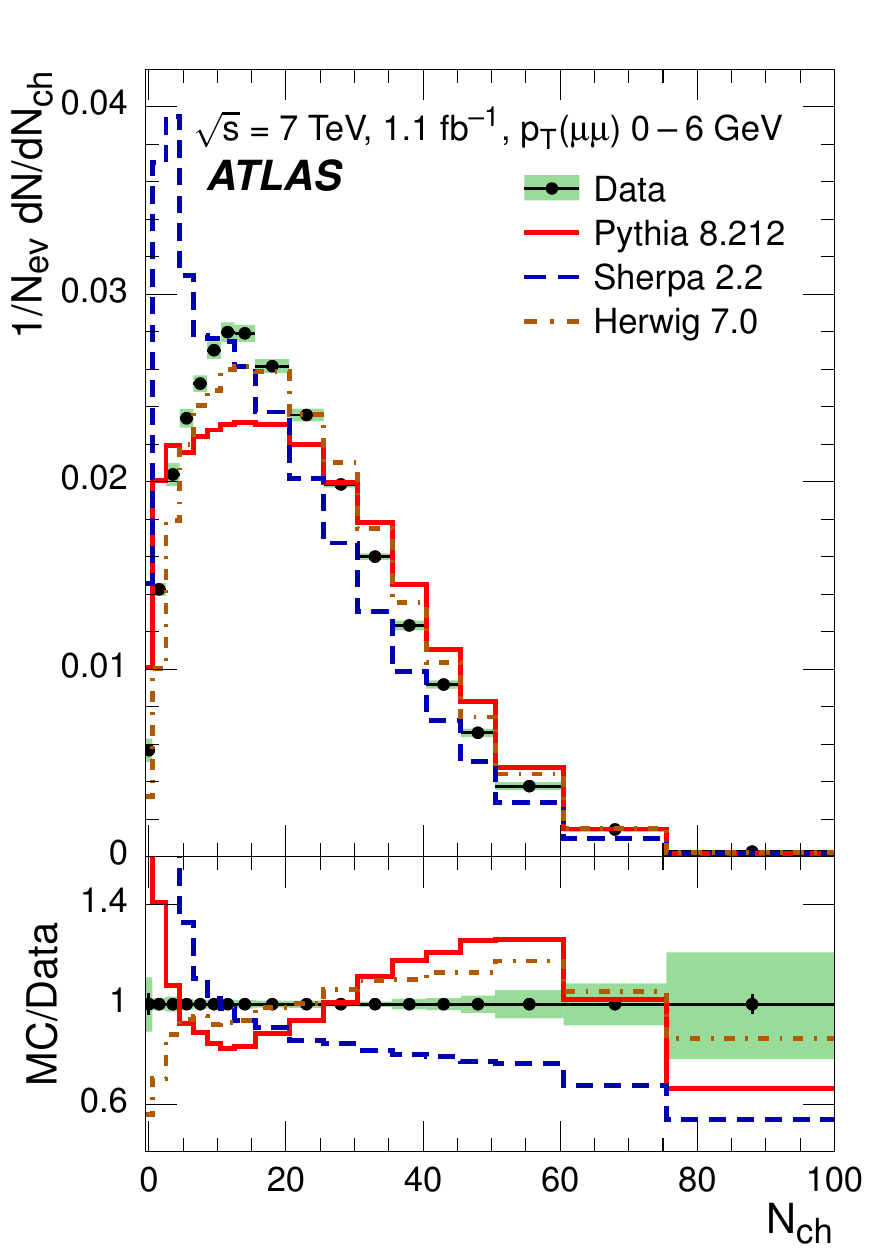}
        \label{fig:results-nch-0_6-Muon-final}
    }
    \subfloat[$N_\text{ch}$, $\zptmumu$: 6--12\;\GeV]{
        \includegraphics[width=.42\textwidth]{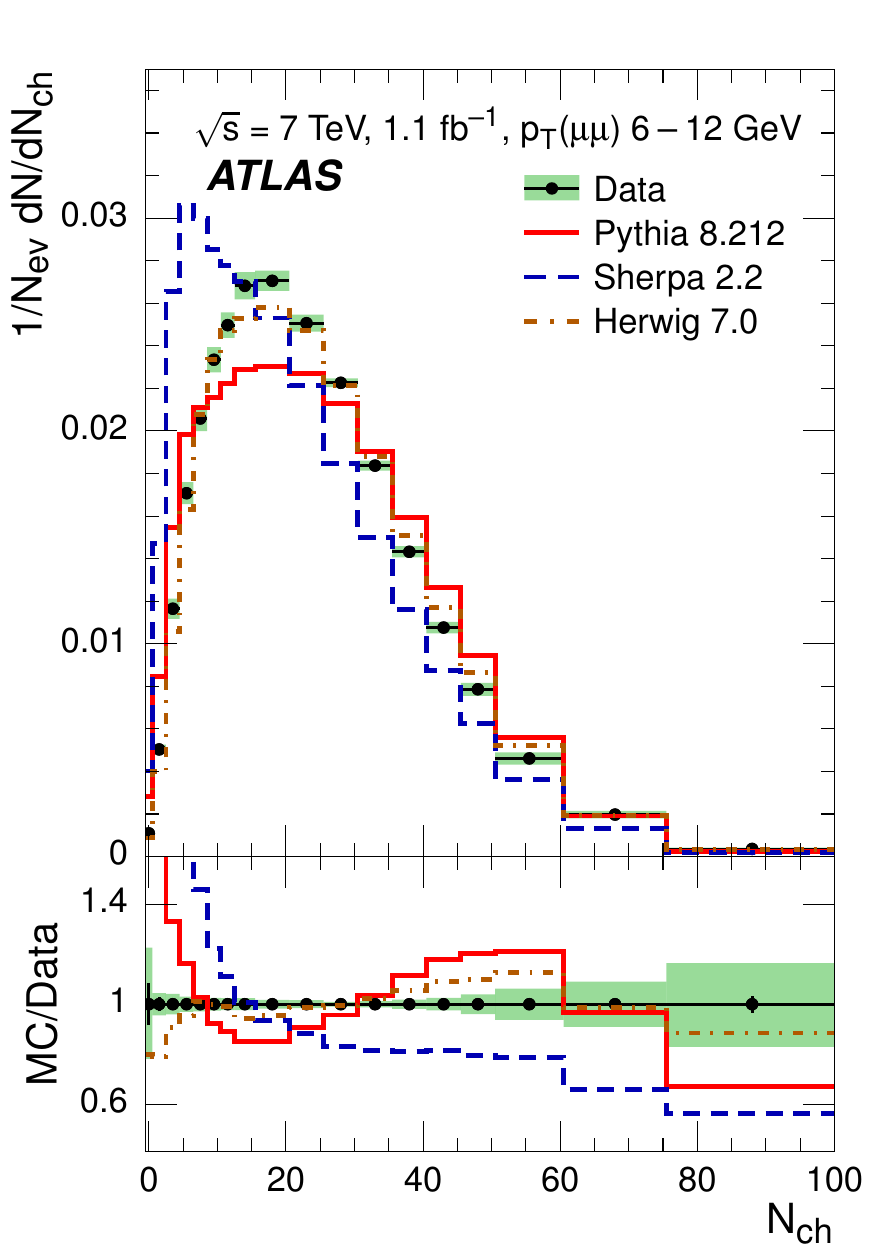}
        \label{fig:results-nch-6_12-Muon-final}
    }
    \\
    \subfloat[$N_\text{ch}$, $\zptmumu$: 12--25\;\GeV]{
        \includegraphics[width=.42\textwidth]{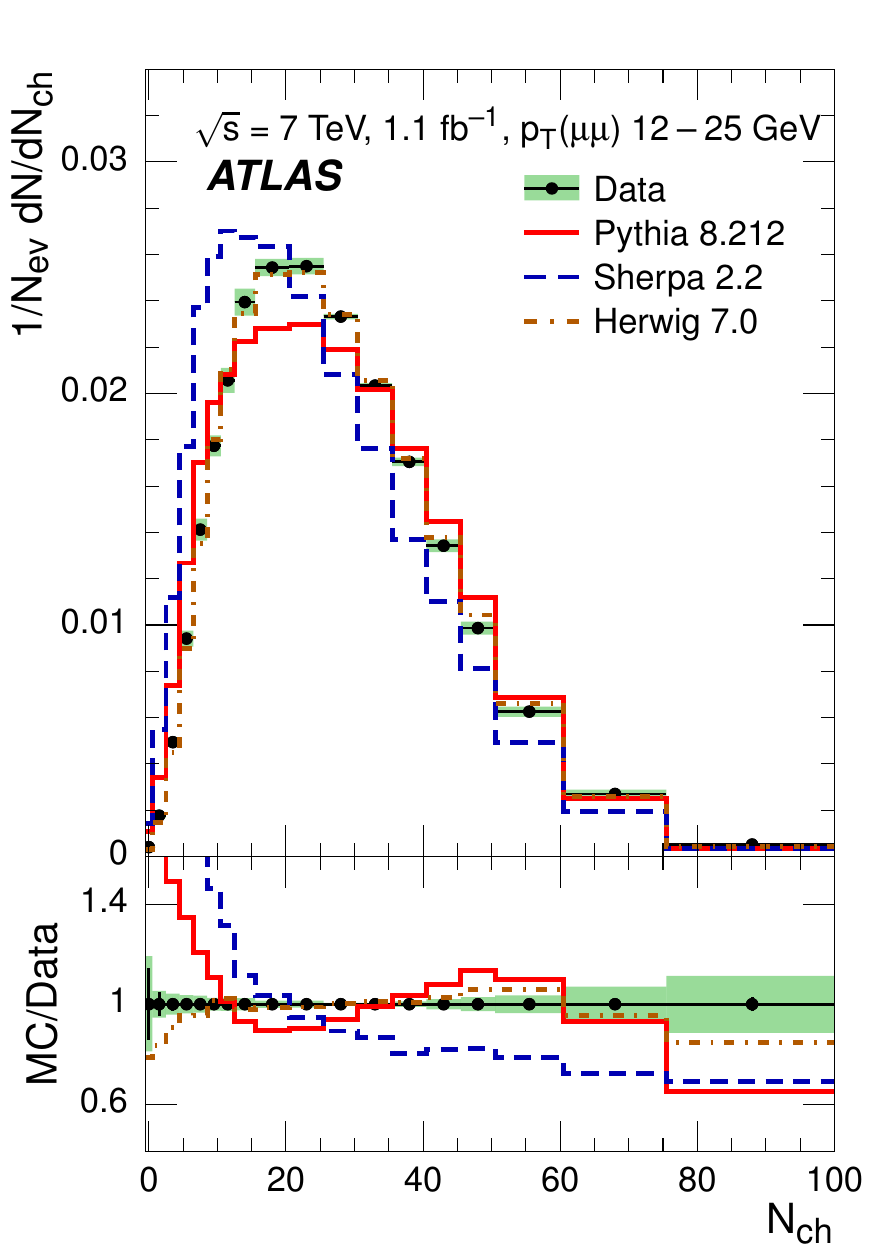}
        \label{fig:results-nch-12_25-Muon-final}
    }
    \subfloat[$N_\text{ch}$, $\zptmumu \ge 25$~\GeV]{
        \includegraphics[width=.42\textwidth]{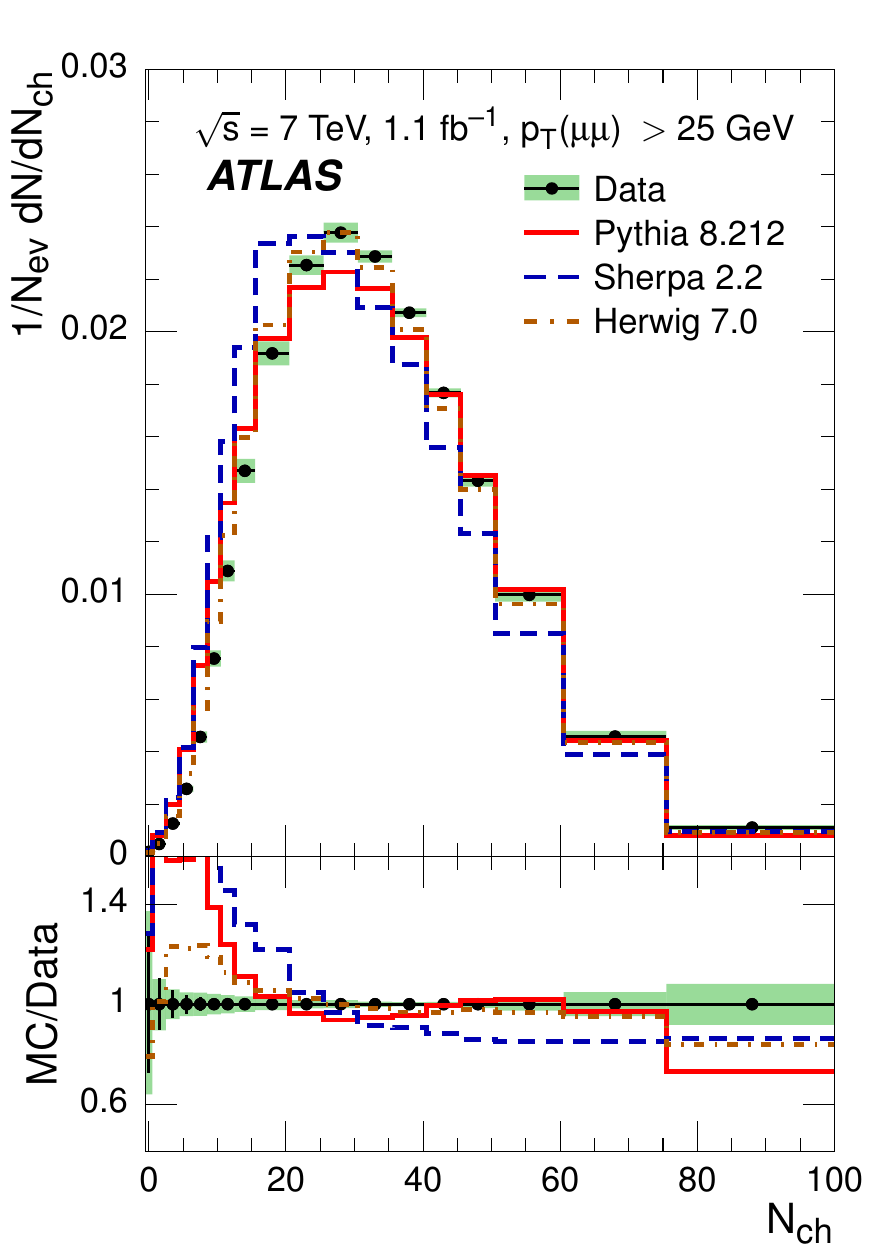}
        \label{fig:results-nch-25_999-Muon-final}
    }
    \caption[Results, $N_\text{ch}$, all phase-spage regions]{Distribution of charged-particle multiplicity, $N_\text{ch}$,
for $Z \rightarrow \mu^{+}\mu^{-}$ with statistical (error bars) and total systematic (band) uncertainties for the
four $\zptmumu$ ranges ((a): 0--6\;\GeV, (b): 6--12\;\GeV, (c): 12--25\;\GeV, (d): $\ge 25$~\GeV) compared to the predictions
from the MC generators \pythiaeight (full line), \sherpa (dashed line), and \herwigseven (dashed-dotted line).
In each subfigure, the top plot shows the observable and the bottom plot shows the ratio of the MC simulation to the data.}
    \label{fig:results-nch-Muon-final}
\end{figure}
\begin{figure}[p]
    \captionsetup[subfigure]{margin=0pt, width=.42\textwidth}
    \centering
    \subfloat[$\sumpt$, $\zptmumu$: 0--6\;\GeV]{
        \includegraphics[width=.42\textwidth]{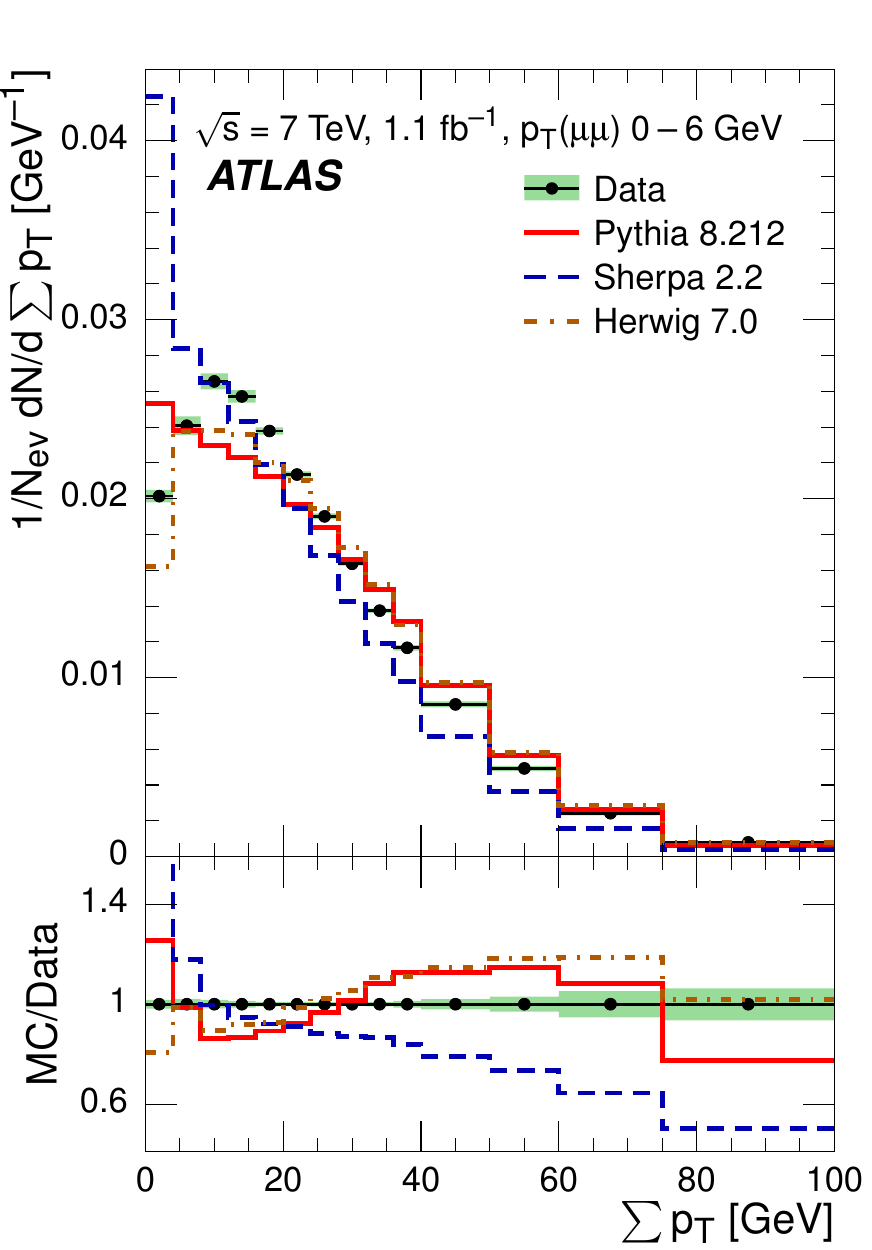}
        \label{fig:results-sumpt-0_6-Muon-final}
    }
    \subfloat[$\sumpt$, $\zptmumu$: 6--12\;\GeV]{
        \includegraphics[width=.42\textwidth]{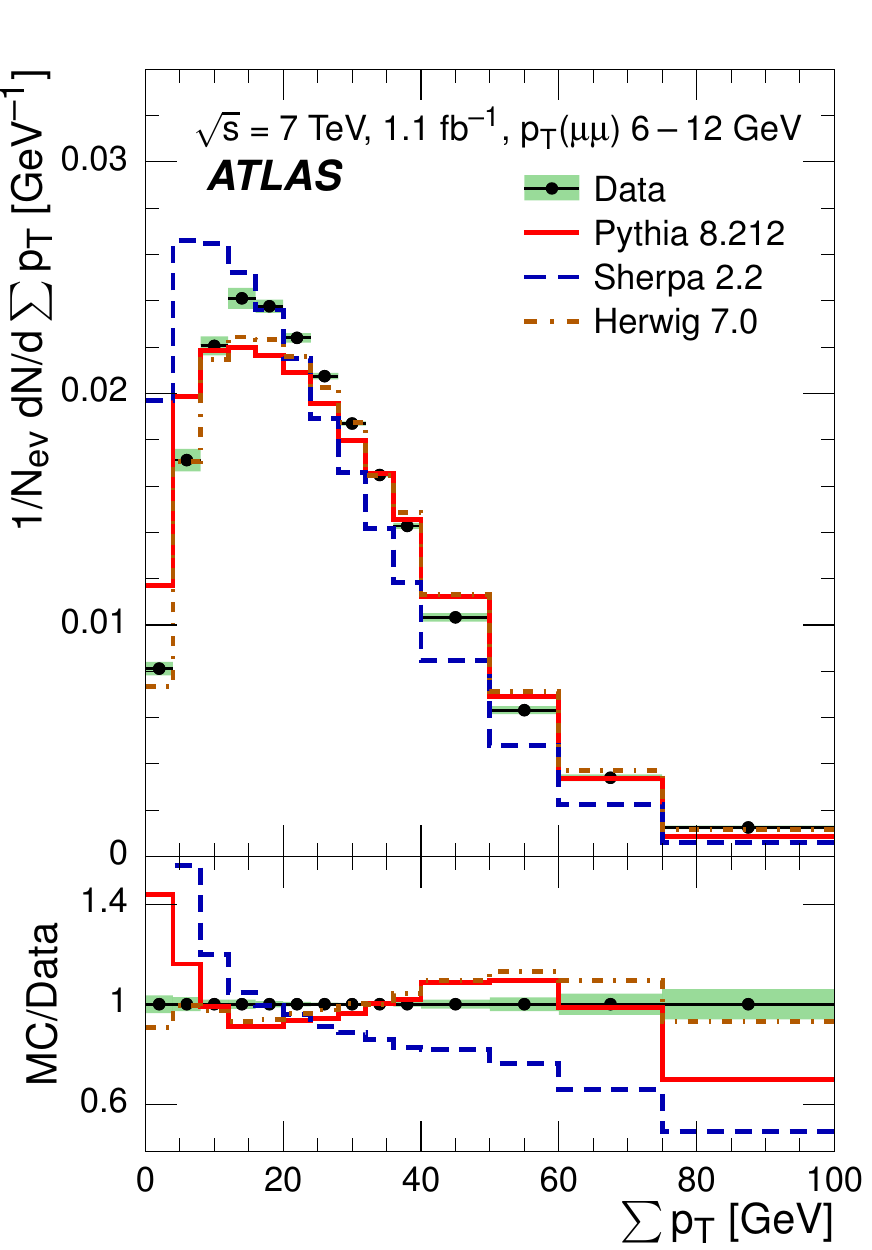}
        \label{fig:results-sumpt-6_12-Muon-final}
    }
    \\
    \subfloat[$\sumpt$, $\zptmumu$: 12--25\;\GeV]{
        \includegraphics[width=.42\textwidth]{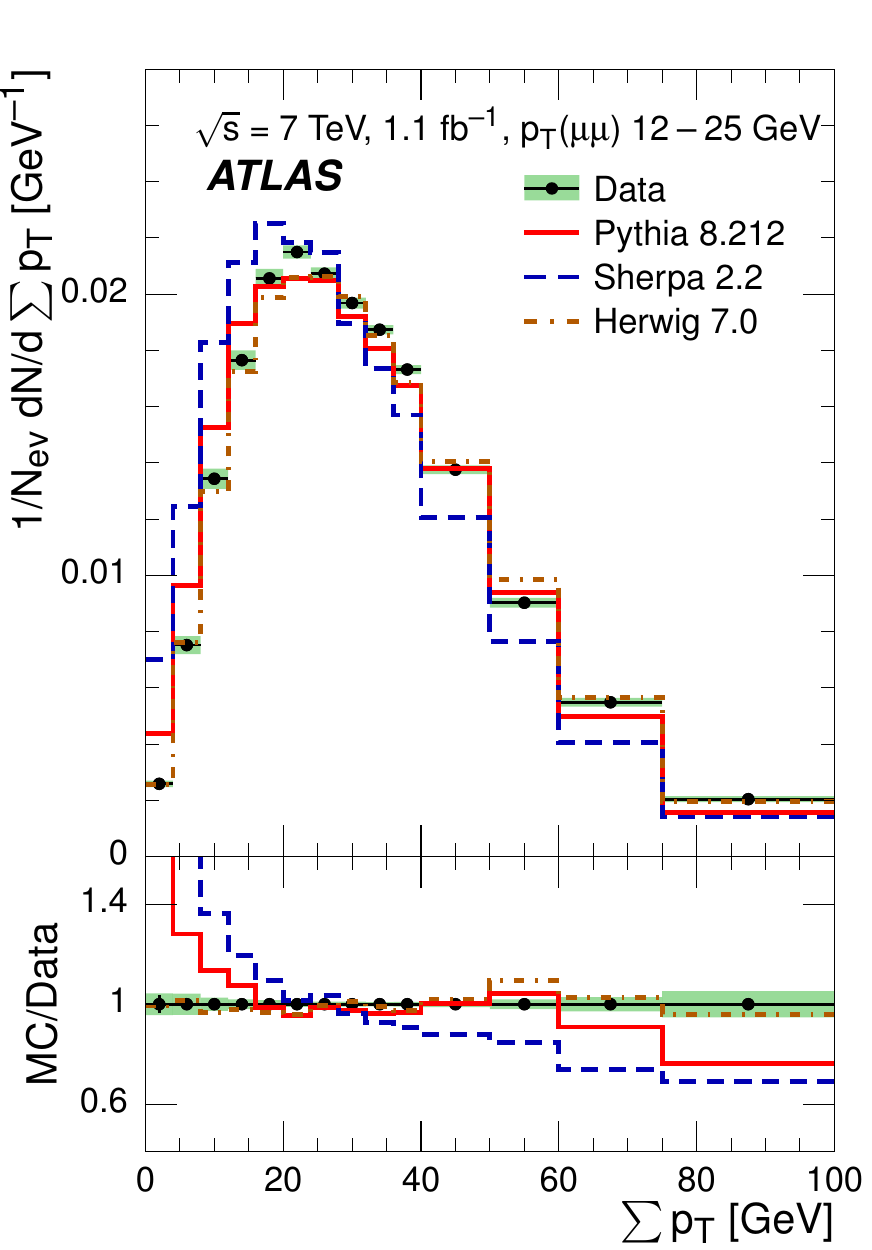}
        \label{fig:results-sumpt-12_25-Muon-final}
    }
    \subfloat[$\sumpt$, $\zptmumu \ge 25$~\GeV]{
        \includegraphics[width=.42\textwidth]{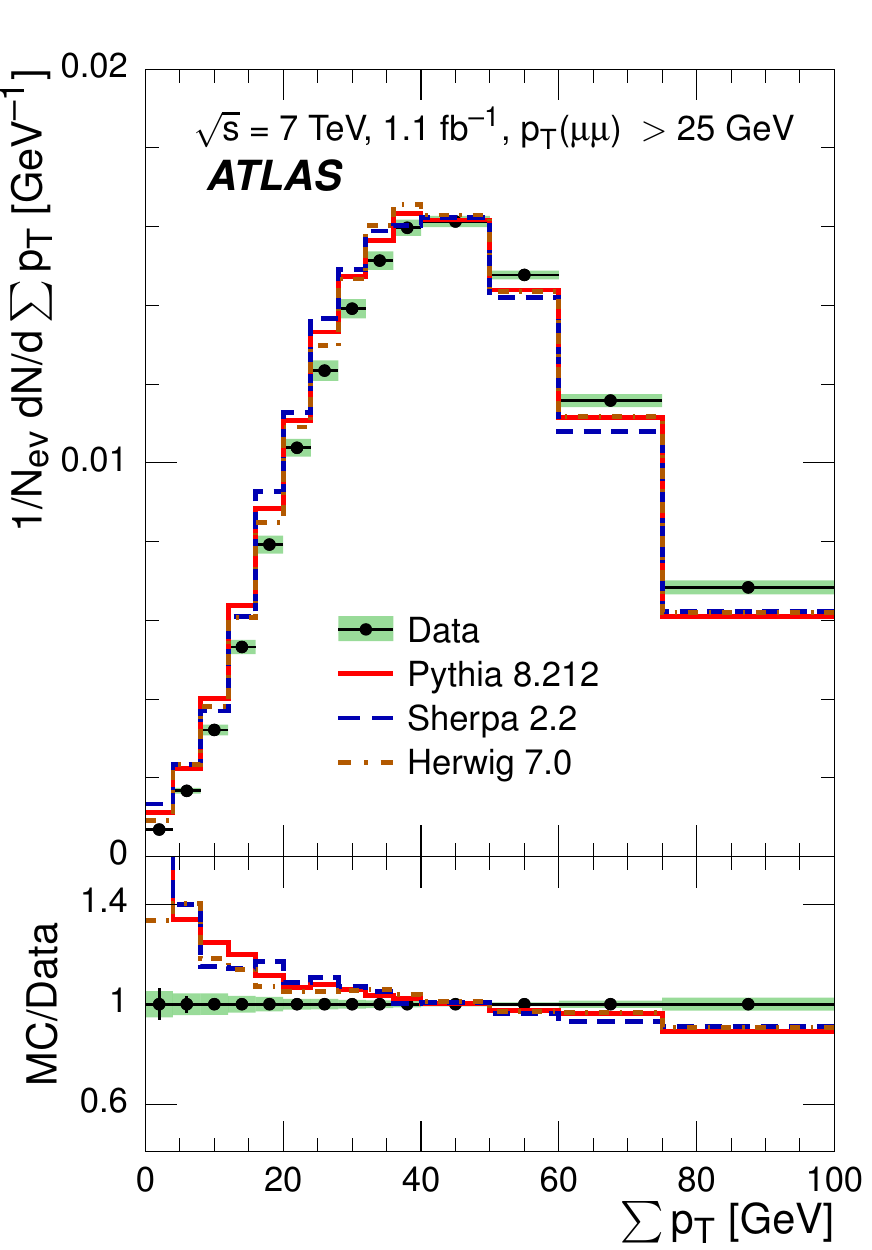}
        \label{fig:results-sumpt-25_999-Muon-final}
    }
    \caption[Results, $\sumpt$, all phase-spage regions]{Summed transverse momenta $\sumpt$ distribution of charged particles
for $Z \rightarrow \mu^{+}\mu^{-}$ with statistical (error bars) and total systematic (band) uncertainties for the
four $\zptmumu$ ranges ((a): 0--6\;\GeV, (b): 6--12\;\GeV, (c): 12--25\;\GeV, (d): $\ge 25$~\GeV) compared to the predictions
from the MC generators \pythiaeight (full line), \sherpa (dashed line), and \herwigseven (dashed-dotted line).
In each subfigure, the top plot shows the observable and the bottom plot shows the ratio of the MC simulation to the data.}
    \label{fig:results-sumpt-Muon-final}
\end{figure}
\begin{figure}[p]
    \captionsetup[subfigure]{margin=0pt, width=.42\textwidth}
    \centering
    \subfloat[Beam thrust, $\zptmumu$: 0--6\;\GeV]{
        \includegraphics[width=.42\textwidth]{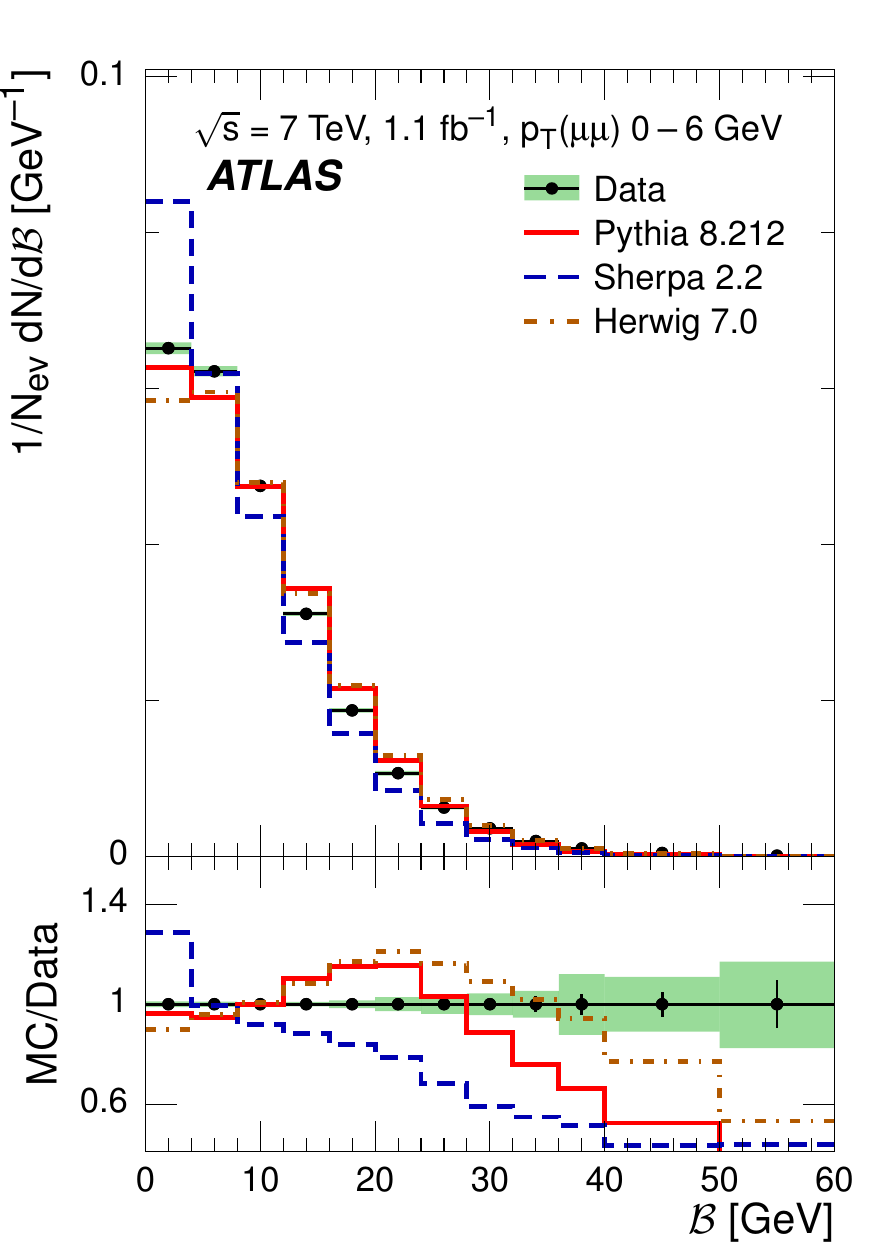}
        \label{fig:results-beamthrust-0_6-Muon-final}
    }
    \subfloat[Beam thrust, $\zptmumu$: 6--12\;\GeV]{
        \includegraphics[width=.42\textwidth]{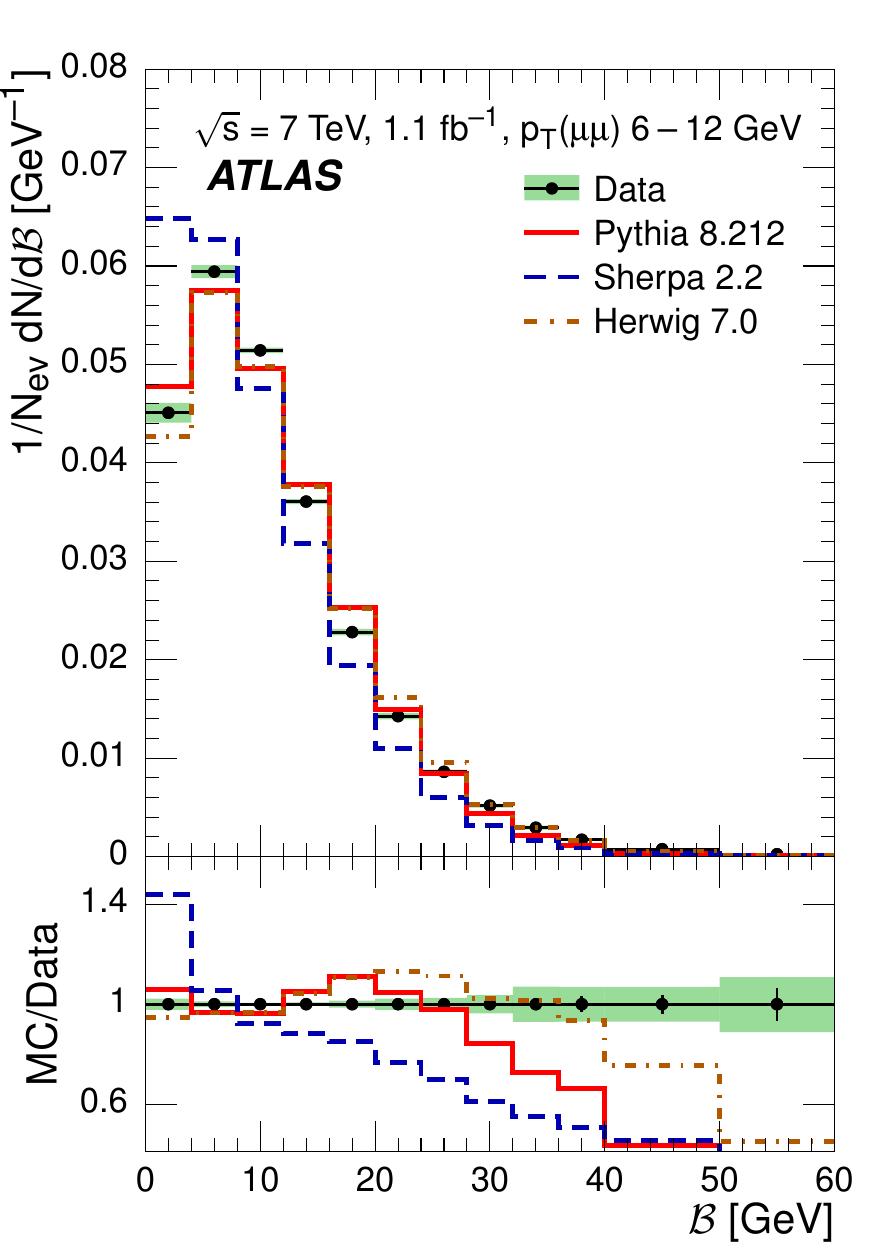}
        \label{fig:results-beamthrust-6_12-Muon-final}
    }
    \\
    \subfloat[Beam thrust, $\zptmumu$: 12--25\;\GeV]{
        \includegraphics[width=.42\textwidth]{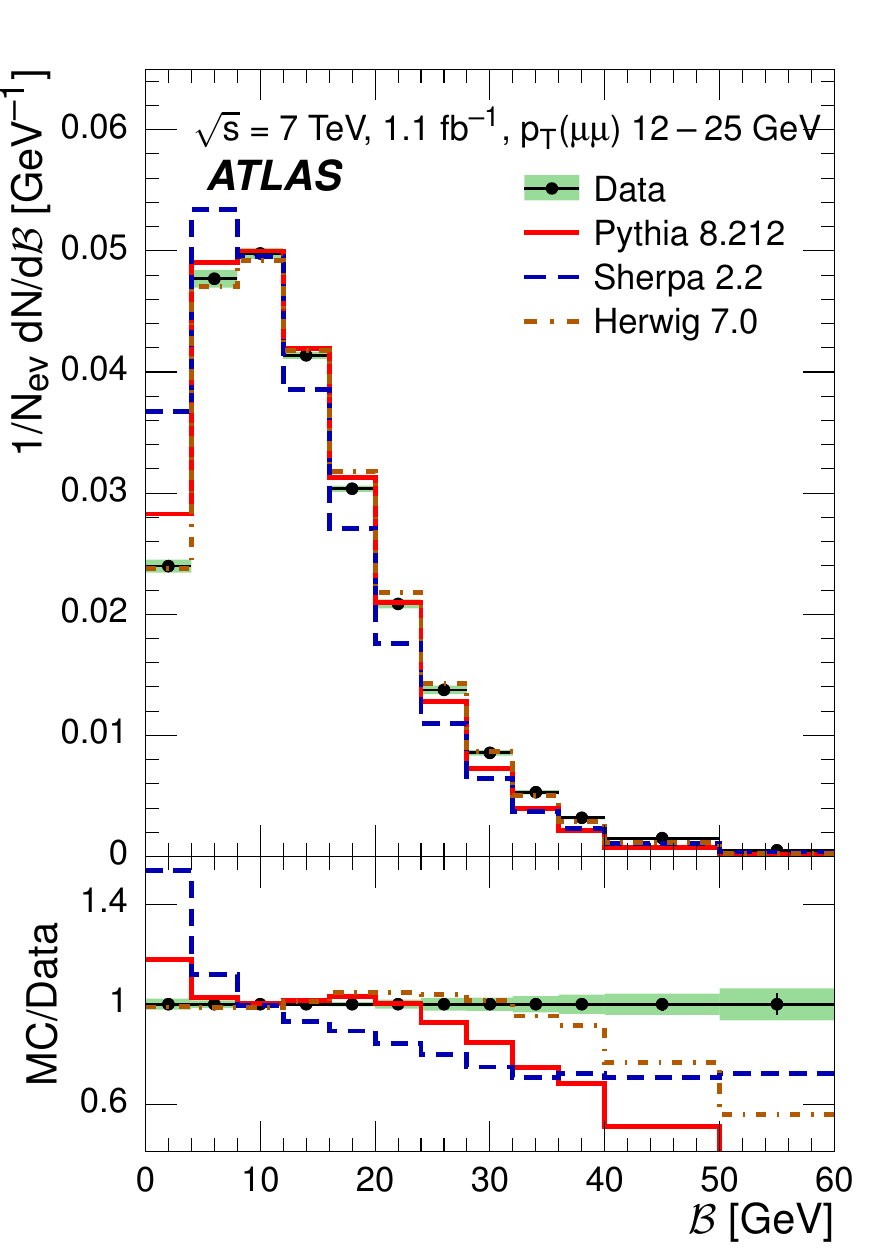}
        \label{fig:results-beamthrust-12_25-Muon-final}
    }
    \subfloat[Beamt thrust, $\zptmumu \ge 25$~\GeV]{
        \includegraphics[width=.42\textwidth]{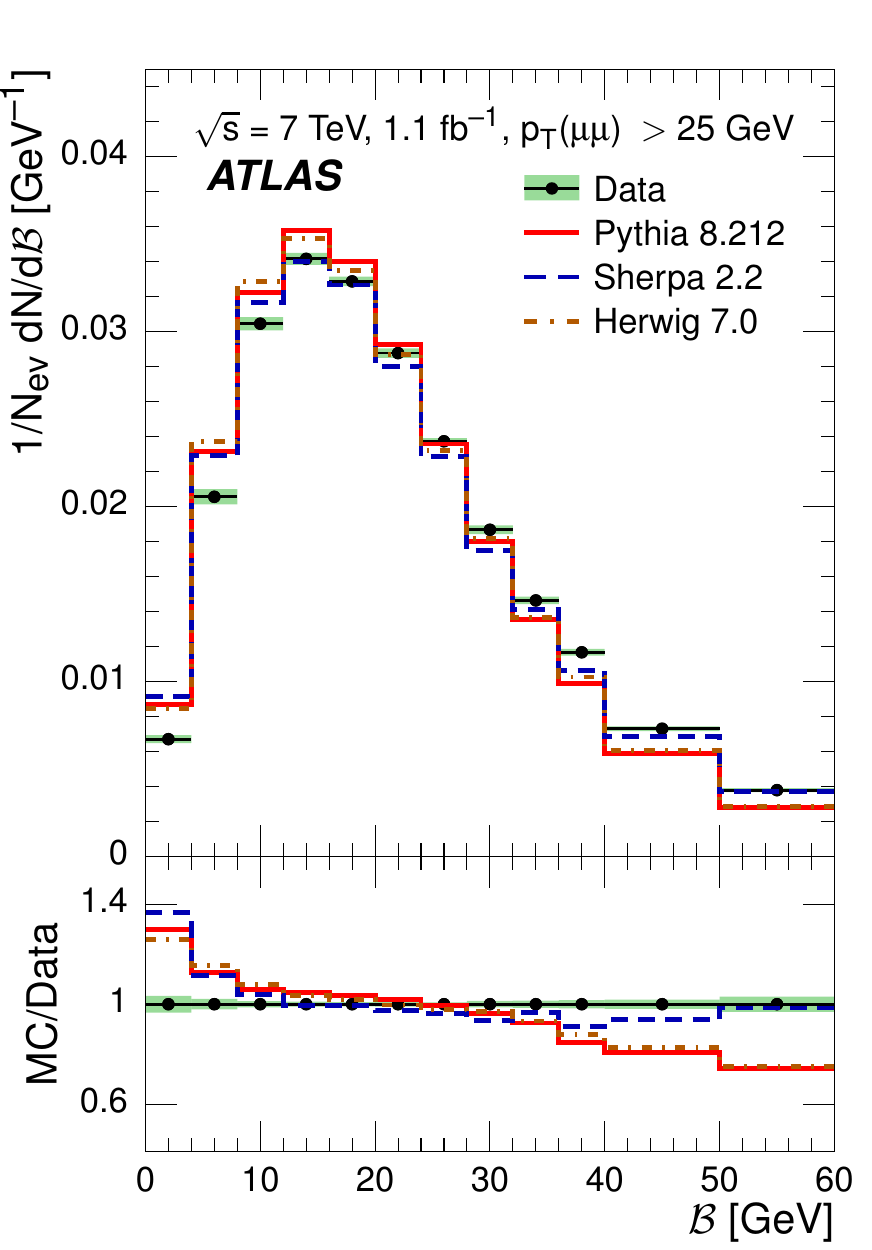}
        \label{fig:results-beamthrust-25_999-Muon-final}
    }
    \caption[Results, beam thrust, all phase-spage regions]{Beam thrust $\mathcal{B}$ distribution of charged particles
for $Z \rightarrow \mu^{+}\mu^{-}$ with statistical (error bars) and total systematic (band) uncertainties for the
four $\zptmumu$ ranges ((a): 0--6\;\GeV, (b): 6--12\;\GeV, (c): 12--25\;\GeV, (d): $\ge 25$~\GeV) compared to the predictions
from the MC generators \pythiaeight (full line), \sherpa (dashed line), and \herwigseven (dashed-dotted line).
In each subfigure, the top plot shows the observable and the bottom plot shows the ratio of the MC simulation to the data.}
    \label{fig:results-beamthrust-Muon-final}
\end{figure}
\begin{figure}[p]
    \captionsetup[subfigure]{margin=0pt, width=.42\textwidth}
    \centering
    \subfloat[Transverse thrust, $\zptmumu$: 0--6\;\GeV]{
        \includegraphics[width=.42\textwidth]{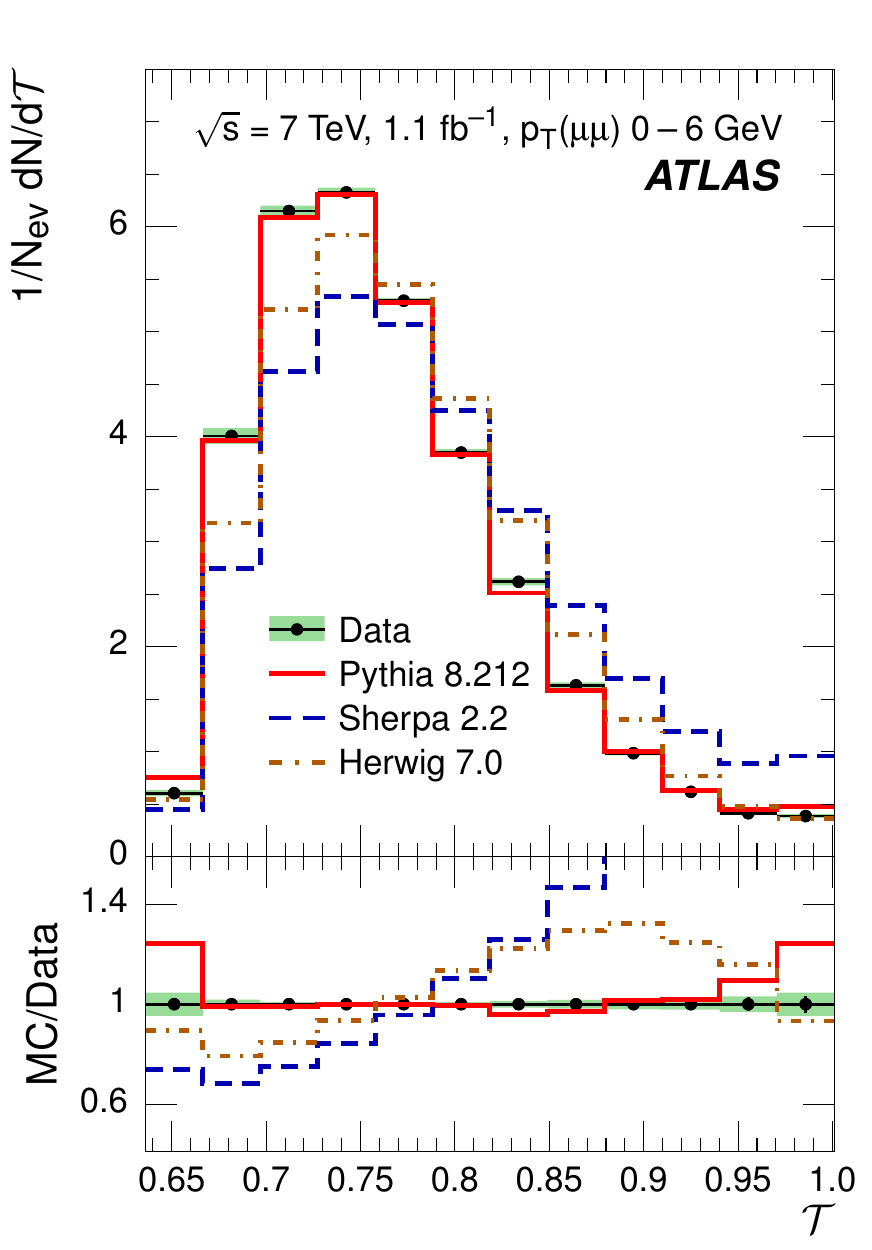}
        \label{fig:results-thrust-0_6-Muon-final}
    }
    \subfloat[Transverse thrust, $\zptmumu$: 6--12\;\GeV]{
        \includegraphics[width=.42\textwidth]{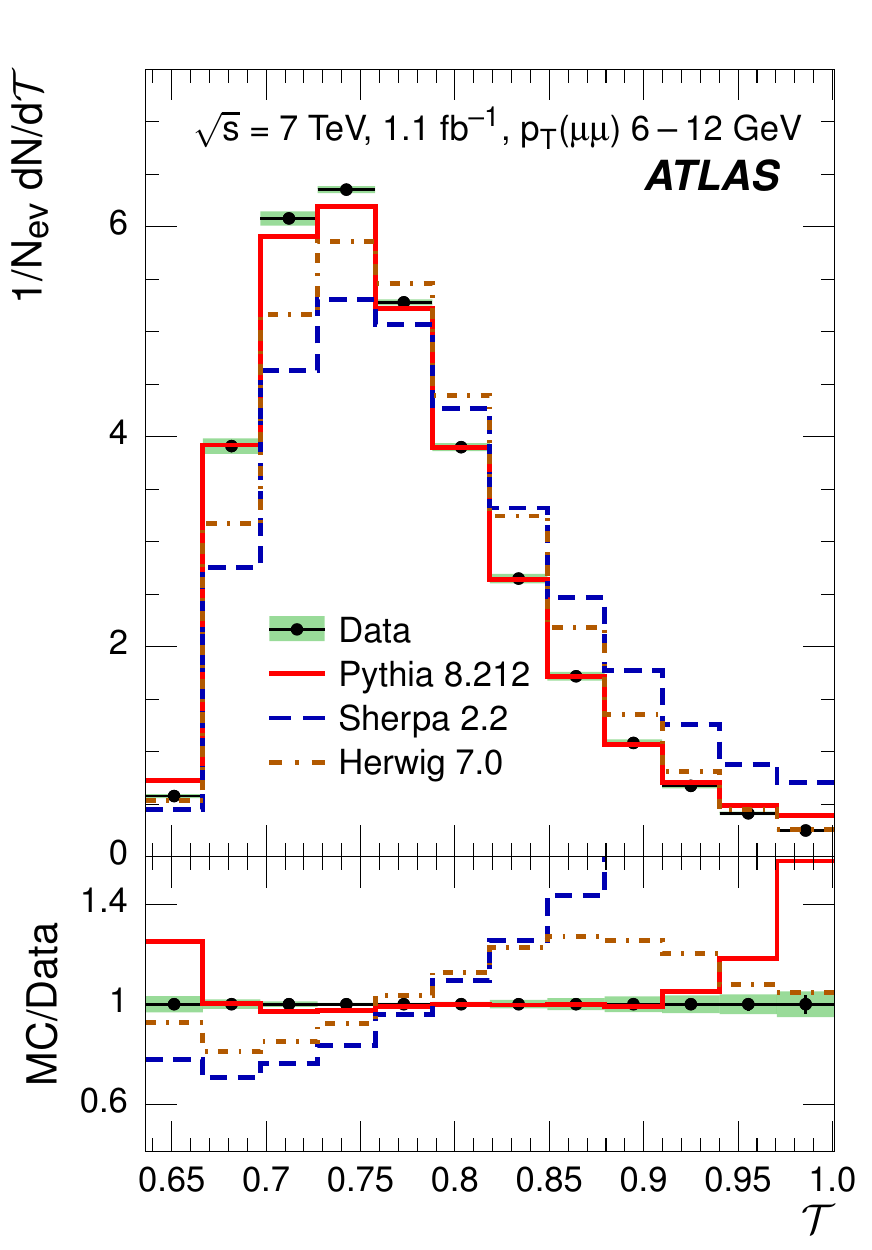}
        \label{fig:results-thrust-6_12-Muon-final}
    }
    \\
    \subfloat[Transverse thrust, $\zptmumu$: 12--25\;\GeV]{
        \includegraphics[width=.42\textwidth]{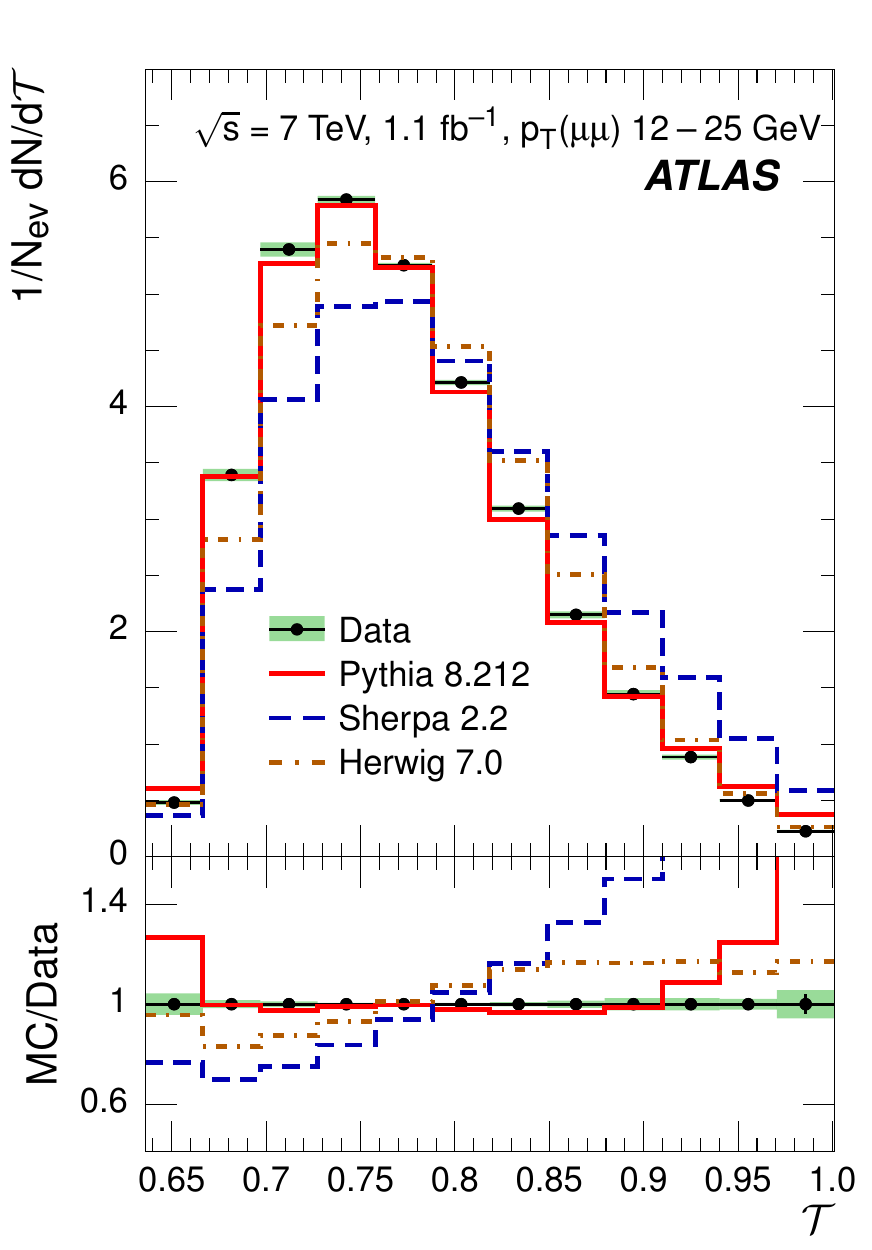}
        \label{fig:results-thrust-12_25-Muon-final}
    }
    \subfloat[Transverse thrust, $\zptmumu \ge 25$~\GeV]{
        \includegraphics[width=.42\textwidth]{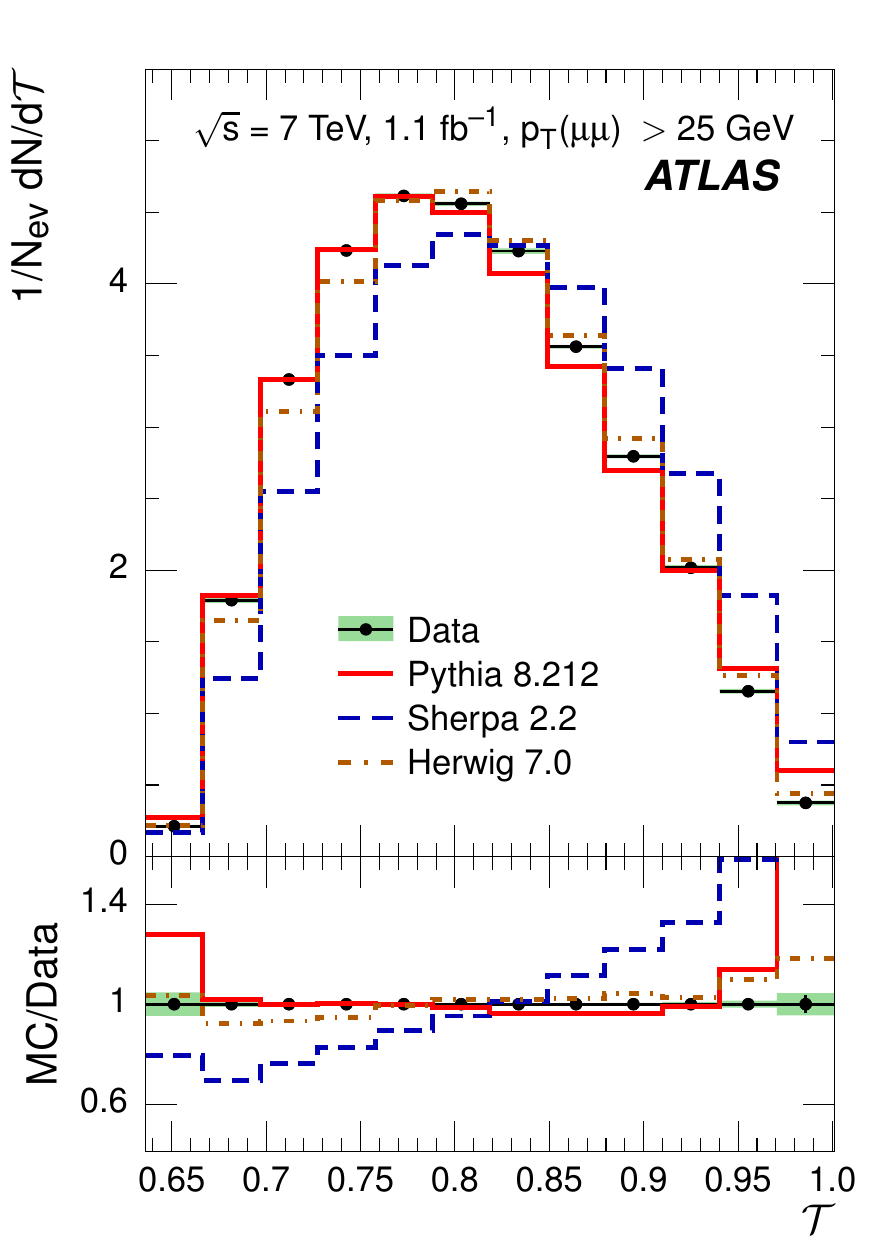}
        \label{fig:results-thrust-25_999-Muon-final}
    }
    \caption[Results, Transverse thrust, all phase-spage regions]{Transverse thrust $\mathcal{T}$ distribution of 
charged particles for $Z \rightarrow \mu^{+}\mu^{-}$ with statistical (error bars) and total systematic (band) uncertainties 
for the four $\zptmumu$ ranges ((a): 0--6\;\GeV, (b): 6--12\;\GeV, (c): 12--25\;\GeV, (d): $\ge 25$~\GeV) compared to the 
predictions from the MC generators \pythiaeight (full line), \sherpa (dashed line), and \herwigseven (dashed-dotted line).
In each subfigure, the top plot shows the observable and the bottom plot shows the ratio of the MC simulation to the data.}
    \label{fig:results-thrust-Muon-final}
\end{figure}
\begin{figure}[p]
    \captionsetup[subfigure]{margin=0pt, width=.42\textwidth}
    \centering
    \subfloat[Spherocity, $\zptmumu$: 0--6\;\GeV]{
        \includegraphics[width=.42\textwidth]{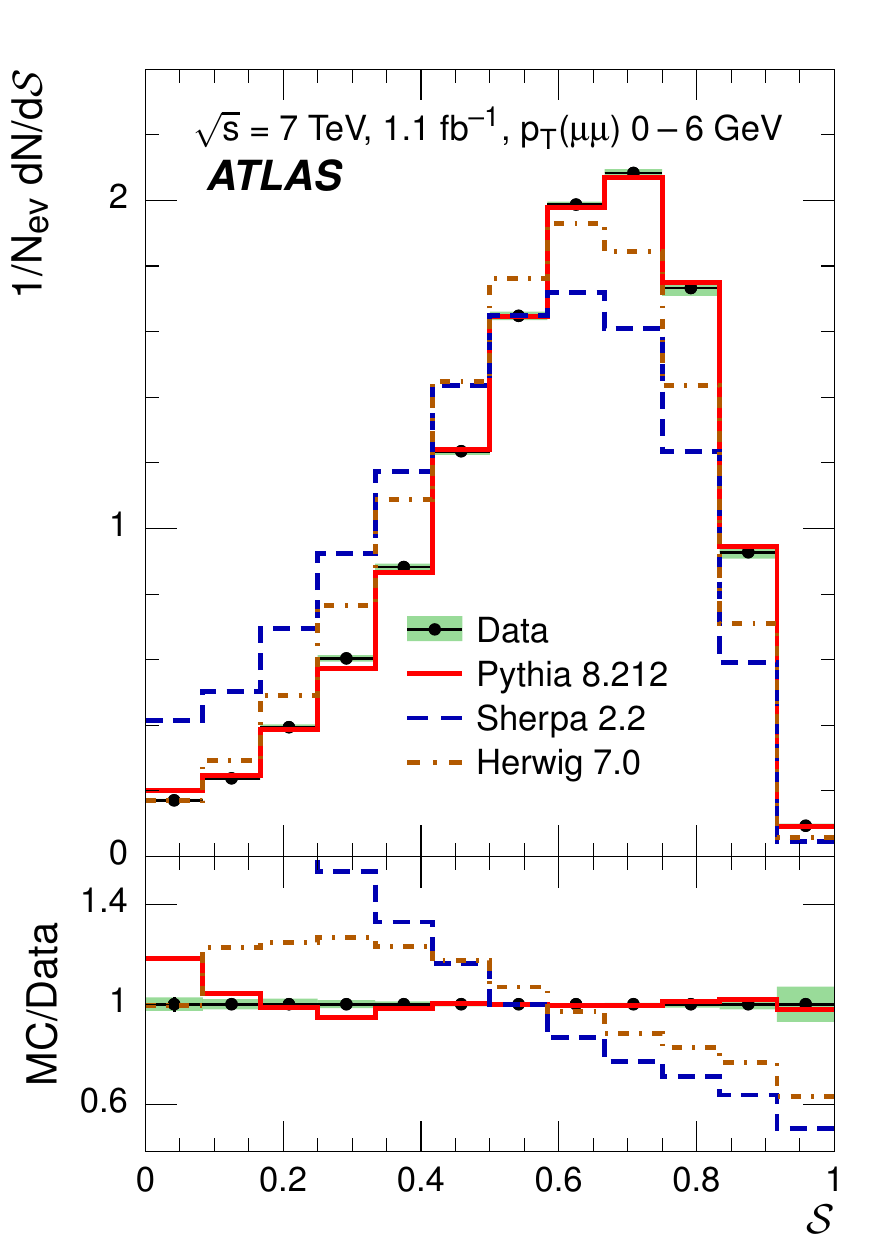}
        \label{fig:results-spherocity-0_6-Muon-final}
    }
    \subfloat[Spherocity, $\zptmumu$: 6--12\;\GeV]{
        \includegraphics[width=.42\textwidth]{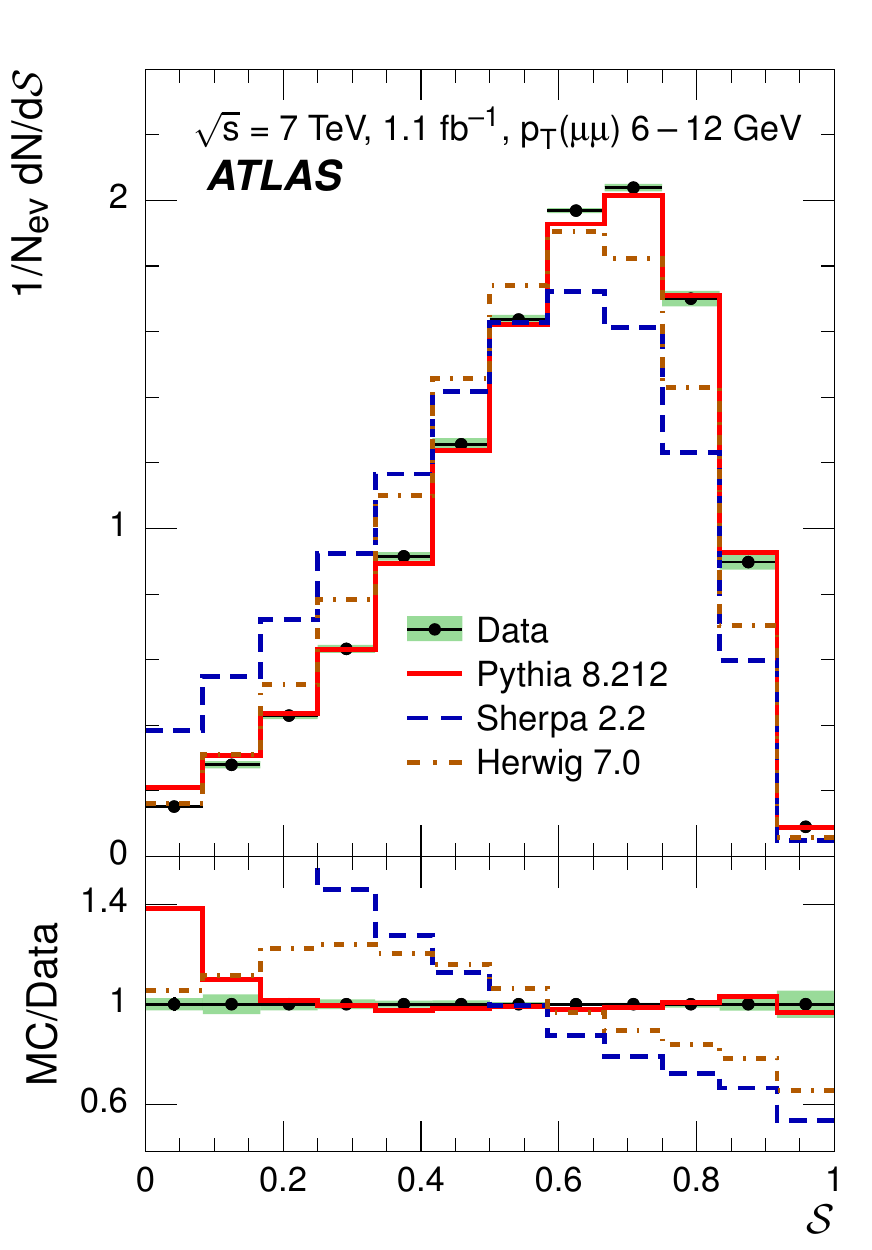}
        \label{fig:results-spherocity-6_12-Muon-final}
    }
    \\
    \subfloat[Spherocity, $\zptmumu$: 12--25\;\GeV]{
        \includegraphics[width=.42\textwidth]{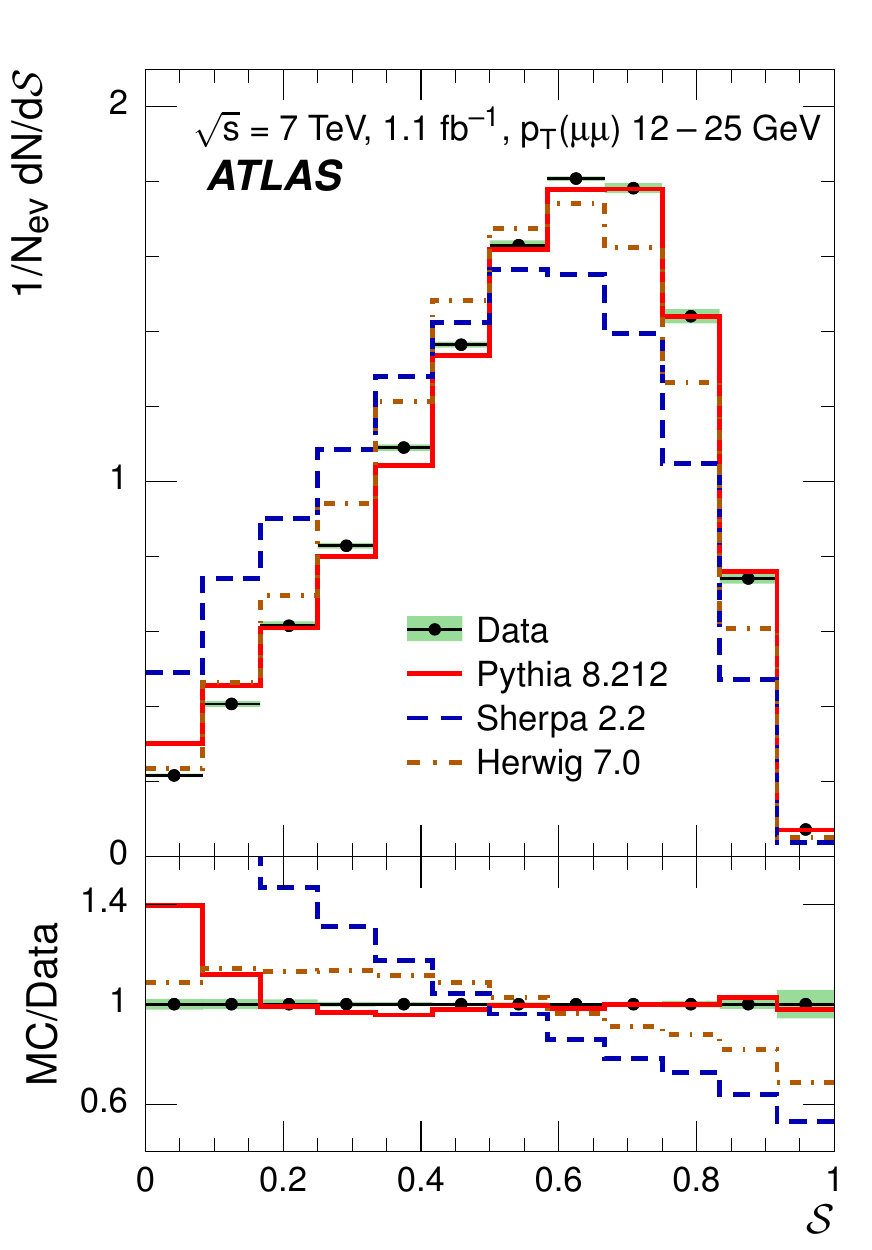}
        \label{fig:results-spherocity-12_25-Muon-final}
    }
    \subfloat[Spherocity, $\zptmumu \ge 25$~\GeV]{
        \includegraphics[width=.42\textwidth]{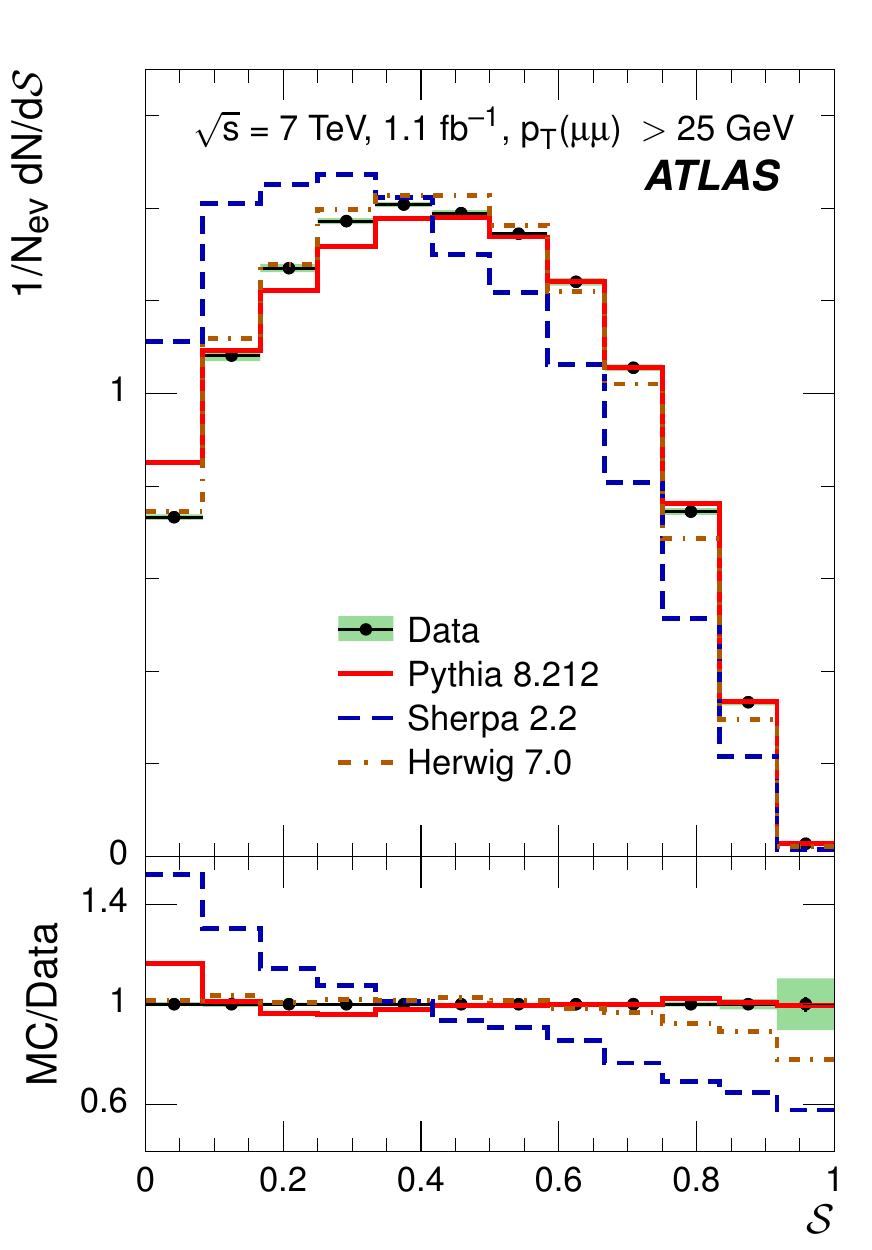}
        \label{fig:results-spherocity-25_999-Muon-final}
    }
    \caption[Results, Spherocity, all phase-spage regions]{Spherocity $\mathcal{S}$ distribution of charged particles
for $Z \rightarrow \mu^{+}\mu^{-}$ with statistical (error bars) and total systematic (band) uncertainties for the
four $\zptmumu$ ranges ((a): 0--6\;\GeV, (b): 6--12\;\GeV, (c): 12--25\;\GeV, (d): $\ge 25$~\GeV) compared to the predictions
from the MC generators \pythiaeight (full line), \sherpa (dashed line), and \herwigseven (dashed-dotted line).
In each subfigure, the top plot shows the observable and the bottom plot shows the ratio of the MC simulation to the data.}
    \label{fig:results-spherocity-Muon-final}
\end{figure}
\begin{figure}[p]
    \captionsetup[subfigure]{margin=0pt, width=.42\textwidth}
    \centering
    \subfloat[$\mathcal{F}$-parameter, $\zptmumu$: 0--6\;\GeV]{
        \includegraphics[width=.42\textwidth]{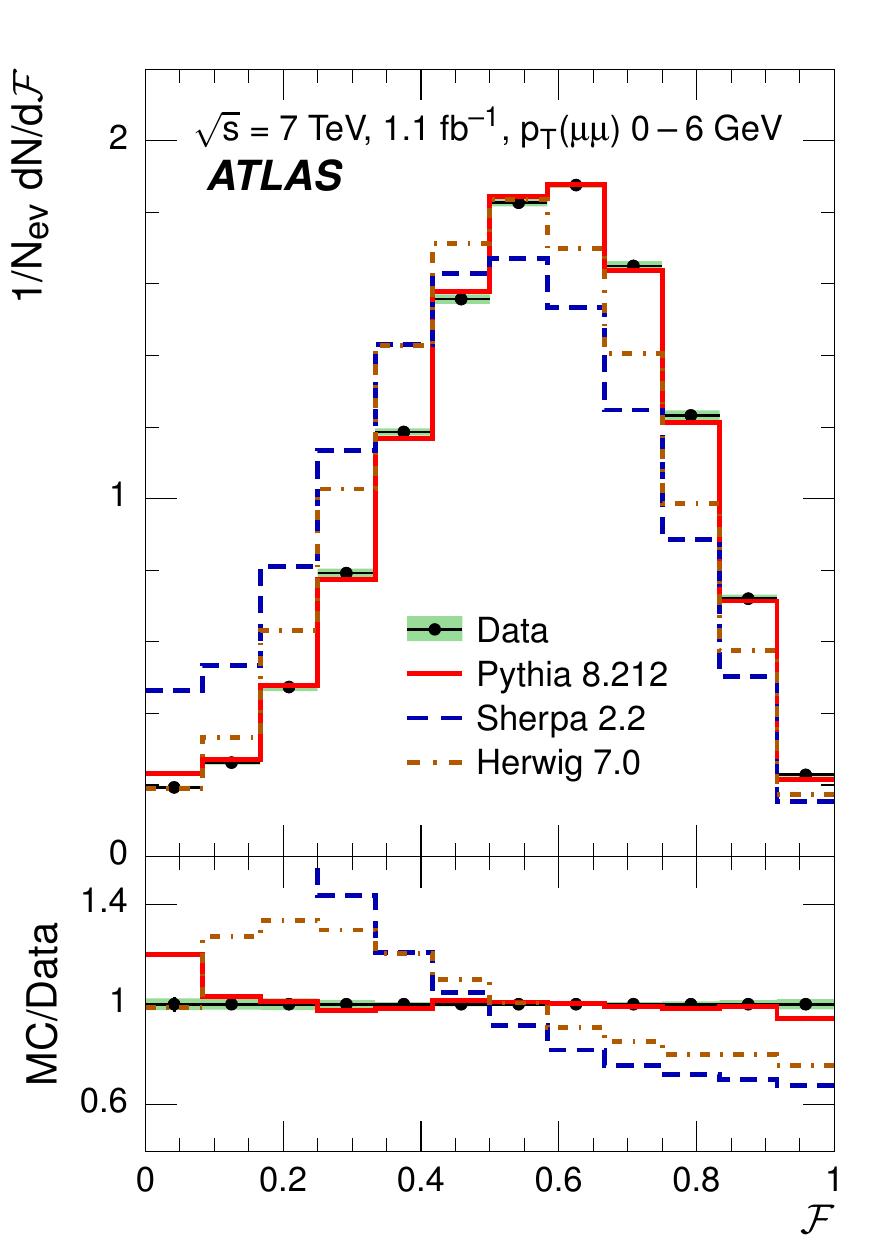}
        \label{fig:results-fparameter-0_6-Muon-final}
    }
    \subfloat[$\mathcal{F}$-parameter, $\zptmumu$: 6--12\;\GeV]{
        \includegraphics[width=.42\textwidth]{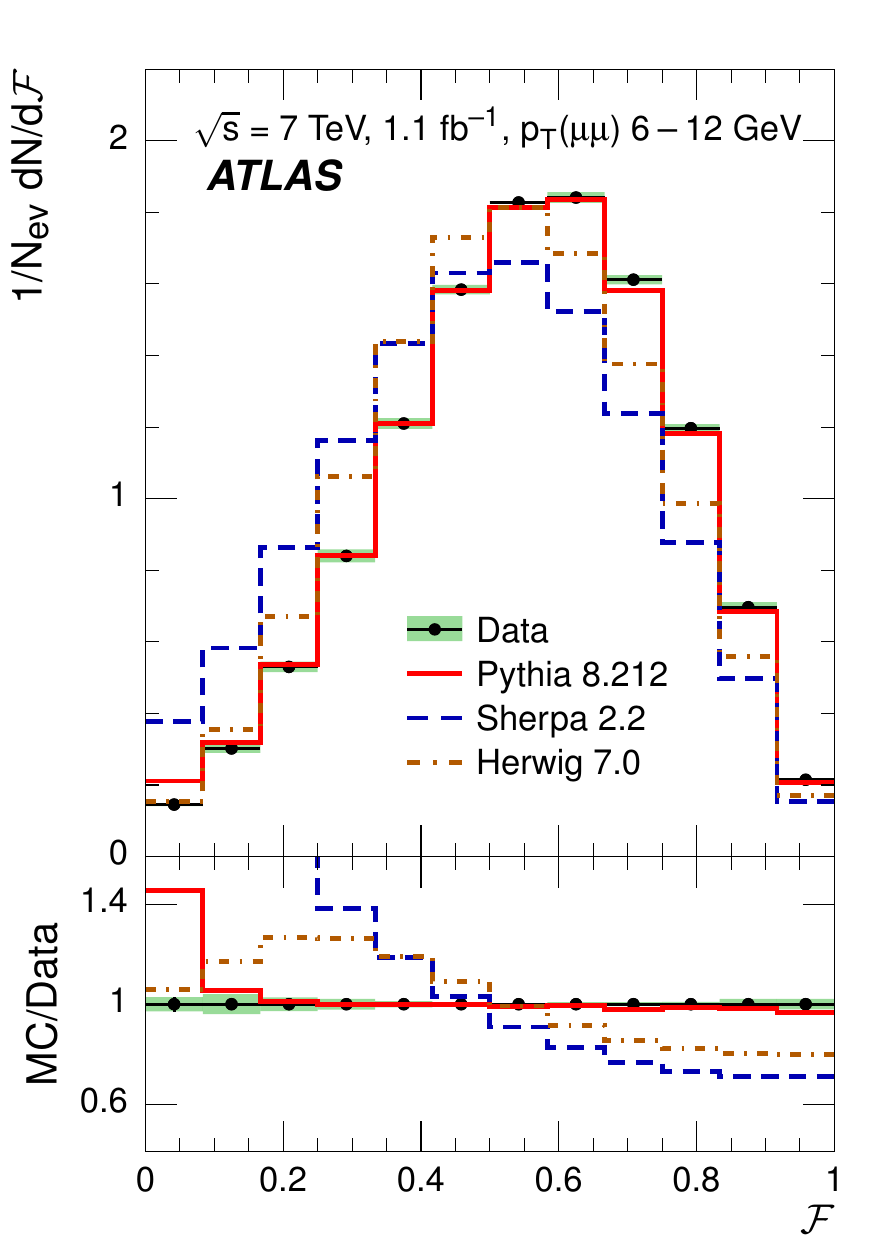}
        \label{fig:results-fparameter-6_12-Muon-final}
    }
    \\
    \subfloat[$\mathcal{F}$-parameter, $\zptmumu$: 12--25\;\GeV]{
        \includegraphics[width=.42\textwidth]{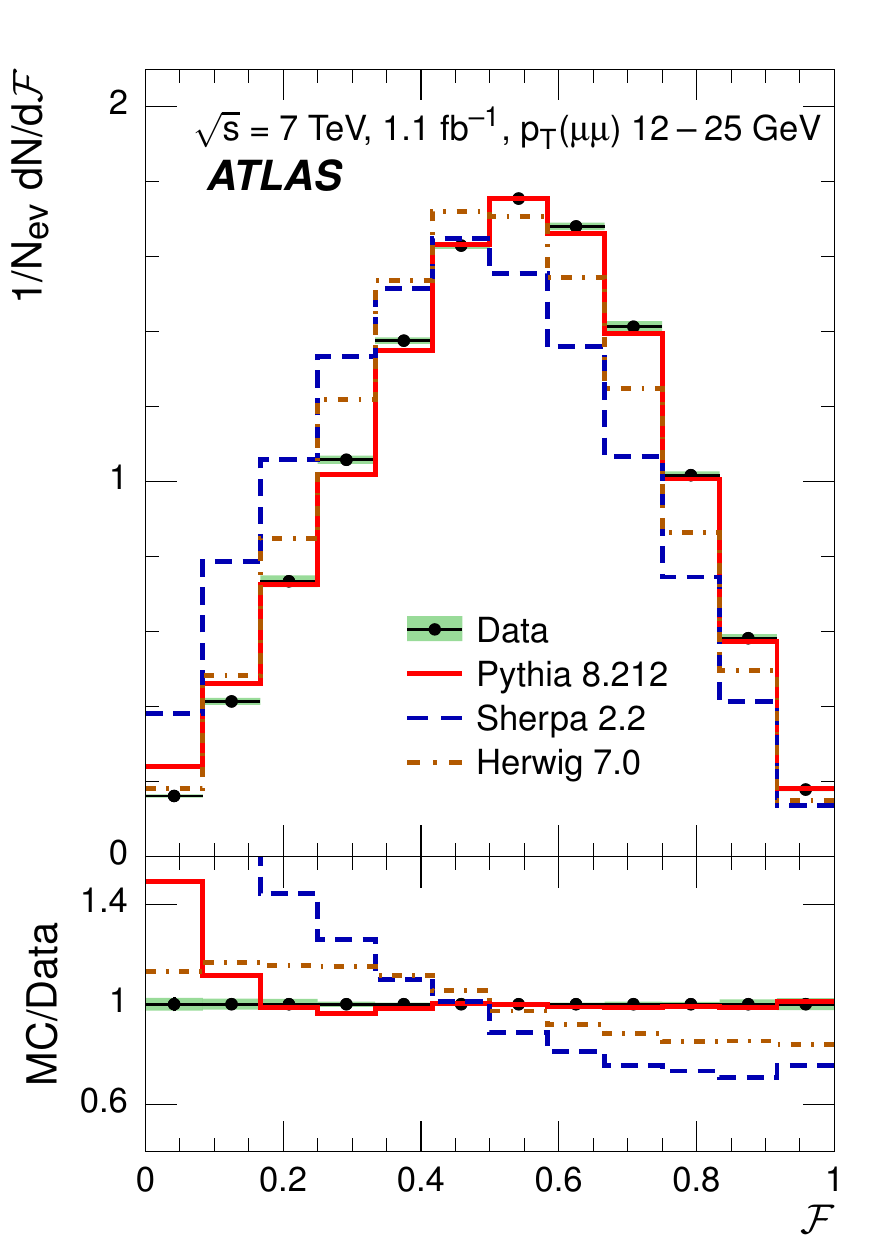}
        \label{fig:results-fparameter-12_25-Muon-final}
    }
    \subfloat[$\mathcal{F}$-parameter, $\zptmumu \ge 25$~\GeV]{
        \includegraphics[width=.42\textwidth]{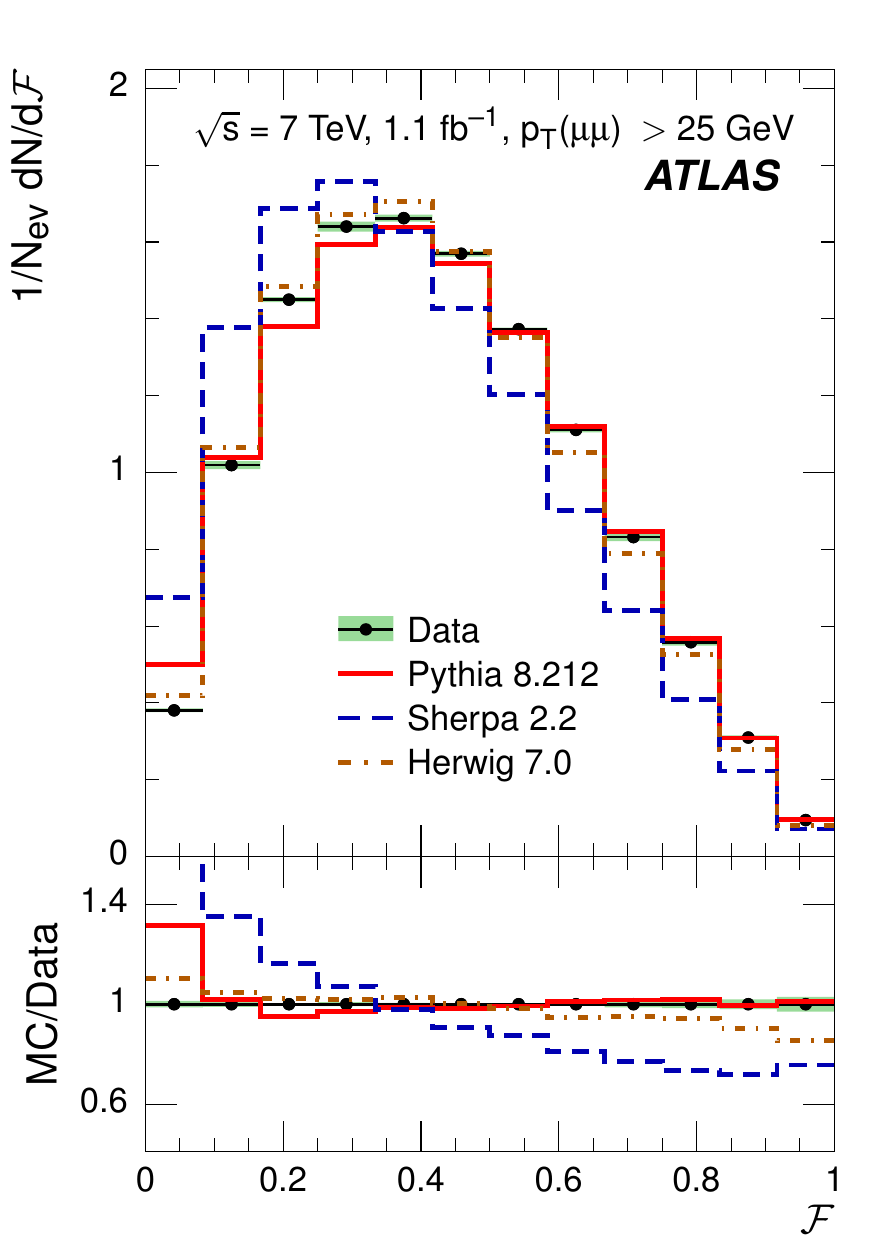}
        \label{fig:results-fparameter-25_999-Muon-final}
    }
    \caption[Results, $\mathcal{F}$-parameter, all phase-spage regions]{$\mathcal{F}$-parameter distribution of charged particles
for $Z \rightarrow \mu^{+}\mu^{-}$ with statistical (error bars) and total systematic (band) uncertainties for the
four $\zptmumu$ ranges ((a): 0--6\;\GeV, (b): 6--12\;\GeV, (c): 12--25\;\GeV, (d): $\ge 25$~\GeV) compared to the predictions
from the MC generators \pythiaeight (full line), \sherpa (dashed line), and \herwigseven (dashed-dotted line).
In each subfigure, the top plot shows the observable and the bottom plot shows the ratio of the MC simulation to the data.}
    \label{fig:results-fparameter-Muon-final}
\end{figure}

\noindent
Figures~\ref{fig:results-nch-Elec-final}--\ref{fig:results-fparameter-Elec-final}
(\ref{fig:results-nch-Muon-final}--\ref{fig:results-fparameter-Muon-final})
show the individual event-shape observables for the electron (muon) channel compared 
to predictions obtained with the most recent versions of three different MC generators 
as described in Section~\ref{sec:montecarlo}: \sherpa version 2.2.0, \herwigseven 
version 7.0, and \pythiaeight version 8.212. In general, \pythiaeight and \herwigseven 
agree better with the data than does \sherpa.

\noindent
The $\zpt<6$~{\GeV} bin is expected to be characterised by low jet activity from
the hard matrix element and hence should be particularly sensitive to UE characteristics. 
In this case, \pythiaeight shows very good agreement with the data in the event-shape 
observables that are not very sensitive to the number of charged particles 
($\mathcal{T}$, $\mathcal{S}$, and $\mathcal{F}$-parameter). 
The observables that depend explicitly on the number of charged particles (\nch, $\sumpt$, $\mathcal{B}$)
are less well described, with none of the generators succeeding fully. 
In this case, the best agreement is observed for \herwigseven while \pythiaeight still 
performs better than \sherpa. Low {\nch} and $\sumpt$ values represent a challenging 
region for all three generators: 
while \pythiaeight and \sherpa overestimate the data, \herwigseven significantly 
underestimates the measurements. This region might be particularly sensitive to the 
way beam-remnant interactions are modelled in the MC generators. Similar observations 
can be made for $\zpt$ ranges 6--12\;{\GeV} and 12--25\;{\GeV}.
At low values of $\mathcal{B}$, the observable in which tracks with
larger $|\eta^\text{trk}|$ values contribute less to the sum of the 
track transverse momenta, better agreement of the generator predictions 
with the data is observed than at low $\sumpt$.

\noindent
At $\zpt \ge 25$~{\GeV} the event is expected to contain at least one jet of high transverse
momentum recoiling against the $Z$ boson, which is expected to be well described by 
the hard matrix element. In this case, one still observes significant deviations of the MC generators 
from the measurement, where, depending on the observable, either \herwigseven or \pythiaeight shows 
in general the best agreement. However, all three generators show better agreement with data compared 
to the $\zpt<6$~{\GeV} range. 

\noindent
The observed deviations of MC predictions from the measured observables reveal that MC parameters
tuned to presently measured observables fail to describe more detailed characteristics of the 
UE modelling and the level of disagreement depends on the generator under consideration. 
It has to be seen whether these discrepancies can be reduced by a refined parameter tuning 
when also including the event-shape observables in the tuning or whether further developments 
in the UE modelling are required.

\FloatBarrier
% All figures and tables should appear before the summary and conclusion
% The package placeins provides the macro \FloatBarrier to achieve this
% \FloatBarrier

%-------------------------------------------------------------------------------
\section{Conclusion}
\label{sec:conclusion}
%-------------------------------------------------------------------------------

\noindent
In this paper, event-shape observables sensitive to the underlying event were 
measured in 1.1\;${\rm fb}^{-1}$ integrated luminosity of proton--proton collisions 
collected with the ATLAS detector at the LHC at a centre-of-mass energy of 7 TeV. 
Events containing an oppositely charged electron or muon pair with an invariant mass 
close to the $Z$-boson mass were selected, and the charged particle multiplicity, 
mean transverse momentum, beam thrust, transverse thrust, spherocity, and 
$\mathcal{F}$-parameter were measured, excluding the particles from the $Z$-boson decay.

\noindent
The measured observables were corrected for the effect of pile-up and multijet background,
and then for contributions from non-primary particles, detector efficiency, and resolution 
effects using an unfolding technique. 
The resulting distributions are presented in different regions of the $Z$-boson 
transverse momentum and compared to predictions of the MC event generators \pythiaeight, 
\herwigseven and \sherpa. These comparisons reveal significant deviations of the \sherpa 
predictions from the measured observables. Depending on the observable under 
consideration and the transverse momentum of the $Z$ boson, the data are in much better 
agreement with the \pythiaeight and \herwigseven predictions than with \sherpa. 

\noindent
Typically, all three Monte Carlo generators provide predictions that are in better 
agreement with the data at high $Z$-boson transverse momenta than at low $Z$-boson 
transverse momenta and for the observables that are less sensitive to the number of charged 
particles in the event (transverse thrust, spherocity, and $\mathcal{F}$-parameter). 
The Monte Carlo generator predictions show significant differences from the data at low values 
of \nch, $\sumpt$, and beam thrust in certain regions of the $Z$-boson transverse momentum.
The measured event-shape observables are therefore expected to provide valuable insight
into the phenomenon of the underlying event and new information for the tuning of current
underlying-event models and the development of new models for high-precision
measurements to be performed at the LHC at $\sqrt{s}=13$ TeV.

%-------------------------------------------------------------------------------
\section*{Acknowledgements}
%-------------------------------------------------------------------------------

% Acknowledgements for papers with collision data
% Version 15-Apr-2016

% Standard acknowledgements start here
%----------------------------------------------
We thank CERN for the very successful operation of the LHC, as well as the
support staff from our institutions without whom ATLAS could not be
operated efficiently.

We acknowledge the support of ANPCyT, Argentina; YerPhI, Armenia; ARC, Australia; BMWFW and FWF, Austria; ANAS, Azerbaijan; SSTC, Belarus; CNPq and FAPESP, Brazil; NSERC, NRC and CFI, Canada; CERN; CONICYT, Chile; CAS, MOST and NSFC, China; COLCIENCIAS, Colombia; MSMT CR, MPO CR and VSC CR, Czech Republic; DNRF and DNSRC, Denmark; IN2P3-CNRS, CEA-DSM/IRFU, France; GNSF, Georgia; BMBF, HGF, and MPG, Germany; GSRT, Greece; RGC, Hong Kong SAR, China; ISF, I-CORE and Benoziyo Center, Israel; INFN, Italy; MEXT and JSPS, Japan; CNRST, Morocco; FOM and NWO, Netherlands; RCN, Norway; MNiSW and NCN, Poland; FCT, Portugal; MNE/IFA, Romania; MES of Russia and NRC KI, Russian Federation; JINR; MESTD, Serbia; MSSR, Slovakia; ARRS and MIZ\v{S}, Slovenia; DST/NRF, South Africa; MINECO, Spain; SRC and Wallenberg Foundation, Sweden; SERI, SNSF and Cantons of Bern and Geneva, Switzerland; MOST, Taiwan; TAEK, Turkey; STFC, United Kingdom; DOE and NSF, United States of America. In addition, individual groups and members have received support from BCKDF, the Canada Council, CANARIE, CRC, Compute Canada, FQRNT, and the Ontario Innovation Trust, Canada; EPLANET, ERC, FP7, Horizon 2020 and Marie Sk{\l}odowska-Curie Actions, European Union; Investissements d'Avenir Labex and Idex, ANR, R{\'e}gion Auvergne and Fondation Partager le Savoir, France; DFG and AvH Foundation, Germany; Herakleitos, Thales and Aristeia programmes co-financed by EU-ESF and the Greek NSRF; BSF, GIF and Minerva, Israel; BRF, Norway; Generalitat de Catalunya, Generalitat Valenciana, Spain; the Royal Society and Leverhulme Trust, United Kingdom.

The crucial computing support from all WLCG partners is acknowledged
gratefully, in particular from CERN and the ATLAS Tier-1 facilities at
TRIUMF (Canada), NDGF (Denmark, Norway, Sweden), CC-IN2P3 (France),
KIT/GridKA (Germany), INFN-CNAF (Italy), NL-T1 (Netherlands), PIC (Spain),
ASGC (Taiwan), RAL (UK) and BNL (USA) and in the Tier-2 facilities
worldwide.
%----------------------------------------------

%
%The \texttt{atlaslatex} package contains the acknowledgements that were valid 
%at the time of the release you are using. These can be found in the
%\texttt{acknowledgements} subdirectory.
%When your ATLAS papers or CONF note is ready to be published,
%download the latest set of acknowledgements from:\\
%\url{https://twiki.cern.ch/twiki/bin/view/AtlasProtected/PubComAcknowledgements}
%
%The supporting notes for the analysis should also contain a list of contributors.
%This information should usually be included in \texttt{mydocument-metadata.tex}.
%The list should be printed either here or before the table of contents.

\printbibliography

%%-------------------------------------------------------------------------------
%\clearpage
%\appendix
%\part*{Appendix}
%\addcontentsline{toc}{part}{Appendix}
%\input{chapters/appendix}
%%-------------------------------------------------------------------------------

%In a paper, an appendix is used for technical details that would otherwise disturb the flow of the paper.
%Such an appendix should be printed before the Bibliography.

%-------------------------------------------------------------------------------
% If you use biblatex and either biber or bibtex to process the bibliography 
% just say \printbibliography here
%\printbibliography
% If you want to use the traditional BibTeX you need to use the syntax below.
%\bibliographystyle{bibtex/bst/atlasBibStyleWoTitle}
%\bibliography{mydocument,bibtex/bib/ATLAS}
%-------------------------------------------------------------------------------

%-------------------------------------------------------------------------------
% Print the list of contributors to the analysis
% The argument gives the fraction of the text width used for the names
%-------------------------------------------------------------------------------
\clearpage
%\PrintAtlasContribute{0.30}

%-------------------------------------------------------------------------------
\clearpage
%\appendix
%\par*{Auxiliary material}
%\addcontentsline{toc}{part}{Auxiliary material}
%-------------------------------------------------------------------------------

 \newpage % ATLAS Collaboration author list
% Data extracted on 20-Jan-2016 for paper reference STDM-2014-07
%\documentclass[11pt]{article}
%\usepackage{a4wide}\begin{document}
\begin{flushleft}
{\Large The ATLAS Collaboration}

\bigskip

G.~Aad$^\textrm{\scriptsize 86}$,
B.~Abbott$^\textrm{\scriptsize 113}$,
J.~Abdallah$^\textrm{\scriptsize 151}$,
O.~Abdinov$^\textrm{\scriptsize 11}$,
B.~Abeloos$^\textrm{\scriptsize 117}$,
R.~Aben$^\textrm{\scriptsize 107}$,
M.~Abolins$^\textrm{\scriptsize 91}$,
O.S.~AbouZeid$^\textrm{\scriptsize 137}$,
N.L.~Abraham$^\textrm{\scriptsize 149}$,
H.~Abramowicz$^\textrm{\scriptsize 153}$
% ...
\bigskip
\\
% ...
$^{11}$ Institute of Physics, Azerbaijan Academy of Sciences, Baku, Azerbaijan\\
% ...
$^{86}$ CPPM, Aix-Marseille Universit{\'e} and CNRS/IN2P3, Marseille, France\\
% ...
$^{91}$ Department of Physics and Astronomy, Michigan State University, East Lansing MI, United States of America\\
% ...
$^{107}$ Nikhef National Institute for Subatomic Physics and University of Amsterdam, Amsterdam, Netherlands\\
% ...
$^{113}$ Homer L. Dodge Department of Physics and Astronomy, University of Oklahoma, Norman OK, United States of America\\
% ...
$^{117}$ LAL, Univ. Paris-Sud, CNRS/IN2P3, Universit{\'e} Paris-Saclay, Orsay, France\\
% ...
$^{137}$ Santa Cruz Institute for Particle Physics, University of California Santa Cruz, Santa Cruz CA, United States of America\\
% ...
$^{149}$ Department of Physics and Astronomy, University of Sussex, Brighton, United Kingdom\\
% ...
$^{151}$ Institute of Physics, Academia Sinica, Taipei, Taiwan\\
% ...
$^{153}$ Raymond and Beverly Sackler School of Physics and Astronomy, Tel Aviv University, Tel Aviv, Israel\\
% ...
\end{flushleft}

%\end{document}
% Created with xml2latex.py

\end{document}